\documentclass[usenatbib]{mn2e}
\usepackage{epsfig,times,natbib,lscape,aas_macros}
\usepackage{aas_macros}

\title[FR-I environments with {\it XMM-Newton}]{An {\it XMM-Newton} study of the environments, particle content and impact of low-power radio galaxies}
\author[J. H. Croston et al.]  {J. H. Croston$^{1}$\thanks{Email:
j.h.croston@herts.ac.uk}, M.J. Hardcastle$^{1}$,
M. Birkinshaw$^2$, D.M. Worrall$^{2}$, and R.A. Laing$^{3}$\\$^1$ School of Physics, Astronomy and
Mathematics, University of Hertfordshire, College Lane, Hatfield,
Hertfordshire AL10 9AB\\$^2$ H. H. Wills Physics Laboratory, University of
Bristol, Tyndall Avenue, Bristol BS8 1TL\\$^{3}$ European Southern
Observatory,  Karl-Schwarzschild-Stra\ss e 2, D-85748
Garching-bei-M\"unchen, Germany \\}
 
\pagerange{\pageref{firstpage}--\pageref{lastpage}}
\pubyear{2008}
\begin{document}

\maketitle

\label{firstpage}

\begin{abstract}

We present a detailed study of the environments of a sample of nine
low-power (Fanaroff \& Riley type I) radio galaxies, based on new and
archival {\it XMM-Newton} observations. We report new detections of
group-scale environments around three radio galaxies, 3C\,296,
NGC\,1044 and 3C\,76.1. As with previous studies, we find that FR-I
radio galaxies inhabit group environments ranging over nearly two
orders of magnitude in bolometric X-ray luminosity; however, we find
no evidence for a tight relationship between large-scale X-ray
environment and radio-source properties such as size, radio
luminosity, and axial ratio. This leads us to conclude that
radio-source evolution cannot be determined entirely by the global
properties of the hot gas. We confirm earlier work showing that
equipartition internal pressures are typically lower than the external
pressures acting on the radio lobes, so that additional non-radiating
particles must be present or the lobes must be magnetically dominated.
We present the first direct observational evidence that entrainment
may provide this missing pressure, in the form of a relationship
between radio-source structure and apparent pressure imbalance.
Finally, our study provides further support for the presence of an
apparent temperature excess in radio-loud groups compared to the group
population as a whole. For five of eight temperature excesses, the
energy required to inflate the radio lobes is comparable to the energy
required to produce this excess by heating of the group gas; however,
in three cases the current radio source appears too weak to produce
the temperature excess. It remains unclear whether the temperature
excess can be directly associated with the current phase of AGN
activity, or whether it is instead either a signature of previous AGN
activity or simply an indicator of the particular set of group
properties most conducive to the growth of an FR-I radio galaxy.

\end{abstract}


\section{Introduction}

The dynamics of low-power (FR-I: \citealt{FR74}), twin-jet radio
galaxies are thought to be controlled by interactions between the
initially highly relativistic jets and the surrounding hot-gas medium,
which act to decelerate the jets significantly on scales of typically
a few kpc \citep[e.g.][]{bic94,lb02a}. On scales of hundreds of kpc,
low-power radio galaxies exhibit a wide range of morphologies, from
narrowly collimated plumes to large, rounded lobes
\citep[e.g.][]{lai93}. In order to understand how such structures are
produced, and to investigate the evolution of FR-I sources on large
scales, it is necessary to have good constraints on the hot-gas
environments that confine their radio lobes. {\it ROSAT} observations
first established that the brightest and most well-studied FR-I radio
galaxies usually reside in group-scale environments
\citep[e.g.][]{wb94,kb99,wb00}. With {\it Chandra} and {\it
XMM-Newton} it is possible to study radio-galaxy environments and
jet/environment interactions in more detail
\citep[e.g.][]{bla01,c03b,fab03}; however, while there have been many
studies of the apparent impact of FR-I sources in clusters, to date
there have been few detailed {\it Chandra} and {\it XMM-Newton}
studies of the environmental properties of nearby, well-studied FR-I
radio galaxies. The role of environmental properties such as
group/cluster richness, gas density and temperature distribution in
determining large-scale radio structure therefore remains uncertain.

Another key uncertainty both in modelling radio-galaxy dynamics and in
establishing their energetic impact is the lack of constraints on
particle content in FR-I radio jets and lobes. In the case of powerful
(FR-II) radio galaxies, two lines of evidence suggest that synchrotron
minimum-energy estimates are roughly valid in the lobes: (1) for cases
where external pressure measurements are available, they are in
reasonable agreement with the internal minimum pressure
\citep[e.g.][]{h02b,c04,bel04}; (2) measurements of inverse-Compton
emission from the lobes of FR-II radio galaxies suggest that their
electron energy densities are close to the value for equipartition
with the magnetic field in the absence of an energetically important
proton population \citep[e.g.][]{c05b,ks05}. However, in low-power
(FR-I) radio galaxies, it has been known for several decades that a
relativistic electron population in equipartition with the magnetic
field cannot provide sufficient pressure to balance that of the
external medium \citep[e.g.][]{mor88,hwb98,wb00}. In \citet{c03b}, we
showed for two FR-Is that the additional pressure cannot be due solely
to relativistic-electron dominance, as this would result in detectable
X-ray inverse-Compton emission, which is not seen, while the presence
of sufficient thermal gas at the temperature of the environment was
also ruled out by the presence of deficits in X-ray surface brightness
at the positions of radio lobes. This is also the case for the sample
of cluster-centre FR-I radio sources recently studied by
\citet{dun05}. Relativistic protons intrinsic to the jets would
require extremely large ratios of proton to electron energy in order
to provide the missing pressure -- such high ratios are implausible as
they would lead to overpressured jets on kpc scales. The most plausible
origin of the required additional pressure is either material that has
been entrained and heated, or magnetic dominance of the lobes;
however, these scenarios are both difficult to test.

Despite these uncertainties about radio-source energetics, our
understanding of the impact of radio galaxies on their environments
has improved dramatically over the past few years. Direct evidence for
heating by FR-I radio galaxies has been found in a number of
individual systems \citep[e.g.][]{kra03,c07,for07}, and statistical
studies have shown that radio-source heating may have a significant
impact in the outer regions of galaxy groups \citep{c05a,jet07}.
Progress in hydrodynamical modelling has shown that a range of heating
mechanisms may be capable of heating the central regions of clusters
to prevent cooling flows \citep[e.g.][]{ba03,bk01,rey02} and
semi-analytical galaxy formation models that take into account
radio-source heating appear to be able to reproduce the properties of
the nearby galaxy population more accurately than has previously been
possible \citep{cro06,bow06}. The growing acceptance that AGN, and
particularly low-power (FR-I) radio galaxies, play an essential role
in galaxy and structure evolution means that it is crucial to
understand which types of radio structures form under what
environmental conditions, and how the environmental impact of these
different structures varies.

In this paper, we present new {\it XMM-Newton} observations of a
sample of low-power (FR-I) radio galaxies, which range in morphology
from tailed or plumed sources to those with lobes or bridges. We then
combine this sample with previous observations and archive data to
investigate the environmental properties of the nearby FR-I
radio-galaxy population, to constrain their particle content and
dynamics, and to study their environmental impact.

Throughout the paper we use a cosmology in which $H_0 = 70$ km
s$^{-1}$ Mpc$^{-1}$, $\Omega_{\rm m} = 0.3$ and $\Omega_\Lambda =
0.7$. Spectral indices $\alpha$ are defined in the sense $S_{\nu}
\propto \nu^{-\alpha}$. Errors quoted on parameters derived from
model fitting correspond to the $1\sigma$ (68 per cent) confidence
range for one parameter of interest, except where otherwise stated.

\section{Data analysis}
\label{data}
We carried out {\it XMM-Newton} observations of a sample of five
low-power (FR-I) radio galaxies, 3C\,296, 3C,76.1, NGC\,1044, 3C\,31
and NGC\,315 to follow up our earlier studies of 3C\,66B, 3C\,449 and
NGC\,6251 \citep{c03b,ev05}. In this section we describe the data
analysis for the five new observations, plus one archive observation
of NGC\,4261 [the AGN emission detected in this observation has been
described by \citet{sam03,gli03} and the extended emission was
discussed briefly in \citet{c05a}]. In Section~\ref{disc} we combine
these results with our previous analysis of 3C\,66B, 3C\,449, and
NGC\,6251 (\citealt{c03b} and \citealt{ev05}). Table~\ref{obs} lists
the observational details for the nine {\it XMM-Newton} observations
discussed in the present paper. For three of the radio galaxies, 3C\,296,
NGC\,315 and NGC\,6251 we also made use of previously published {\it
Chandra} radial surface brightness profiles to improve our constraints
on the inner gas distributions. Details of these profiles are given in
Section~\ref{res}.

\subsection{Radio data}

All of the radio galaxies in our sample apart from NGC\,1044 are
well-studied in the radio, with maps in the literature at multiple
frequencies. Table~\ref{radiomaps} lists the radio maps used in the
figures and analysis that follows.

As existing low-frequency VLA observations of NGC\,1044 did not fully
sample its large-scale structure, we observed it with the VLA at 20 cm
to supplement existing archival data. In combination with these
archival data, the new observations were used to produce a
good-quality map of of the large-scale radio structure for comparison
with X-ray images. The VLA observations were carried out on 2004 March
19 in C configuration and 2004 July 17 in D configuration. The new
data, together with an short archival observation in the C
configuration (AH276) were externally calibrated in the standard
manner using {\sc aips}. Each dataset was then imaged, cleaned and
self-calibrated individually. The two C-configuration datasets were
combined and imaged after first self-calibrating the archive dataset
using a model from the new observations. The D-configuration data were
then self-calibrated using the combined C-configuration model.
Finally, the C- and D-configuration data were combined with
appropriate weights and used to produce the images shown below.

\subsection{{\it XMM-Newton} data}
\label{proc}
The data for the five new observations and one archival dataset not
previously analysed were reprocessed using the {\it XMM-Newton} SAS
version 6.0.0, and the latest calibration files from the {\it
XMM-Newton} website. The pn data were filtered to include only single
and double events (PATTERN $\leq 4$), and FLAG==0, and the MOS data
were filted according to the standard flag and pattern masks (PATTERN
$\leq 12$ and \#XMMEA\_EM, excluding bad columns and rows).

Several of the datasets were badly affected by background flares, and
so GTI filtering was applied in order to ensure that accurate
measurements could be made in low surface-brightness regions.
Filtering was carried out by applying a $\pm 20$ per cent clip to a
lightcurve in the energy range where the effective area for X-rays is
negligible (10 -- 12 keV for MOS, 12 -- 14 keV for pn). In one case,
NGC\,315, initial analysis revealed significant contamination from
soft proton flares, and so for this source a second stage of filtering
was applied using the full energy range of 0.3 -- 10 keV and an
off-source region, so as to remove any additional soft flares, again
by applying a $\pm 20$ per cent cut.

\begin{table*}
\caption{Details of the new, archival and previously published {\it
  XMM-Newton} observations discussed in this paper.}
\label{obs}
\vskip 10pt
\begin{tabular}{lrrrrrrr}
\hline
Object&$z$&Ang. scale&$N_{H}$&ObsID&Date&Exposure\\
&&(kpc/arcsec)&($10^{20}$ cm$^{-2}$)&&&MOS1,MOS1,pn (s)\\
\hline
NGC\,315&0.016&0.326&5.88&0305290201&2005-07-02&15236,14984,12580\\
3C\,31&0.017&0.346&5.39&0305290101&2005-08-03&14734,14745,10819\\
3C\,66B&0.0215&0.435&8.36&0002970201&2002-02-05&17915,17862,14692$^{1}$\\
NGC\,1044&0.021&0.425&8.76&0201860301&2004-07-18&13674,13839,13565\\
3C\,76.1&0.032&0.639&10.8&0201860201&2004-08-02&17762,18165,10961\\
NGC\,4261&0.0073&0.150&1.55&0056340101&2001-12-16&26150,26191,16257\\
3C\,296&0.024&0.484&1.86&0201860101&2004-08-05&19584,20315,14729\\
NGC\,6251&0.0244&0.492&5.58&0056340201&2002-03-26&23725,24715,8074$^{1}$\\
3C\,449&0.0171&0.348&11.8&0002970101&2001-12-09&20865,20837,16708$^{1}$\\
\hline
\end{tabular}
\begin{minipage}{17cm}
$^{1}$ The exposure times quoted for the three previously published
  datasets are livetimes after GTI-filtering using hard-band
  lightcurves, similar to the method used for the new observations, as
  described in \citet{c03b} and \citet{ev05}.
\end{minipage}
\end{table*}

\begin{table*}
\caption{Details of the new and previously published radio maps used
  in the analysis presented in this paper. The radio map for NGC\,4261
  was made from archival VLA data.}
\label{radiomaps}
\vskip 10pt
\begin{tabular}{lrrrr}
\hline
Object&Frequency (GHz)&Observatory&Date observed&Reference/Proposal ID\\
\hline
NGC\,315&1.365&VLA&2001 March 10&\citet{lai06b}\\
3C\,31&1.636&VLA&1995 April 28&\citet{lai08}\\
3C\,66B&1.425&VLA&1994 November 10&\citet{h96}\\
NGC\,1044&1.425&VLA&2004 March 19&AC712, AH276\\
3C\,76.1&1.477&VLA&1987 November 25&\citet{lea98}\\
NGC\,4261&1.550&VLA&1984 April 23&AP77\\
3C\,296&1.452&VLA&1987 November 25&\citet{lea98}\\
NGC\,6251&0.337&WSRT&1988 May 12&\citet{lea98}\\
3C\,449&0.609&WSRT&-&\citet{lea98}\\
\hline
\end{tabular}
\end{table*}

To enable the use of double background subtraction techniques for
spectral and spatial analysis (see Section~\ref{xray}), the events
lists were vignetting corrected using the {\sc sas} task {\it
evigweight}. Filter-wheel closed datasets (Pointecouteau et al.,
private communication) for particle background subtraction were
processed and filtered in the same way as the source datasets: the
events lists for each camera were first filtered using the same FLAG
and PATTERN filters as the data files, and weighted using {\it
evigweight}. Weighting the particle background events lists allows
this background component to be subtracted correctly from the particle
component in the source files, which must be weighted up in order that
the vignetting correction is applied correctly to the source and
background X-ray emission in the target observation. Appropriate
background files were then generated for each source using {\it
attcalc} to recast the events list to the correct physical
coordinates. The resulting background file appropriate to each source
events file was used as part of the double subtraction process in the
analysis described below. A scaling factor was determined for each
background dataset to account for differences in the normalisation of
the particle and instrumental background between the source and
filter-wheel closed datasets. The scaling factors were calculated by
comparing the 10--12 keV (MOS) or 12--14 keV (pn) count rates for the
source and background datasets. The background data products were
scaled by this factor before carrying out background subtraction of
spectra or profiles.

In the case of the pn camera, it is also necessary to take into
account the contribution of out-of-time events: 6.3 percent of
detected pn counts (in full-frame mode) are recorded with an incorrect
detector co-ordinate in the $y$ direction. These events produce an
additional position-dependent background component that can be
important in the analysis of low surface brightness emission. We
therefore generated an out-of-time events list for the pn camera using
{\it epchain}, which was scaled appropriately and subtracted as an
additional background component in the analysis that follows.

\subsection{X-ray analysis}
\label{xray}
\subsubsection{Imaging and spectral analysis}

In order to make pictures of the extended emission and to make
overlays on the radio emission we extracted images in the 0.5 - 5 keV
energy range from the three EPIC cameras for each dataset using {\it
evselect} (these images were not used in our quantitative analysis).
We then combined the images to produce a single image for each source,
weighting the two MOS images by a factor chosen to increase the MOS
sensitivity to be equivalent to that of the pn data (this is necessary
in order to remove chip-gap artefacts from the image). The combined
image was then exposure-corrected using an exposure map made by
summing appropriate maps for each camera obtained using {\it eexpmap}.
This method is similar to that used by \citet{boh07}. The images were
not corrected for vignetting as this excessively weights up the
unvignetted particle background at large radii. Point sources were
then identified and removed from the combined image and Gaussian
smoothing was applied with a range of kernel sizes so as to emphasize
structure on different size scales. X-ray images and radio overlays
for each radio-galaxy environment are shown in Figs.~\ref{315im} to
\ref{296im}.

We carried out spectral analysis for several regions of each
environment, as described in more detail in the following sections.
Spectra were extracted using the {\sc sas} {\it evselect} task, with
vignetting correction included by applying the event weights. We used
a double subtraction technique that makes uses of filter-wheel closed
observations, which provide high signal-to-noise information about the
particle and instrumental background as a function of spatial
position\footnote{See the {\it XMM-Newton} background analysis website
at
http://xmm.vilspa.esa.es/external/xmm\_sw\_cal/background/index.shtml}.
To account for variations in the level of particle background between
our observations and the filter-wheel closed datasets, the background
events were scaled by the ratio of 10-12 keV (MOS) or 12-14 keV (pn)
counts over the full field of view for the source and background
datasets. Spectra were then extracted for source and local background
extraction regions (typically a surrounding annulus) from both target
and background events lists, with the background events scaled as
explained above. The spectra extracted from the background events list
were used as backgrounds for the corresponding target observation
spectra. Appropriate response files were generated using {\it rmfgen}
and {\it arfgen}. The target spectra for both the source and local
background regions were grouped to a minimum of 20 counts per bin
after background subtraction. The local background spectrum from the
target observation was then used to determine the Galactic and cosmic
X-ray background spectrum for our observation. By using the scaled
filter-wheel closed spectrum for the same region as a background, the
particle background is completely removed. We therefore fitted an
X-ray background model to the local background spectrum consisting of
two {\it mekal} models to account for emission from the Galactic
bubble and a power-law model absorbed by the Galactic $N_{H}$ in the
direction of the target (see Table~\ref{properties}) to account for
the cosmic X-ray background. The {\it mekal} temperatures were allowed
to vary, but the power-law index was fixed at $\Gamma = 1.41$
\citep{lum02}. The normalisations of all three components were allowed
to vary. In all cases this model provided a good fit to the background
spectrum of our target observations. For each source spectrum, we used
the corresponding filter-wheel closed spectrum as a background
spectrum, to account for particle background, and a fixed X-ray
background model consisting of the best-fitting model for the
particular observation, with the normalisations of each component
fixed at the best-fitting values scaled to the appropriate area for
the source extraction region. This method gave results consistent with
other methods (double background subtraction using the background
template files of Read \& Ponman 2003 and local background
subtraction), but in many cases the more accurate subtraction of both
particle and X-ray background components afforded by this method
enabled us to achieve significantly better constrained temperatures
and abundances. Spectral fitting results are summarized in
Table~\ref{spectra} and discussed in detail in Section~\ref{res}.

\subsubsection{Spatial analysis}

Analysis of the gas distributions in each object was carried out using
radial surface brightness profiles, extracted in concentric annuli
with point sources and chip gaps masked out. In all cases where
resolved jets have been detected with {\it Chandra}, they are within
the core radius of the PSF so that they can be considered to be part
of the central point source, which is modelled as part of the surface
brightness profile analysis. Vignetting correction and double
background subtraction using filter-wheel closed datasets was applied
in the same manner as for the spectral analysis. The three surface
brightness profiles were fitted separately with models including
point-source components, a single $\beta$ model \citep[e.g.][]{bw93},
and in some cases a model consisting of a line-of-sight projection of
the sum of two $\beta$ models in gas density, hereafter referred to as
a {\it projb} model. The {\it projb} model is defined by:
\begin{equation}
n(r) = n_0 \left[ \left(1 +
  \frac{r^{2}}{r_{c,in}^{2}}\right)^{-3\beta_{in}/2} + N \left(1 +
  \frac{r^{2}}{r_{c}^{2}}\right)^{-3\beta / 2} \right]
\end{equation}
where $N$ is the relative normalisation of the two $\beta$ model
components. The corresponding surface brightness profile is determined
via numerical integration of the following expression:
\begin{equation}
S_{X} (R) \propto \int_{-\infty}^{\infty} n^{2}(l,R) dl
\end{equation}
where $r = (l^{2} + R^{2})^{2}$ is the radius, $R$ is measured in the
plane of the sky, and $l$ is along the line of sight. We explored the
three or six dimensional parameter space for the $\beta$ and $projb$
models respectively, using a Markov-Chain Monte Carlo (MCMC)
method\footnote{Our implementation of the MCMC method uses the
Metropolis-Hastings algorithm in a manner very similar to the {\sc
metro} code of \citet{hb04}, but is implemented to run on a cluster of
multiprocessor computers using the Message Passing Interface (MPI).}.
We used the joint $\chi^{2}$ value for the three profiles as the
likelihood estimator and in the sections that follow we quote both the
parameters of the best-fitting model and the Bayesian estimates for
the value of each parameter. Plausible ranges for each parameter were
estimated by carrying out extreme fits and these were used as priors
for the MCMC method. The parameter ranges for each source are given in
Table~\ref{priors}. Uncertainties (formally credible intervals:
\citealt{greg05}) for each parameter are determined from the
1-dimensional projection of the minimal n-dimensional volume that
encloses 68 per cent of the posterior probability distribution as
returned by the MCMC algorithm. The uncertainties determined in this
way correspond to $1\sigma$ errors for 2 or 5 interesting parameters
for the $\beta$ and $projb$ models, respectively. Quantities derived
from the surface brightness model fits, included X-ray luminosity and
pressure at a given radius, are obtained by determining that quantity
for each model fit and then obtaining the Bayesian estimate for the
quantity in question. Uncertainties on the derived quantities
come from the minimal one-dimensional interval that encloses 68 per
cent of the posterior probability space for the given quantity, and so
correspond to $1\sigma$ errors for a single parameter of interest in a
traditional fitting procedure. As seen in Section~\ref{res}, some of
the parameters are strongly correlated, leading to large
uncertainities on the individual model parameters; however, tight
constraints can be obtained on derived quantities such as pressure and
luminosity. Our method does not treat the model normalisation as a
free parameter, instead finding the best normalisation for each model;
however, the uncertainty in model normalisation does not significantly
affect the constraints on model parameters or derived quantities.

Before fitting, each model was convolved with the {\it XMM-Newton} PSF
based on the on-axis parametrization described in the {\it XMM-Newton}
CCF files XRT1\_XPSF\_0006.CCF, XRT2\_XPSF\_0007.CCF and
XRT3\_XPSF\_0007.CCF. For four sources with {\it Chandra}-detected
galaxy-scale environments with core radii too small to be resolved by
{\it XMM-Newton} (NGC\,315, 3C\,296, and NGC\,6251), we included
previously published {\it Chandra} profiles in the combined fits so as
to model the inner profiles as well as possible. Final pn surface
brightness profiles for each radio-galaxy environment with a model
using the best-fitting model parameters, as discussed in the
individual sections below, are shown in Fig.~\ref{sxprofs}.

\section{Results}
\label{res}

\subsection{NGC\,315}

\begin{figure*}
\centerline{\hbox{
\epsfig{figure=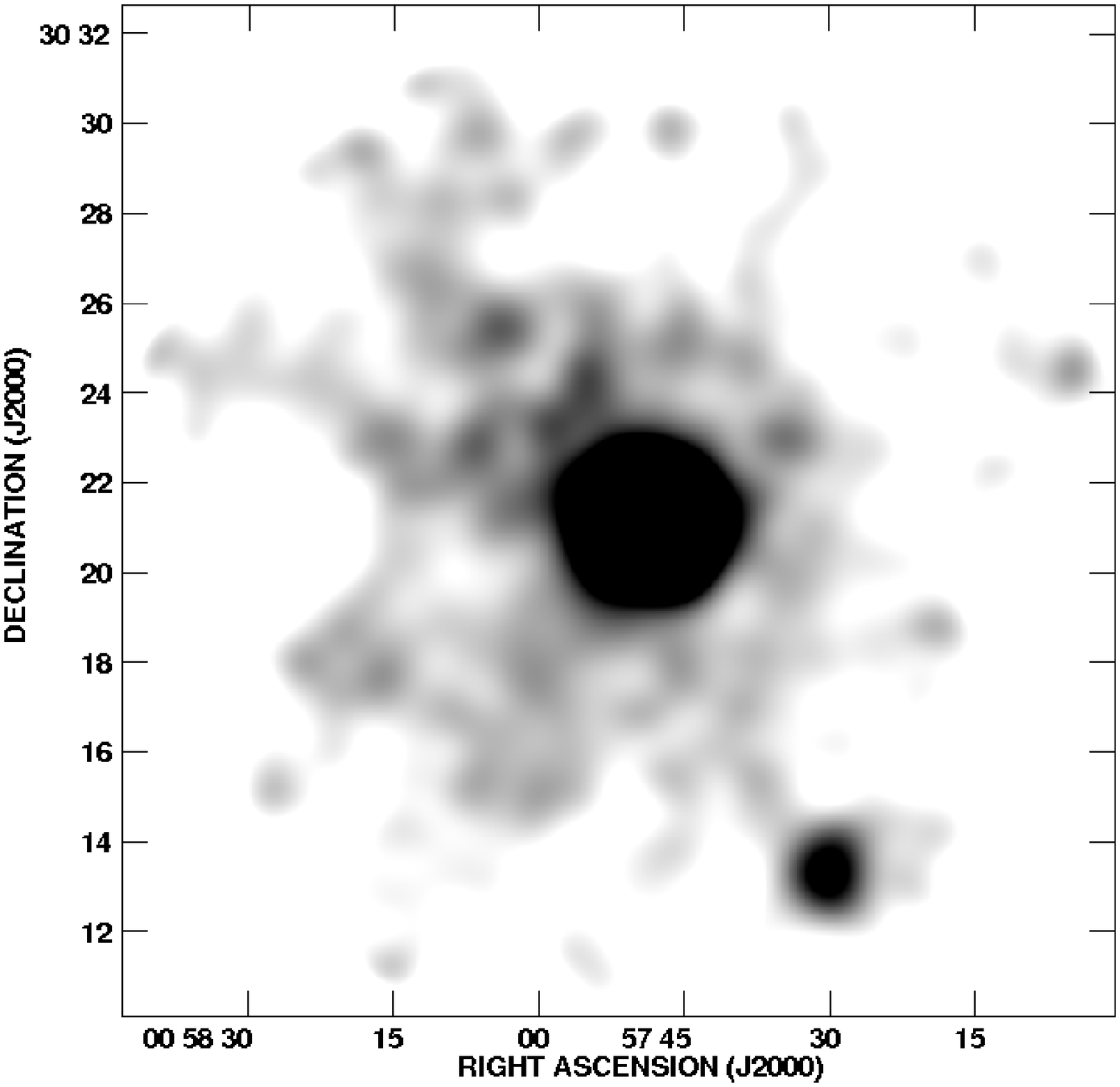, height=7cm}
\epsfig{figure=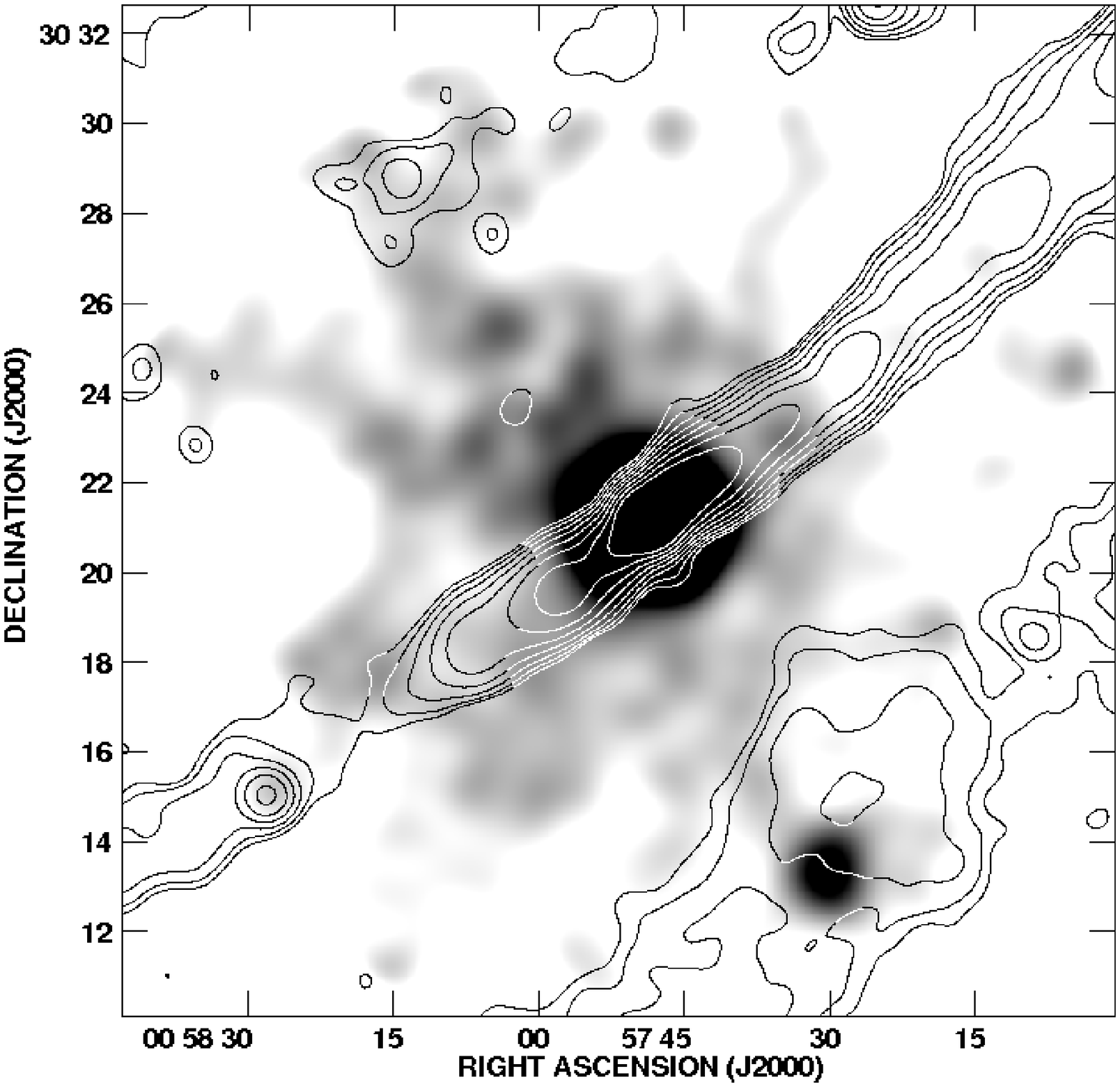,height=7cm}}}
\caption{The X-ray environment of NGC\,315. Images made from the
  combined MOS1, MOS2 and pn events lists in the 0.5 - 5.0 keV energy
  range, with exposure correction to correct for chip gaps but not
  vignetting, as described in the text. Left: smoothed with a Gaussian
  of FWHM 33 arcsec to show the group-scale emission. Right: with
  1.4-GHz radio contours \citep{lai06b} overlaid.}
\label{315im}
\end{figure*}

Images of the environment of NGC\,315 are shown in Fig.~\ref{315im}.
Emission is detected from NGC\,315, the group galaxies NGC\,311 and
NGC\,318, a candidate more distant cluster at $\alpha =
00^{h}57^{m}31^{s}, \delta = +30^{\circ}13'$, which does not appear to
have been identified previously, and the intragroup medium. The X-ray
emission appears somewhat elongated to the north-east in a
direction perpendicular to the radio jets . We measure a total of
$\sim 4000$ 0.3 - 7.0 keV pn counts from the galaxy and group
environment. Although NGC\,315 has previously been studied in detail
with {\it ROSAT} and {\it Chandra}, this is the first detection of an
environment on hundred-kpc scales. Unfortunately this dataset is also
highly flare-contaminated, with only 12 - 15ks usable of a total
observed 50ks (see Table~\ref{properties}).

The results of surface brightness profile fits for NGC\,315 are in
Table~\ref{sxfits}. The inner enviroment of NGC\,315 is well
constrained from {\it Chandra} observations, and so in this case we
carried out a joint fit to four profiles: one for each {\it
XMM-Newton} camera, plus the {\it Chandra} profile of \citet{wo07}. As
shown in Fig.~\ref{sxprofs}, the profile of NGC\,315 does not appear
to flatten into a group-scale component on hundred kpc scales, unlike
the majority of other FR-I environments known to date, and so the
$projb$ model was not required to fit its surface brightness profile.
As shown in Table~\ref{sxfits}, the preferred $\beta$ model parameters
are in agreement at the joint $1\sigma$ level with the parameter
values measured from the {\it Chandra} data alone by \citet{wo07}
($\beta = 0.52\pm0.01$ and $r_{c} = 1.7\pm0.2$ arcsec).

We extracted a spectrum in the inner 60 arcsec to study the core
emission from NGC\,315 and found that the {\it Chandra} best-fitting
model consisting of an absorbed power law with $N_{H} = 7.6 \times
10^{21}$ cm$^{-2}$ and $\Gamma = 1.57$ plus a {\it mekal} model with
$kT = 0.62\pm0.02$ keV and abundance fixed at 0.3 solar was a good fit
($\chi^{2} = 155$ for 125 d.o.f.). The results of fitting to the group
emission in a region between 120 and 300 arcsec are given in
Table~\ref{spectra}.

In order to obtain profiles of gas density and pressure we assumed a
linear ramp in gas temperature obtained by fitting a straight line to
the {\it Chandra} temperature profile of \citet{wo07} and the {\it
XMM-Newton} temperature data point. The temperature was held constant
at the {\it XMM-Newton}-measured value beyond 217 arcsec. Errors on
profiles take into account both the uncertainty in model parameters
and the the uncertainty in temperature.

We were also able to extract spectra for the group member galaxy
NGC\,311. The spectra were well fitted by a {\it mekal} model with $kT
= 1.4^{+0.6}_{-0.3}$ keV with abundance fixed at 0.3 times solar
($\chi^{2} = 1.2$ for 3 d.o.f.) or by a power law with $\Gamma =
2.6\pm0.5$ ($\chi^{2} = 1.6$ for 3 d.o.f.). The data are insufficient
for distinguishing between the two models; however, the temperature of
the thermal fit is higher than would be expected for a galaxy
atmosphere, and so it is likely that some of the emission is
associated with nuclear activity or X-ray binaries.

\subsection{3C\,31}

\begin{figure*}
\centerline{\hbox{
\epsfig{figure=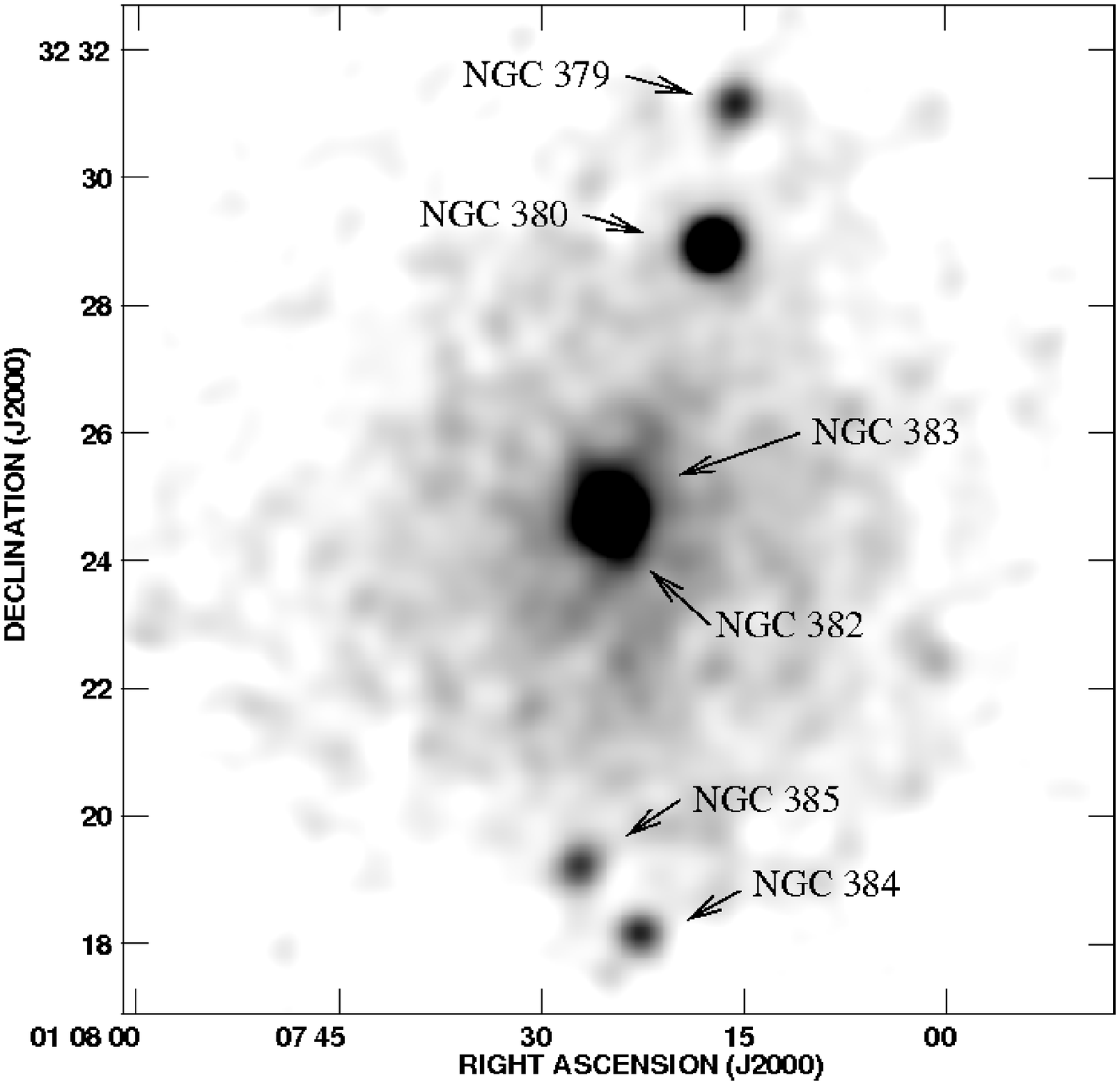,height=7cm}
\epsfig{figure=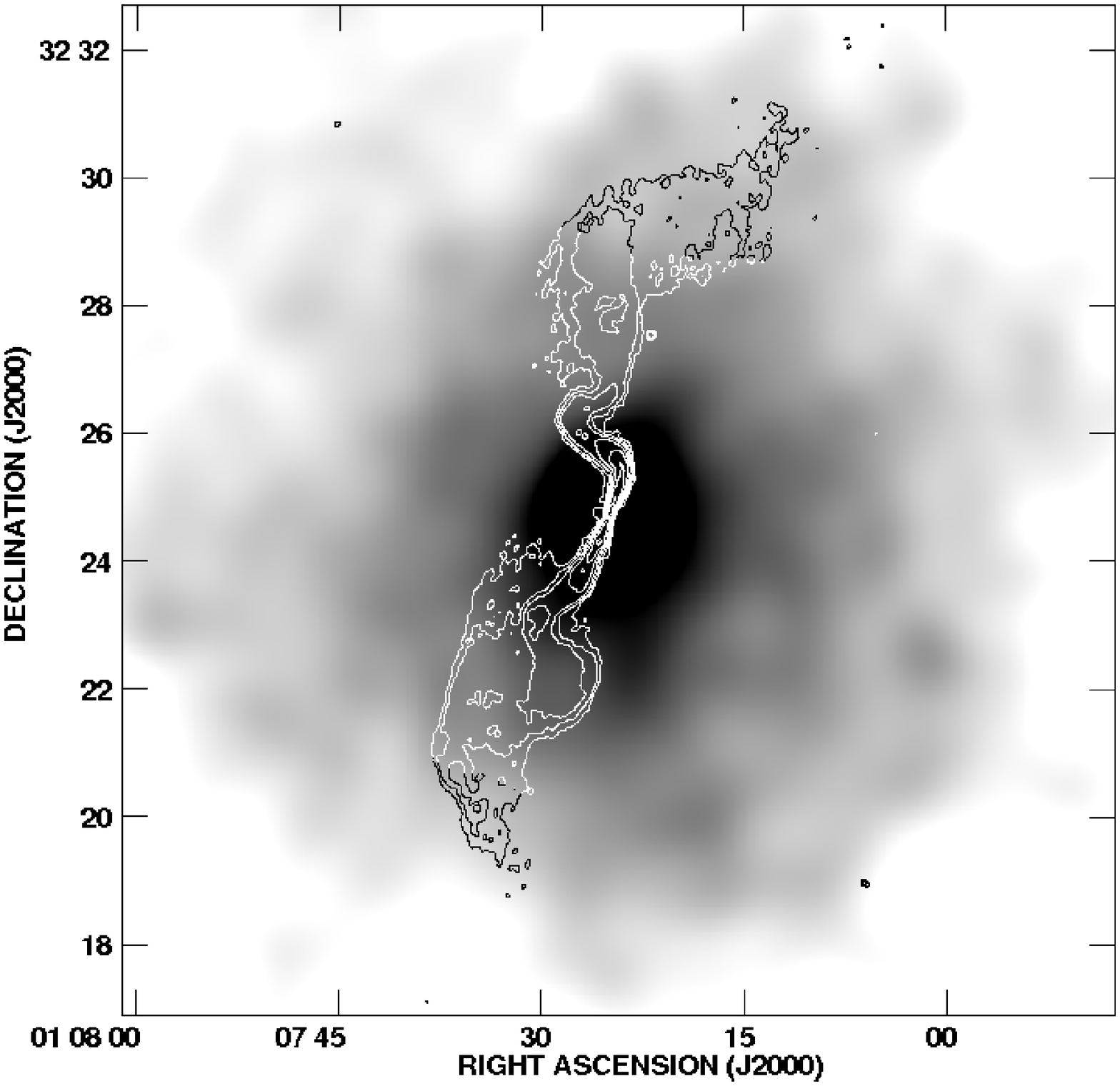,height=7cm}}}
\caption{The X-ray environment of 3C\,31. Images made from the
  combined MOS1, MOS2 and pn events lists in the 0.5 - 5.0 keV energy
  range, with exposure correction to correct for chip gaps but not
  vignetting, as described in the text. Left: smoothed with a Gaussian
  of FWHM 13 arcsec to highlight the emission associated with several
  of the NGC\,383 group galaxies. The central galaxy is NGC\,383, with
  a slight elongation to the south west due to emission from its
  companion galaxy NGC\,382; right: smoothed with a Gaussian of FWHM
  26 arcsec to highlight the group-scale emission, and with galaxy
  emission removed, overlaid with 1.6-GHz radio contours
  \citep{lai08}.}
\label{3c31im}
\end{figure*}

Fig.~\ref{3c31im} shows images of the environment of 3C\,31 from our
new {\it XMM-Newton} observation. Emission is detected from several
group galaxies as well as from the intragroup medium. Unfortunately,
the observations of 3C\,31 were severely flare-contamined making
quantitative analysis extremely difficult. After preliminary analysis
we concluded that it was not possible to estimate reliably the level
of background for this observation and so we did not carry out either
spectral or spatial analysis. 3C\,31 has been studied in detail with
{\it ROSAT} \citep{kb99} and {\it Chandra} \citep{h02b}. For
comparisons with the other radio galaxies in this sample, we therefore
made use of the results of those two studies. A forthcoming
reobservation with {\it XMM-Newton} will allow a much more detailed
study of 3C\,31's environment and environmental interactions.

\subsection{NGC\,1044}
\begin{figure*}
\centerline{\vbox{\hbox{
\epsfig{figure=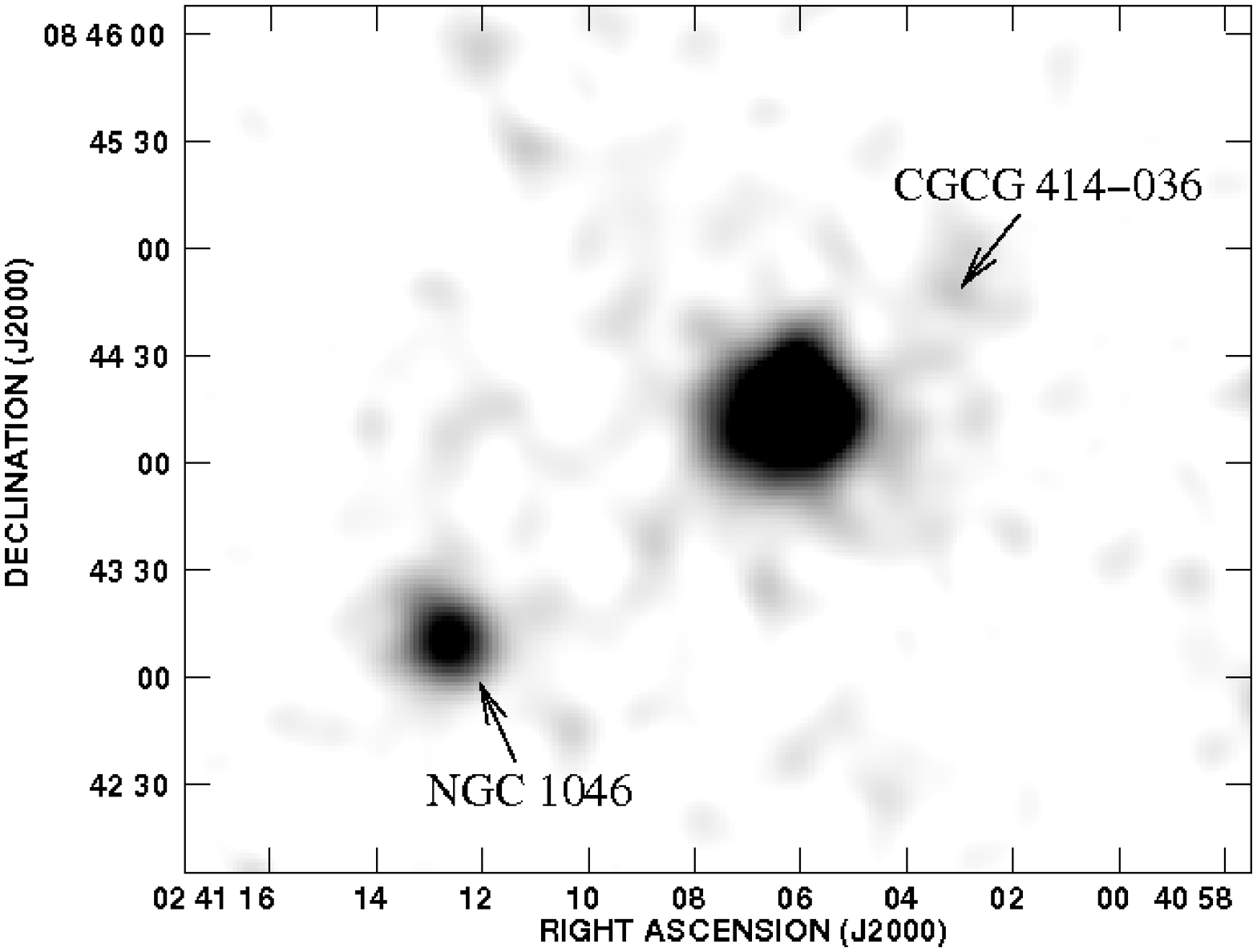,height=7cm}
\epsfig{figure=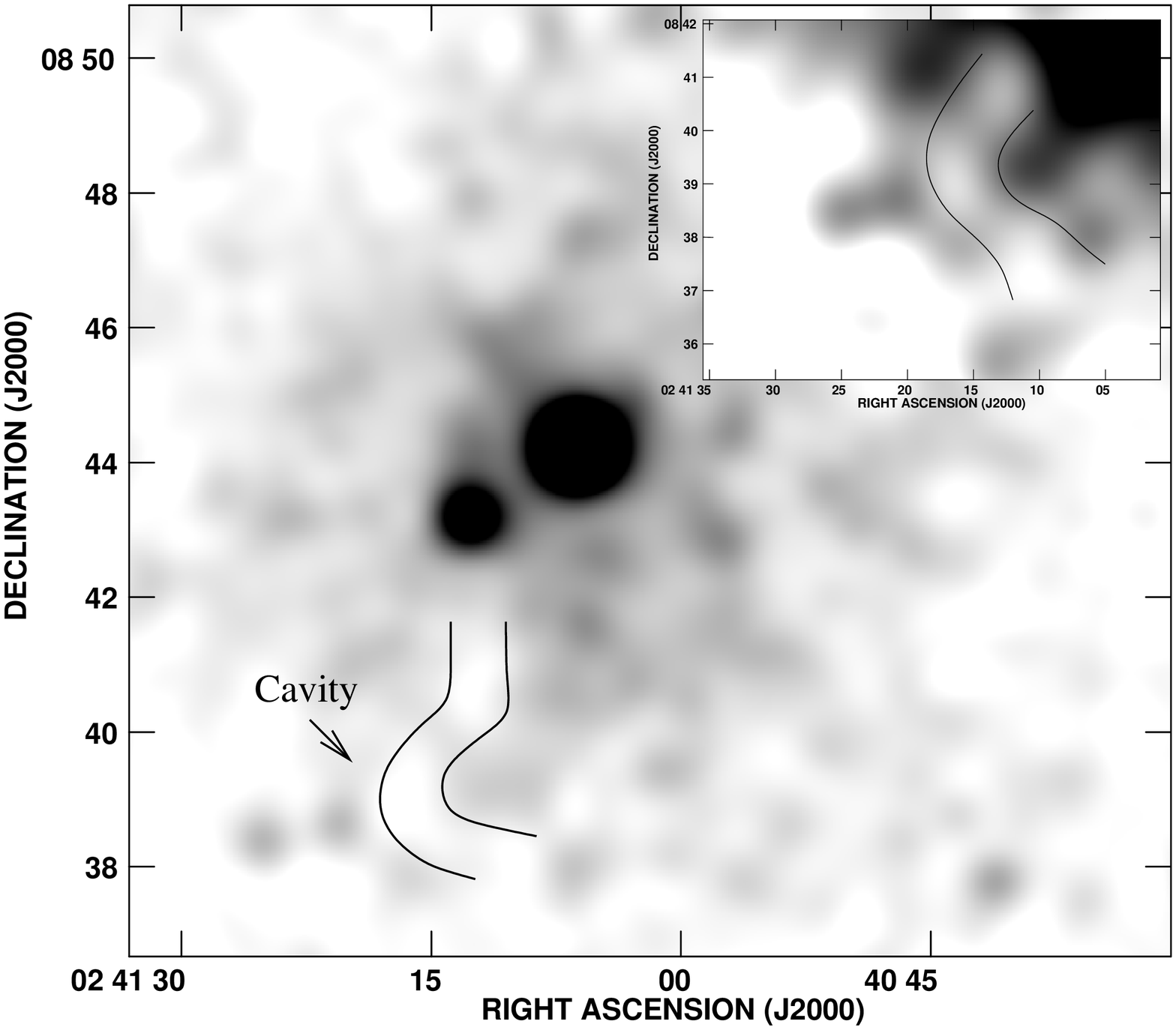,height=7cm}}
\centerline{\hbox{
\epsfig{figure=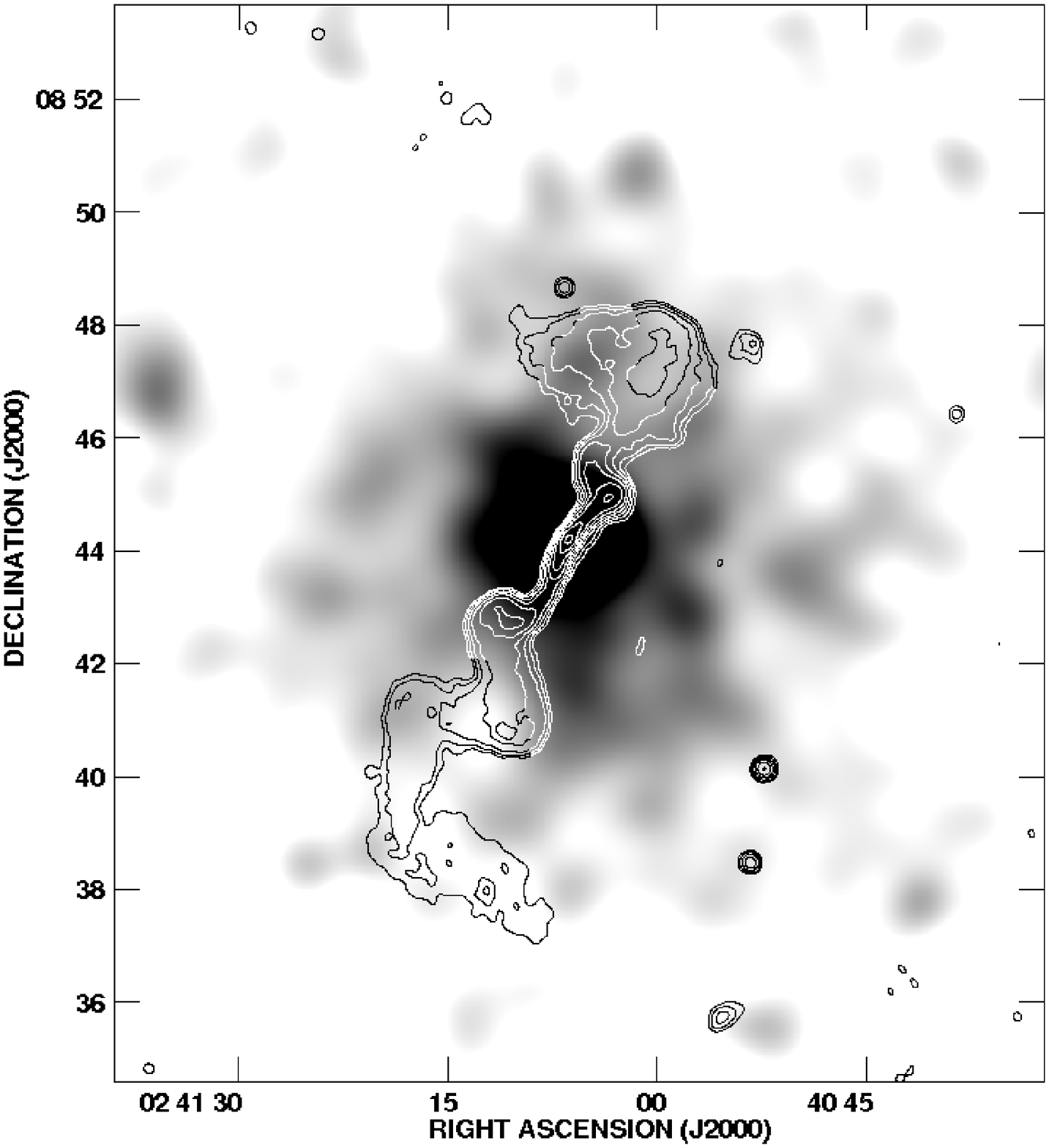,height=8cm}
\epsfig{figure=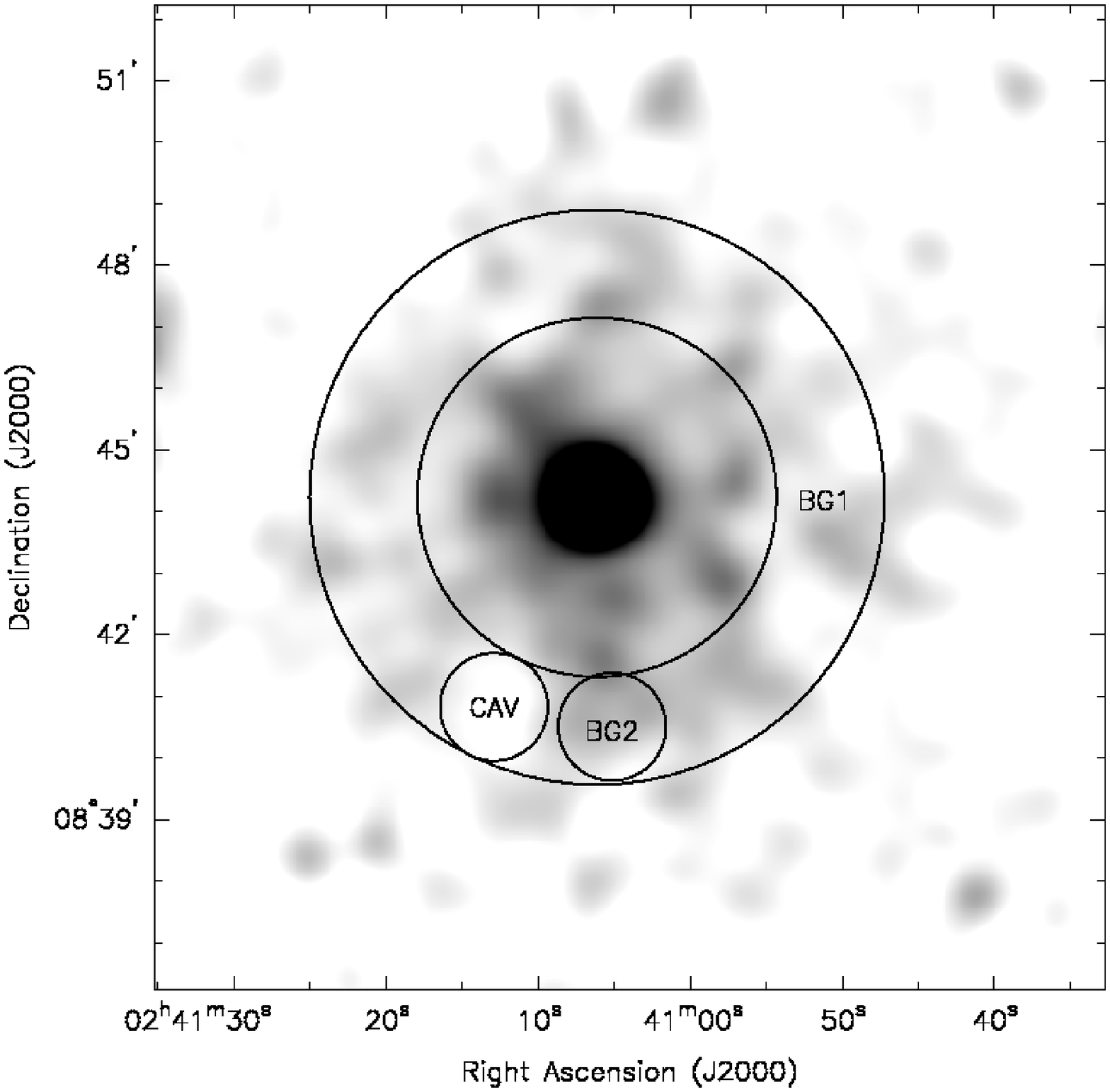,height=8cm}}}}}
\caption{The X-ray environment of NGC\,1044. Images made from the
  combined MOS1, MOS2 and pn events lists in the 0.5 - 5.0 keV energy
  range, with exposure correction to correct for chip gaps but not
  vignetting, as described in the text. Top left: smoothed with a
  Gaussian of FWHM 13 arcsec to highlight the emission associated with
  several of the NGC\,1044 group galaxies; top right: the same image
  smoothed with a Gaussian of FWHM 20 arcsec and with galaxy emission
  removed to show the group-scale emission and cavity to the S; an
  inset shows a close-up of the cavity region from an image smoothed
  with a Gaussian of FWHM 26 arcsec; bottom left: image smoothed with
  a Gaussian of FWHM 26 arcsec to highlight the group emission, with
  1.4-GHz radio contours overlaid illustrating that the southern radio
  plume fills the cavity shown in the middle panel; bottom right:
  regions used to investigate significance of cavity detection.}
\label{1044im}
\end{figure*}

The environment of NGC\,1044 is shown in Fig.~\ref{1044im}. This
dataset was relatively free of flare contamination, and a total of
$\sim 7840$ 0.3 - 7.0 keV pn counts are detected from the environment.
Fig.~\ref{1044im} reveals a cavity at the position of the SE lobe. We
tested the statistical significance of the surface brightness deficit
by comparing the total (MOS1 + MOS2 + pn) surface brightness in a
circle of radius 53 arcsec defined using the radio map, an annular
background centred on the nucleus of NGC\,1044 extending from the
inner to the outer edge of the cavity region, but excluding the
south-east quadrant, and a background of the same size rotated about
the centre of the environment. The cavity has a surface brightness of
$0.0247\pm0.0017$ counts arcsec$^{-2}$, compared to a surface
brightness of $0.0315\pm0.0005$ counts arcsec$^{-2}$ in the annular
background region, and a surface brightness of $0.0398\pm0.0022$
counts arcsec$^{-2}$ in the same-sized background region. The
errors on surface brightness measurements were obtained by combining
in quadrature source and background errors determined from the
Poissonian errors on the counts in each region. The deficit is
significant at a level $>3\sigma$.

NGC\,1044 has not previously been observed in the X-ray, and so its
inner (galaxy-scale) environment is not well-constrained. Our new
observations provide the first detection of an X-ray environment for
this radio galaxy. The surface brightness profile for NGC\,1044
(Fig.~\ref{sxprofs}) shows a characteristic two-component shape
indicating the presence of a central AGN or galaxy-scale component as
well as a large-scale group gas component. A point-source plus single
$\beta$ model is a poor fit to the profiles, and so we used a
point-source plus $projb$ model, with model fitting results given in
Table~\ref{sxfits}. Results of a spectral fit to the group-scale
emission in an annulus between 100 and 400 arcsec are given in
Table~\ref{spectra}.

Although the surface brightness profile does not show evidence for a
dominant AGN component, we extracted spectra from a source-centred
circle of one arcminute radius to investigate the contributions of AGN
and galaxy-scale gas-related emission. Consistent with the surface
brightness profile results we found that a single power-law model did
not provide an acceptable fit to the data. We obtained a good fit
($\chi^{2} = 40$ for 37 d.o.f.) for a single {\it mekal} model with
$kT = 0.82\pm0.05$ keV and $Z = 0.15^{+0.6}_{-0.05} Z_{\sun}$. A
somewhat better fit was obtained for a two-component {\it mekal} plus
power-law model ($\chi^{2} = 32.3$ for 36 d.o.f.) with $kT =
0.77^{+0.05}_{-0.04}$ keV, abundance fixed at 0.5 times solar and
$\Gamma = 1.91^{+0.34}_{-0.42}$. The single {\it mekal} model gives an
unabsorbed 0.1 - 10 keV luminosity of $(1.3\pm0.2) \times 10^{41}$
ergs s$^{-1}$, which is high for a galaxy atmosphere, whereas we
obtain a more reasonable luminosity of $7.5 \times 10^{40}$ erg
s$^{-1}$ for the combined fit. The combined fits gives a 1-keV flux
density of $6\pm1$ nJy for the power-law component, which is
consistent with the nuclear X-ray emission predicted by the relation
of \citet{ev06} for a source of NGC\,1044's core radio flux density
($<0.08$ Jy at 1.5 GHz).

We used the surface brightness profile models and the measured
temperature to determine profiles of gas density and pressure by using
the best-fitting $projb$ model parameters to obtain an emission
measure profile. As we were unable to obtain a temperature profile for
this group, we assumed the gas is isothermal with the spectral
parameters of the best fit to the group spectrum ($kT = 1.3$ keV and
$Z = 0.12$ solar) in order to obtain gas density and pressure
profiles. Errors on the density and pressure profiles were determined
by combining in quadrature the $1\sigma$ confidence range from the
dispersion in profile fits with the uncertainty in temperature as for
the other sources.

Of the NGC\,1044 group galaxies, X-ray emission is clearly detected
from NGC\,1046 ($z = 0.0199$) and CGCG 414$-$036 ($z=0.0206$), as shown
in Fig.~\ref{1044im}. We were able to extract spectra for NGC\,1046,
and fitted {\it mekal}, single power-law and {\it mekal} plus
power-law models. All three models gave good fits to the spectra;
however, the best-fitting single {\it mekal} model ($\chi^{2} = 4.2$
for 6 d.o.f.) required low abundance ($Z < 0.04 Z_{\sun}$) and the
best-fitting power-law model ($\chi^{2} = 1.8$ for 7 d.o.f.) required
a steep photon index ($\Gamma = 3.2^{+0.4}_{-0.5}$). We obtained the
lowest reduced $\chi^{2}$ and most physically plausible parameters for
the two-component fit ($\chi^{2} = 0.8$ for 5 d.o.f.) with $kT =
0.2^{+0.3}_{-0.1}$ keV and $\Gamma = 2.3^{+1.3}_{-0.7}$ and abundance
fixed at 0.5 times solar. We conclude that there is likely to be some
non-thermal emission associated with this galaxy, either associated
with X-ray binaries or nuclear activity, in addition to thermal
emission from a galaxy atmosphere. There were insufficient counts
associated with CGCG414$-$036 to extract a useful spectrum.

\subsection{3C\,76.1}

\begin{figure*}
\centerline{\hbox{
\epsfig{figure=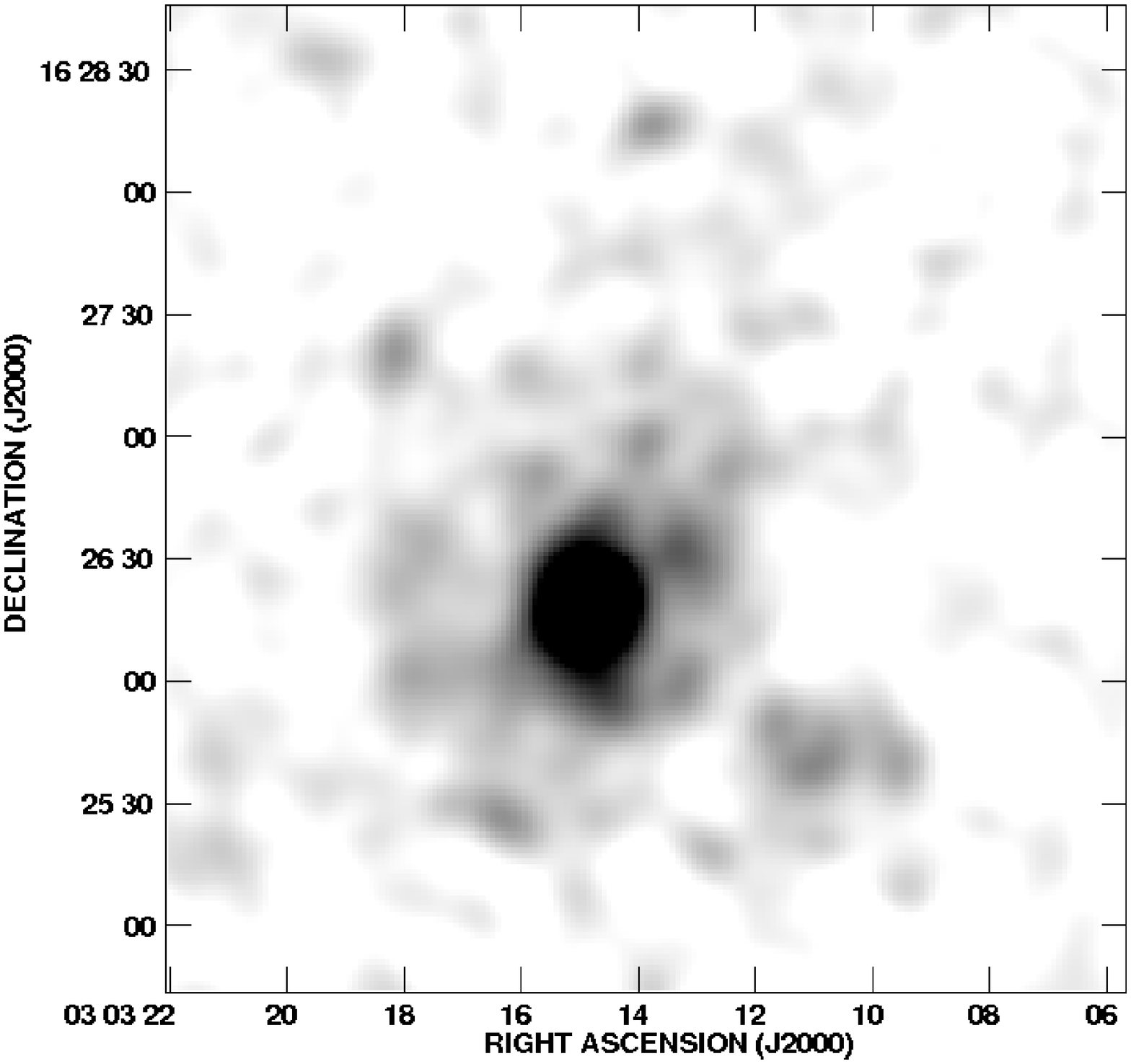,height=7cm}
\epsfig{figure=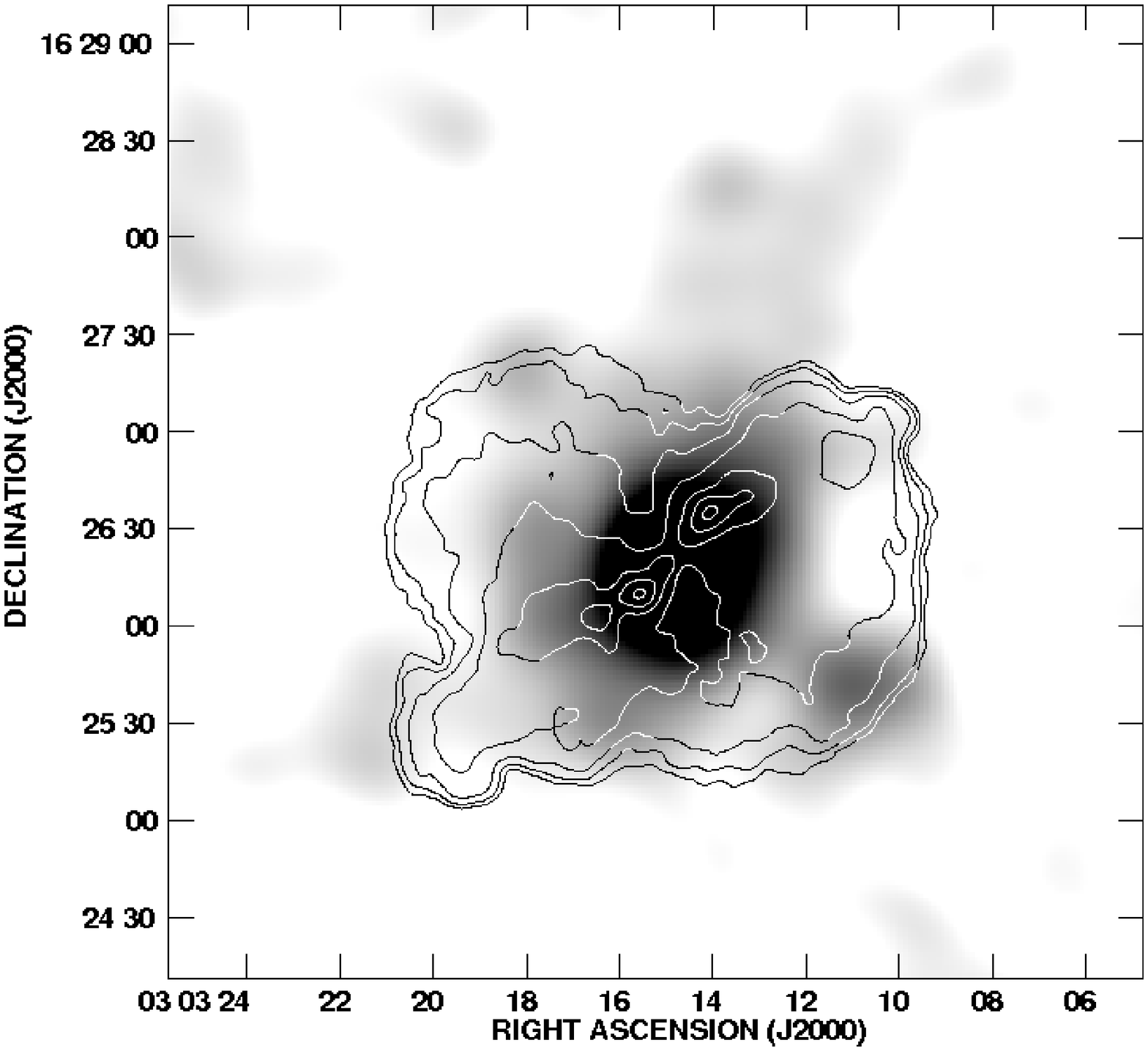, height=7cm}}}
\caption{The X-ray environment of 3C\,76.1. Images made from the
  combined MOS1, MOS2 and pn events lists in the 0.5 - 5.0 keV energy
  range, with exposure correction to correct for chip gaps but not
  vignetting, as described in the text. Left: image smoothed with a
  Gaussian of FWHM 5.2 arcsec; right: smoothed wih a Gaussian of FWHM
  10.4 arcsec and with 1.4-GHz radio contours overlaid.}
\label{3c76.1im}
\end{figure*}

This observation is the first pointed X-ray observation of 3C\,76.1.
As shown in Fig.~\ref{3c76.1im}, the environment of 3C\,76.1 is
considerably poorer than those of the other radio galaxies in this
sample. The X-ray emission appears elongated in the north-south
direction, perpendicular to the radio lobes. We detect a total of
$\sim 385$ 0.3 - 7.0 keV pn counts from the group environment.

To confirm the presence of extended emission, we fitted a
point-source-only model to this surface brightness profile (which is
shown in Fig~\ref{sxprofs}). This did not give an acceptable
$\chi^{2}$, whereas we were able to obtain good fits with a
point-source plus $\beta$ model, as listed in Table~\ref{sxfits}. As
the profiles are inconsistent with a point-source only model, and the
$\beta$-model fit is good, we conclude that 3C\,76.1's environment has
been firmly detected for the first time. As discussed below, the point
source component appears to originate partly from a central AGN and
partly from an unresolved galaxy-scale atmosphere.

We list the results of a spectral fit to the group emission in the
region between 30 and 200 arcsec in Table~\ref{spectra}. We used a
fixed abundance as the abundance value tended to unrealistically high
values if left free. We used the fitted surface brightness profile
model and the measured temperature to determine profiles of gas
density and pressure. As for NGC\,1044, we assumed the gas to be
isothermal with the spectral parameters of the best fit to the group
spectrum ($kT = 0.91$ keV and $Z = 0.3$ solar) in order to obtain gas
density and pressure profiles. Errors on the profiles were determined
as for the other sources described above.

To constrain the origin of the unresolved central emission, we
extracted a spectrum in the central 30 arcsec using local background
to subtract off the contribution from the group-scale emission. We
found that neither a power law nor a {\it mekal} model on their own
gave an acceptable fit, and so we fitted a power-law plus {\it mekal}
model, which gave a $\chi^{2}$ value of 15.5 for 12 d.o.f. with $kT =
0.43^{+0.16}_{-0.11}$ keV (for abundance fixed at 0.3 times solar) and
$\Gamma = 0.9^{+0.2}_{-0.3}$. Such a flat power-law index suggests
that an absorbed power-law component, as commonly seen in narrow-line
radio galaxies \citep{hec06} and in some low-power sources such as
NGC\,3801 \citep{c07}, may be present.

\subsection{NGC\,4261}

\begin{figure*}
\centerline{\vbox{\hbox{
\epsfig{figure=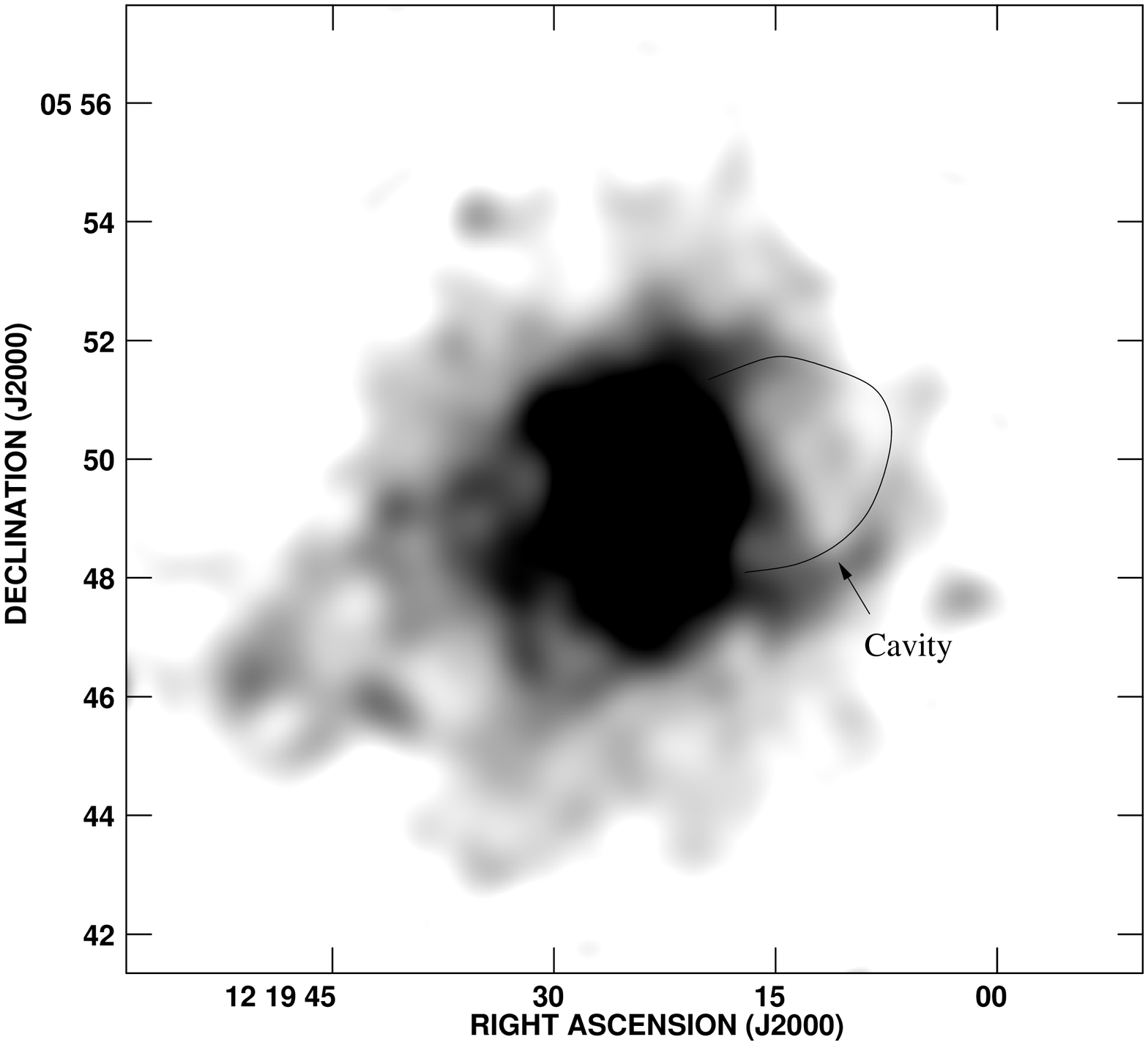,height=7cm}
\epsfig{figure=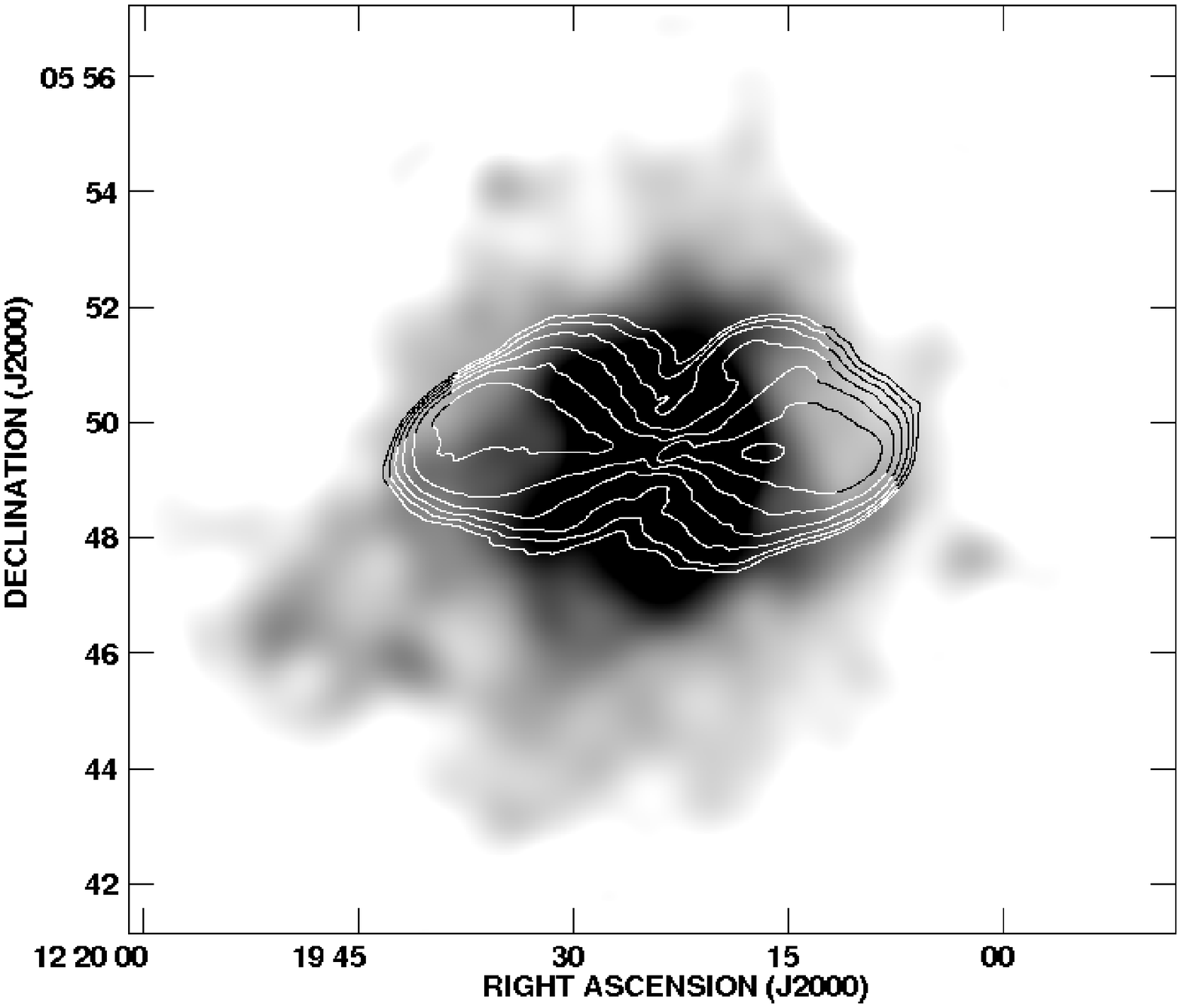, height=7cm}}
\centerline{\hbox{
\epsfig{figure=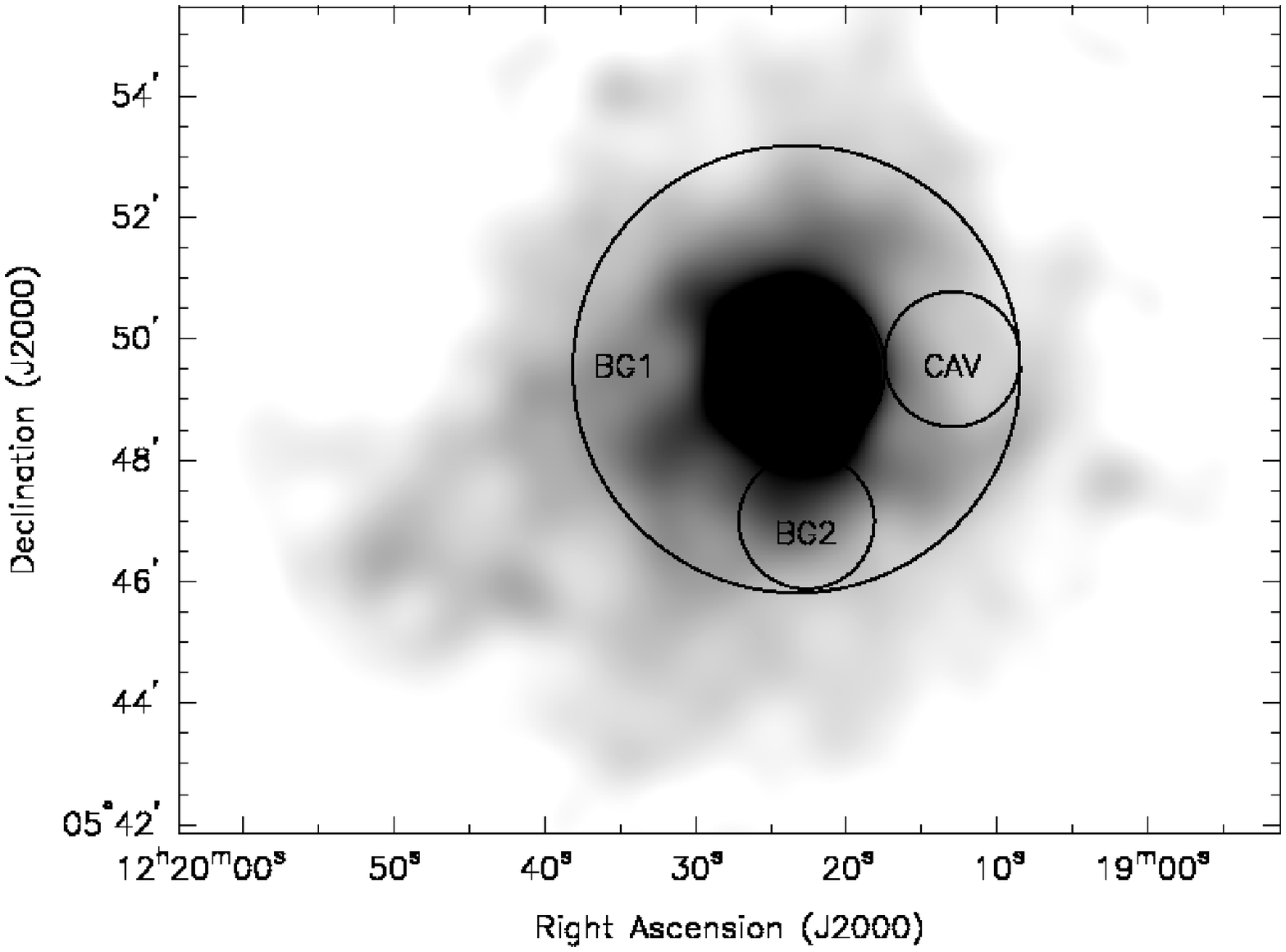,height=7cm}}}}}
\caption{The X-ray environment of NGC\,4261. Images made from the
  combined MOS1, MOS2 and pn events lists in the 0.5 - 5.0 keV energy
  range, with exposure correction to correct for chip gaps but not
  vignetting, as described in the text. Top left: smoothed with a
  Gaussian of FWHM 26 arcsec and with galaxy emission removed to show
  the group-scale emission and cavity to the W; top right: smoothed
  with a Gaussian of 33 arcsec and with 1.4-GHz radio contours
  overlaid illustrating that the W lobe coincides with the cavity in
  the group-scale gas; bottom: regions used to investigate
  significance of cavity detection.}
\label{4261im}
\end{figure*}

The environment of NGC\,4261 is shown in Fig.~\ref{4261im}. The inner
environment of this radio galaxy as observed by {\it Chandra} was
discussed by \citet{zez05} and \citet{c05a}, where we also presented
an image of the {\it XMM-Newton} data but not a detailed analysis, and
by \citet{jet07}. We detect a total of $\sim 33,000$ 0.3 - 7.0 keV
pn counts from the source, including a contribution from the central
AGN. The dataset is not significantly contaminated by flares.

Fig.~\ref{4261im} reveals a cavity at the position of the W lobe. We
 tested the statistical significance of the surface brightness deficit
 by comparing the total (MOS1 + MOS2 + pn) surface brightness in a
 circle of radius 67 arcsec defined using the radio map with an
 annular background region centred on the nucleus of NGC\,4261
 extending from the inner to the outer edge of the cavity region, but
 excluding the south-east quadrant, and with a background of the same
 size as the cavity region but rotated about the centre of the group
 to the south. The cavity has a surface brightness of
 $0.0942\pm0.0027$ counts arcsec$^{-2}$, compared to a surface
 brightness of $0.1081\pm0.0011$ counts arcsec$^{-2}$ in the annular
 background region, and a surface brightness of $0.1434\pm0.003$
 counts arcsec$^{-2}$ in the same-sized background. Hence the deficit
 is significant at a level between 3 and 8$\sigma$. The errors on
 the surface brightness measurements were obtained as for NGC\,1044.
 There is no statistically significant large-scale cavity on the
 eastern side of the source; however, there is evidence for a smaller
 region of lower surface brightness deficit close to the nucleus and
 coincident with the inner parts of the eastern radio lobe.

As shown in Fig.~\ref{sxprofs} the surface brightness profile shows a
two-component form with a large-scale component in addition to a
galaxy-scale inner environment \citep[e.g.][]{zez05}. We therefore
fitted the point-source plus $projb$ model as for NGC\,1044. The
results are listed in Table~\ref{sxfits}. Table~\ref{spectra} gives
the results of spectral fitting to the group-scale emission between 60
and 600 arcsec. We were concerned that because the physical scales we
are sampling in NGC\,4261 are considerably smaller than in the other
sources in our sample (due to its lower redshift) contamination from
X-ray binaries in the galaxy could be a problem ($D_{25} = 4.1$ arcmin
for NGC\,4261, whereas it is $<2$ arcmin for most of the other
galaxies -- here $D_{25}$ is the optical radius of the galaxy at a
$B$-band surface brightness of 25 mag arcsec$^{2}$), and so we
included a thermal bremsstrahlung component with $kT = 5.0$ keV to
account for an X-ray binary population; however, for the global
spectrum this component was not significant. We investigated whether a
temperature gradient is present by extracting spectra in several
annuli. For the outer two bins the normalization of the included
thermal bremsstrahlung component was negligible, consistent with the
expectation of no contribution from binaries on those scales. In the
inner two regions (which are both within the $D_{25}$ radius of the
galaxy) the bremsstrahlung component contained a significant fraction
of the total flux, corresponding to a total X-ray binary 0.3 - 8 keV
luminosity of $\sim 3 \times 10^{40}$ erg s$^{-1}$. This is consistent
with the expectation for NGC\,4261 based on the mean $L_{X}/L_{B}$
relation of Kim \& Fabbiano (1994). We found no evidence of a
significant temperature gradient when the X-ray binary contribution
is accounted for.

We also extracted spectra from a source-centred circle of one
arcminute radius in order to investigate the AGN and galaxy-scale
X-ray emission. We found that neither a single {\it mekal} nor a
single power-law model provided an acceptable fit. We then fitted a
{\it mekal} plus power-law model, which gave a reasonable fit
statistic ($\chi^{2} = 503$ for 408 d.o.f.) for $kT =
0.65^{+0.01}_{-0.02}$ keV and abundance fixed at 0.5 times solar;
however, the best-fitting power-law index is very flat: $\Gamma=
0.57\pm0.11$. We therefore tried fitting the best-fitting model of
\citet{zez05} (their Model 5). If we fix the column densities, power
law indices and Raymond-Smith temperature to their best-fitting
values, we cannot obtain an acceptable fit; however, allowing the
temperature to vary gives a good fit with $\chi^{2} = 585$ for 406
d.o.f. for a temperature of $kT = 0.69\pm0.01$ keV. The higher
temperature is likely to be due to the larger extraction region used
here, which is likely to include hotter gas from outer regions of the
galaxy atmosphere and/or a contribution from X-ray binaries.

Two other galaxies in the NGC\,4261 group were detected with
sufficient counts to carry out spectral fitting, NGC\,4254 ($z=
0.016$) and PCG 039660 ($z=0.017$). We fitted single {\it mekal} and
power-law fits to each galaxy (the statistics were too poor to allow
fitting of more complex models). For NGC\,4254 we found good fits with
both models, with a slightly better fit statistic ($\chi^{2} = 5.9$
for 4 d.o.f.) for the {\it mekal} model, but a temperature of $>1.3$
keV, which is high for a galaxy atmosphere. The best-fitting power-law
model ($\chi^{2} = 6.6$ for 4 d.o.f.) had $\Gamma =
1.4^{+0.6}_{-0.5}$. It seems likely that the emission from this galaxy
is at least partly non-thermal in origin. For PCG 039660, the {\it
mekal} model was unacceptable, whereas the power-law model was a good
fit ($\chi^{2} = 2.0$ for 2 d.o.f.) with $\Gamma = 1.6^{+0.5}_{-0.6}$,
and so we conclude that its emission is predominantly non-thermal.

\subsection{3C\,296}

\begin{figure*}
\centerline{\vbox{\hbox{
\epsfig{figure=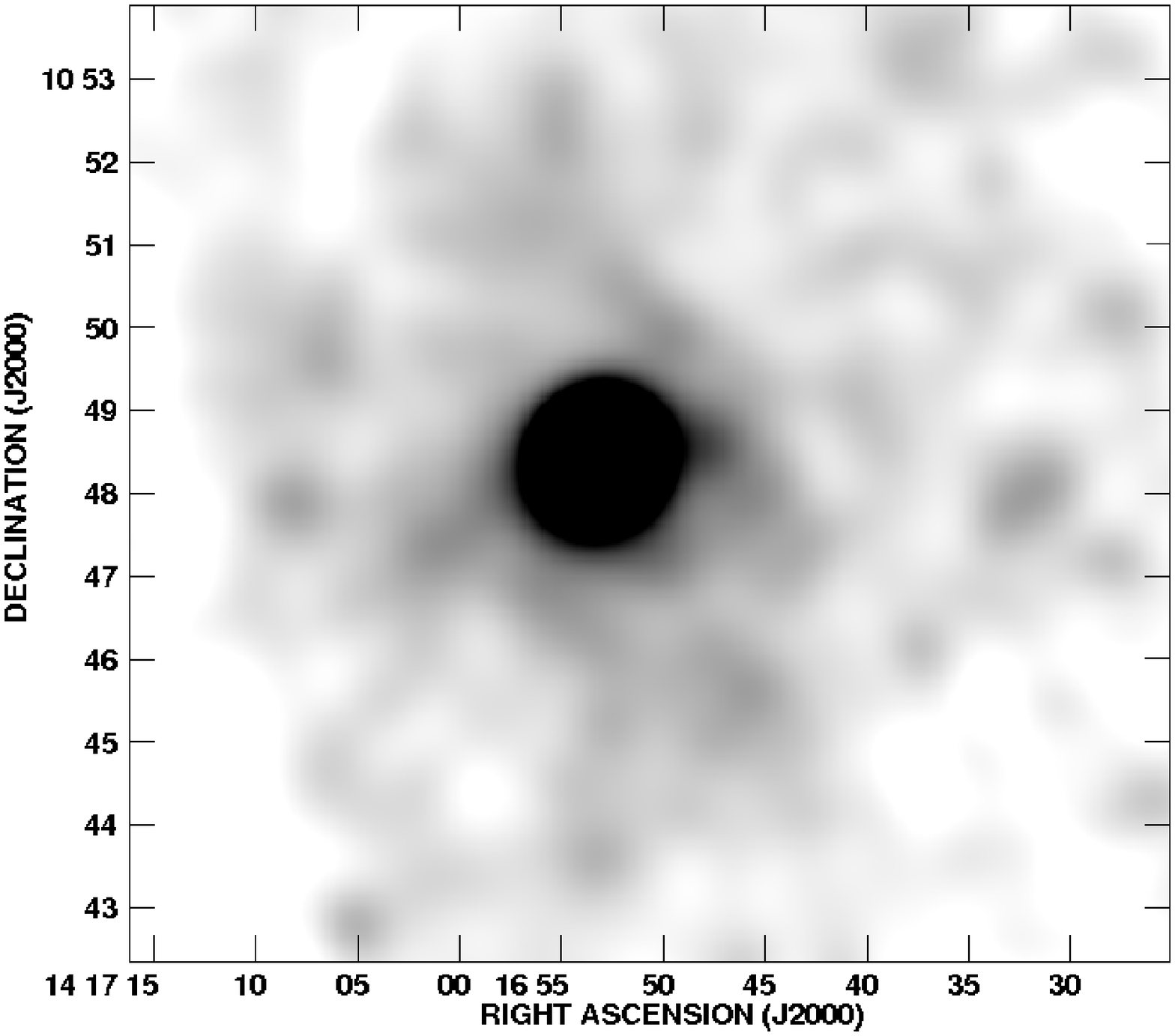, height=7cm}
\epsfig{figure=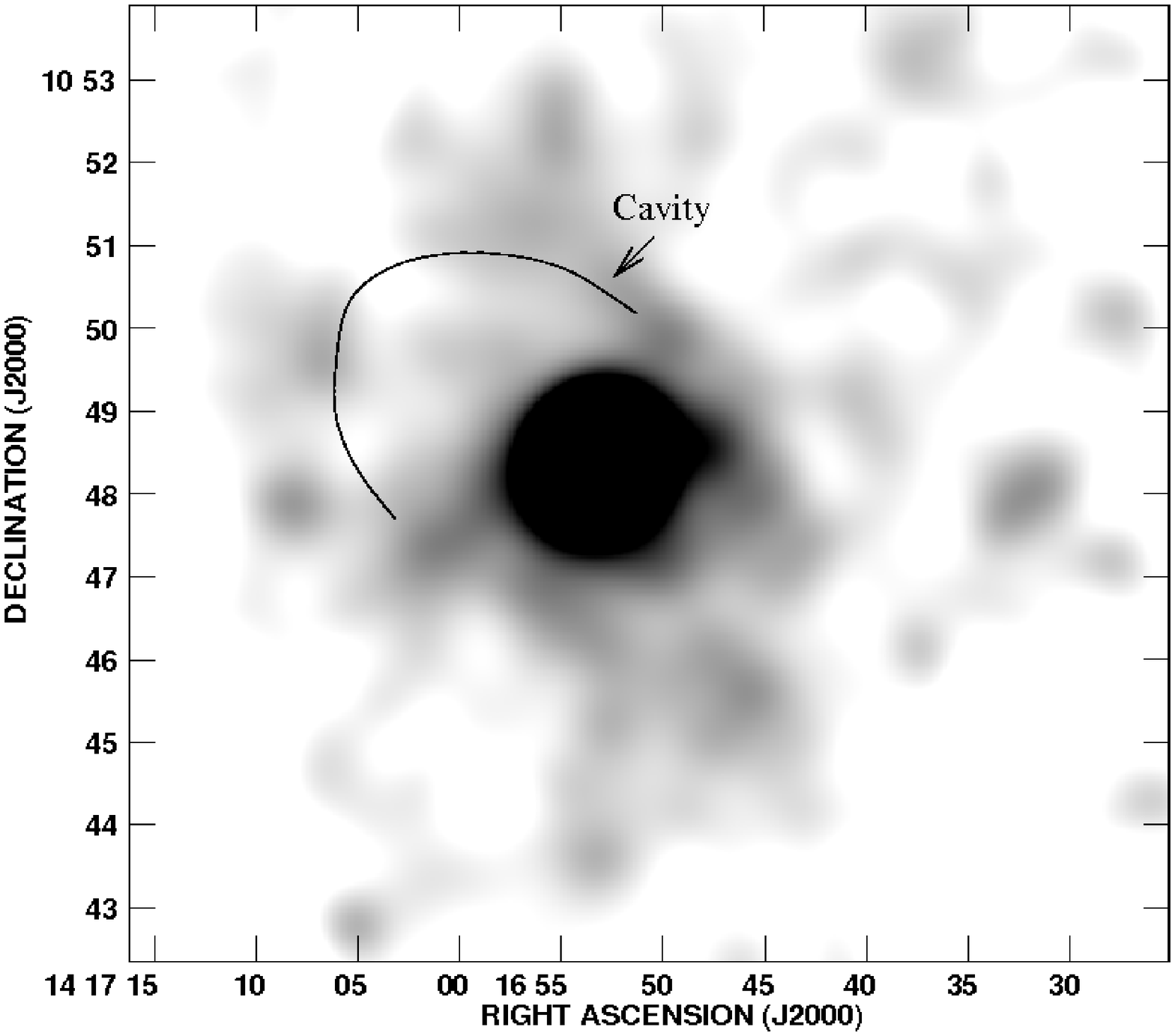,height=7cm}}
\centerline{\hbox{
\epsfig{figure=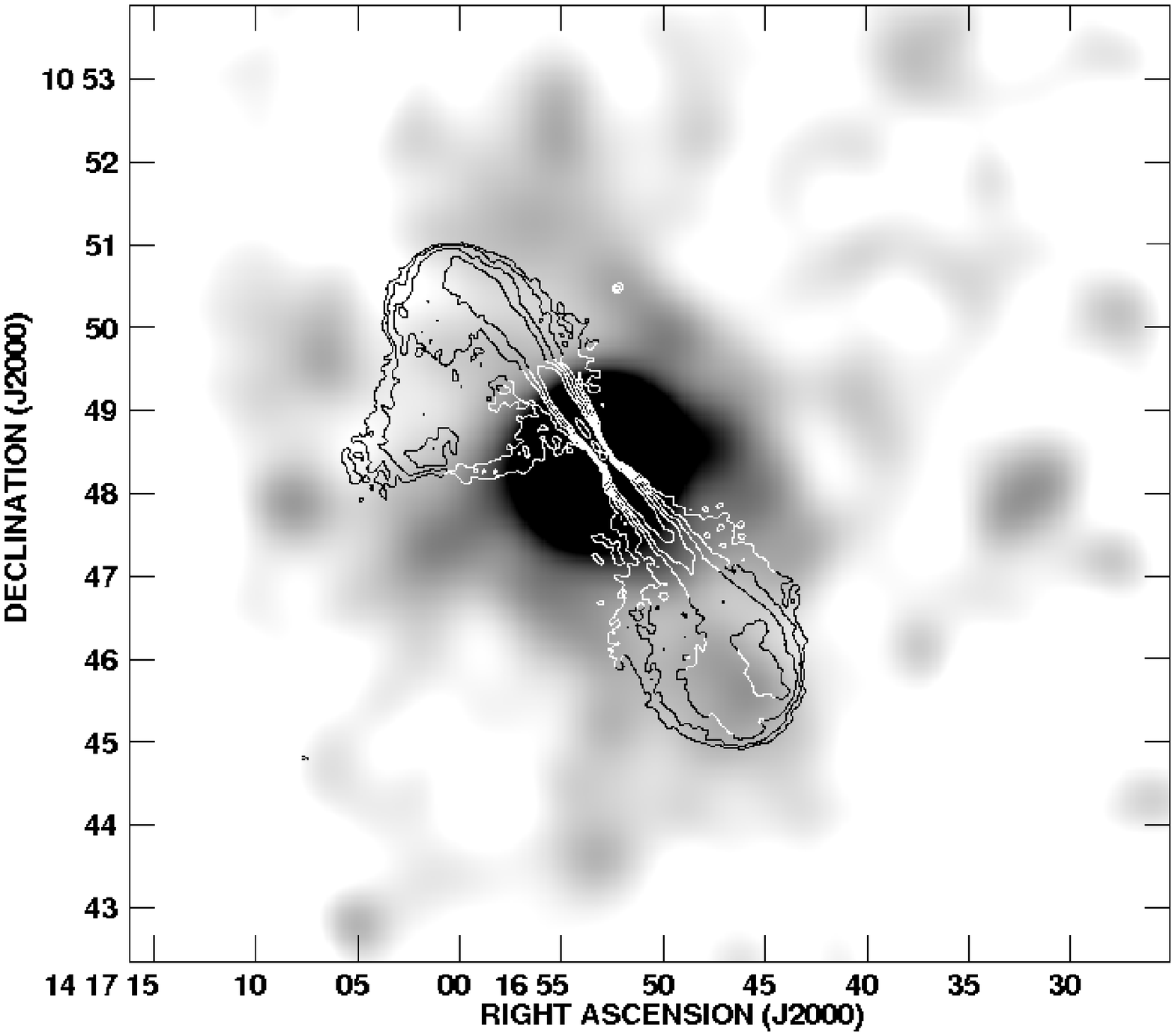,height=7cm}
\epsfig{figure=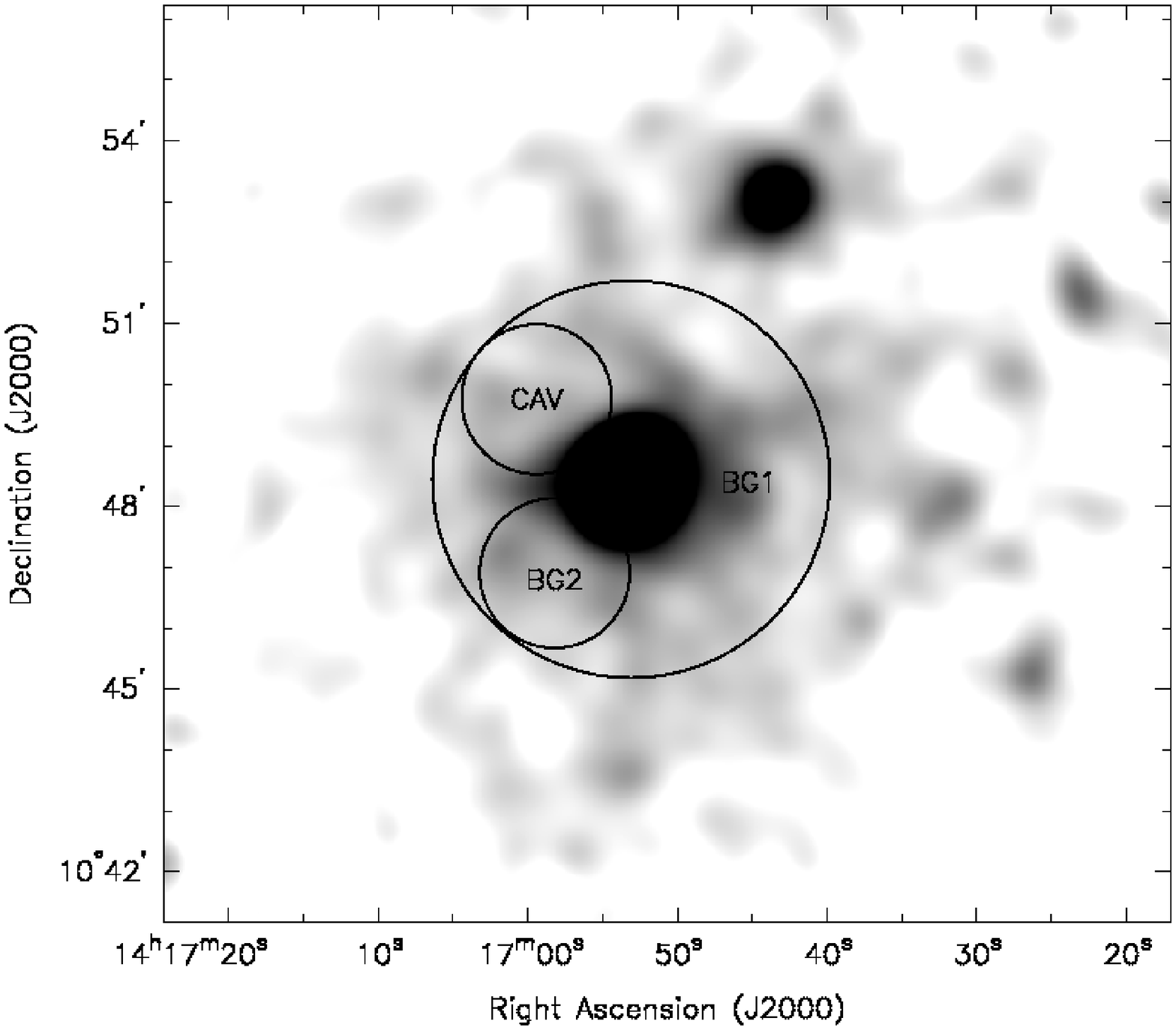,height=7cm}
}}}}
\caption{The X-ray environment of 3C\,296. Images made from the
  combined MOS1, MOS2 and pn events lists in the 0.5 - 5.0 keV energy
  range, with exposure correction to correct for chip gaps but not
  vignetting, as described in the text. Images are smoothed with a
  Gaussian of FWHM 19 arcsec. Top left: group emission; top right:
  cavity to the NE highlighted; bottom left: with 20-cm radio contours
  overlaid illustrating that the northern lobe fills the cavity seen
  in the top right panel; bottom right: regions used to investigate
  significance of cavity detection.}
\label{296im}
\end{figure*}

As shown in the left hand panel of Fig.~\ref{296im}, 3C\,296 possesses
a group environment of moderate X-ray surface brightness. This is the
first detection of a group-scale environment for this radio galaxy. We
measure a total of $\sim 6800$ 0.3 - 7.0 keV pn counts from the
group environment. 

Fig.~\ref{296im} reveals a cavity at the position of the NE lobe. We
 tested the statistical significance of the surface brightness deficit
 by comparing the total (MOS1 + MOS2 + pn) surface brightness in a
 circle of radius 74 arcsec defined using the radio map, in an annular
 background region centred on the nucleus of 3C\,296 extending from
 the inner to the outer edge of the cavity region, but excluding the
 north-east quadrant, and in a same-sized circular region rotated
 about the centre of the group to the south. The cavity has a surface
 brightness of $0.0736\pm0.002$ counts arcsec$^{-2}$, compared to a
 surface brightness of $0.0870\pm0.001$ counts arcsec$^{-2}$ in the
 annular background region, and a surface brightness of
 $0.0927\pm0.002$ counts arcsec$^{-2}$ in the same-sized region. The errors on the surface brightness measurements were obtained as
 for NGC\,1044. Hence the deficit is significant at a level of $\sim
 4.5\sigma$.

\begin{figure*}
\centerline{\vbox{\hbox{
\epsfig{figure=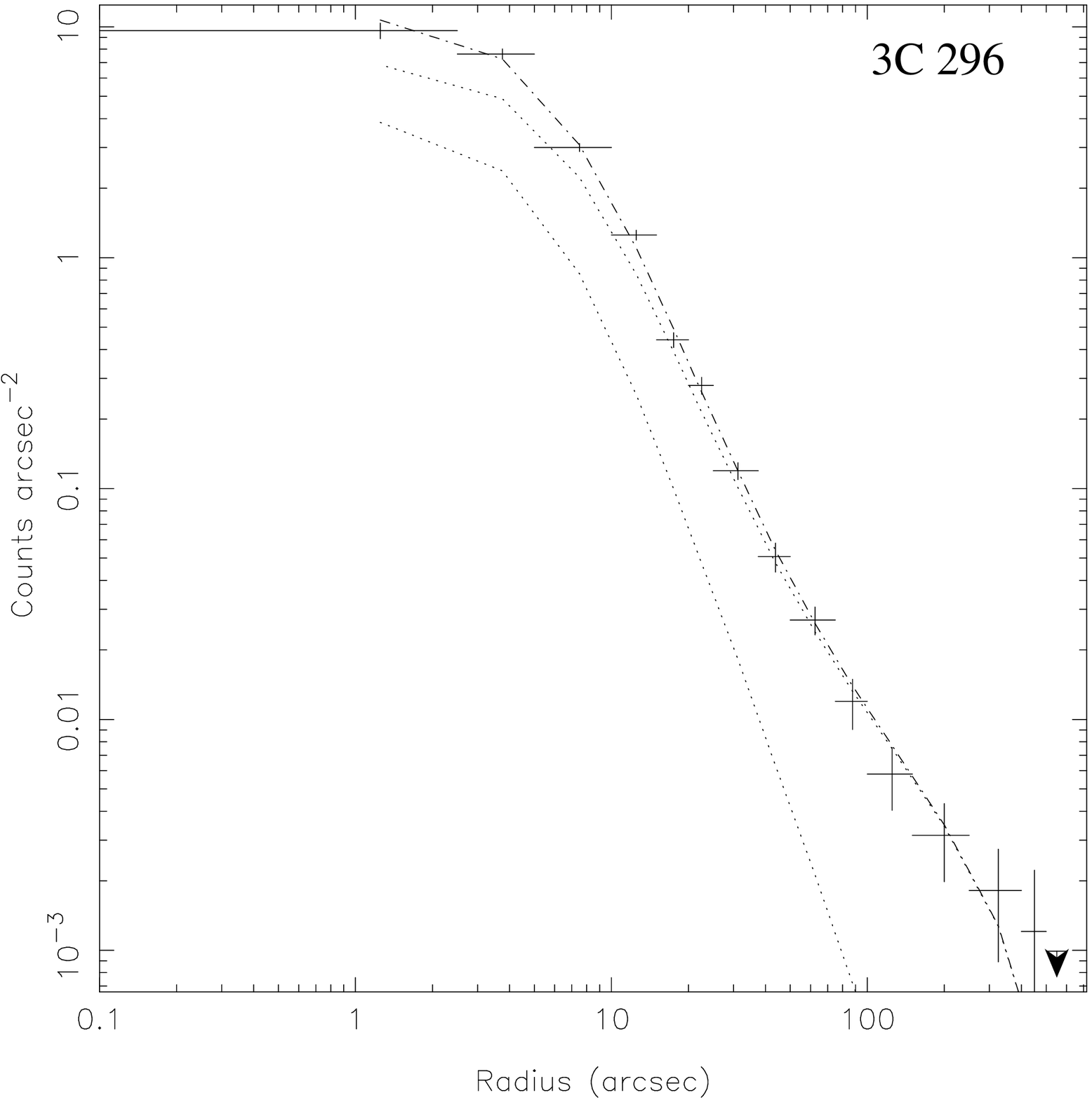,height=5.0cm}
\epsfig{figure=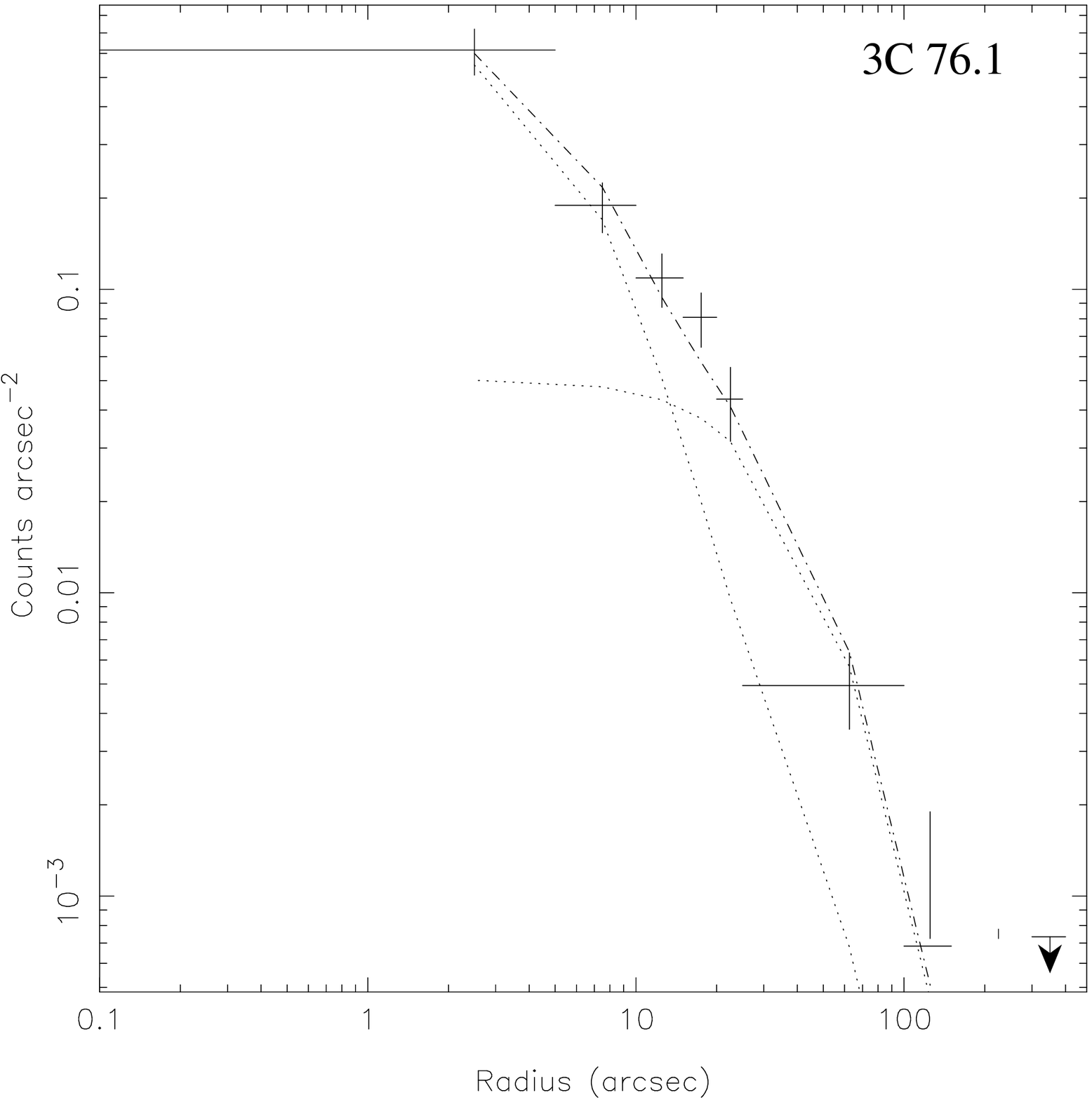,height=5.0cm}
\epsfig{figure=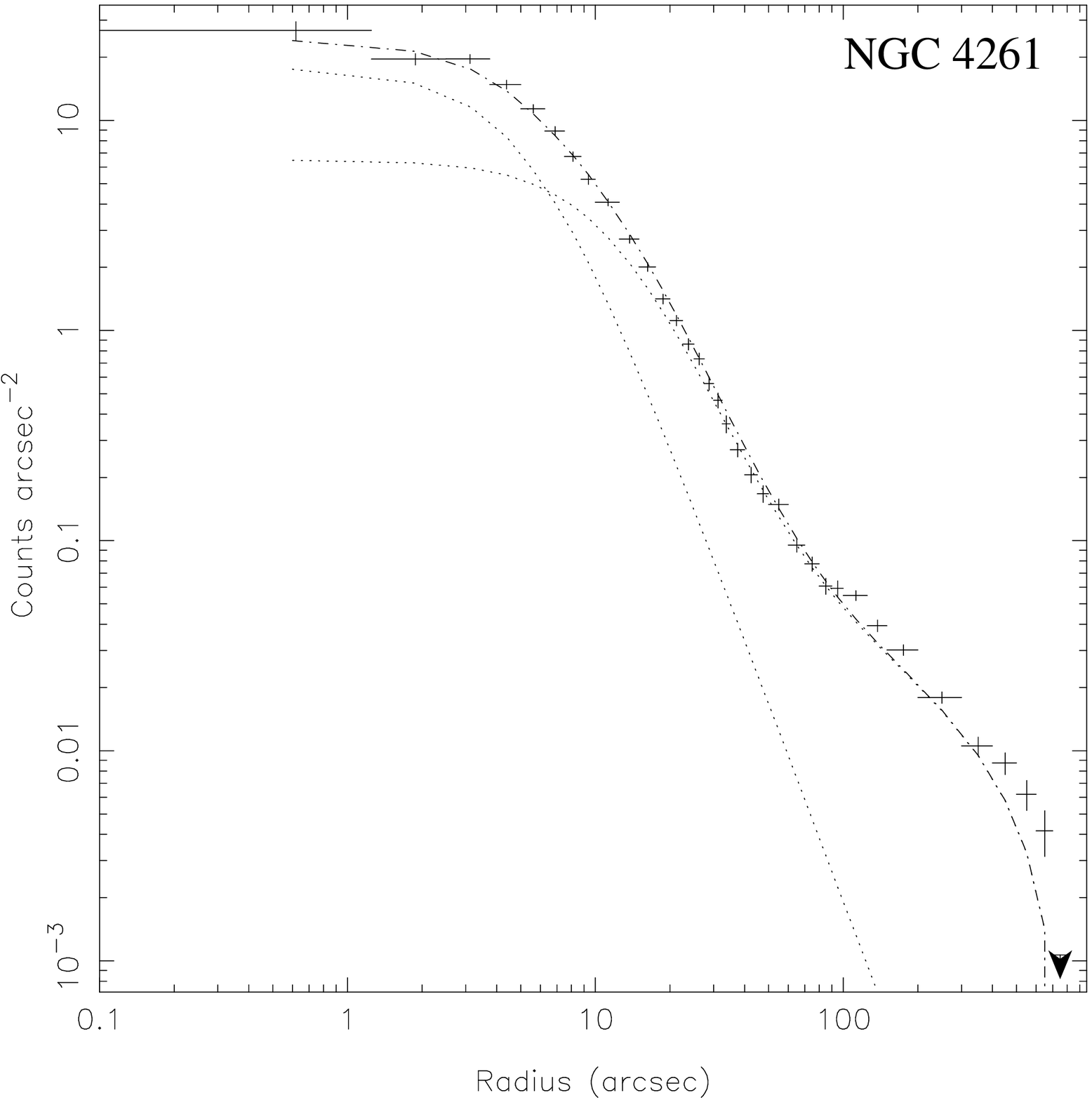,height=5.0cm}}
\hbox{
\epsfig{figure=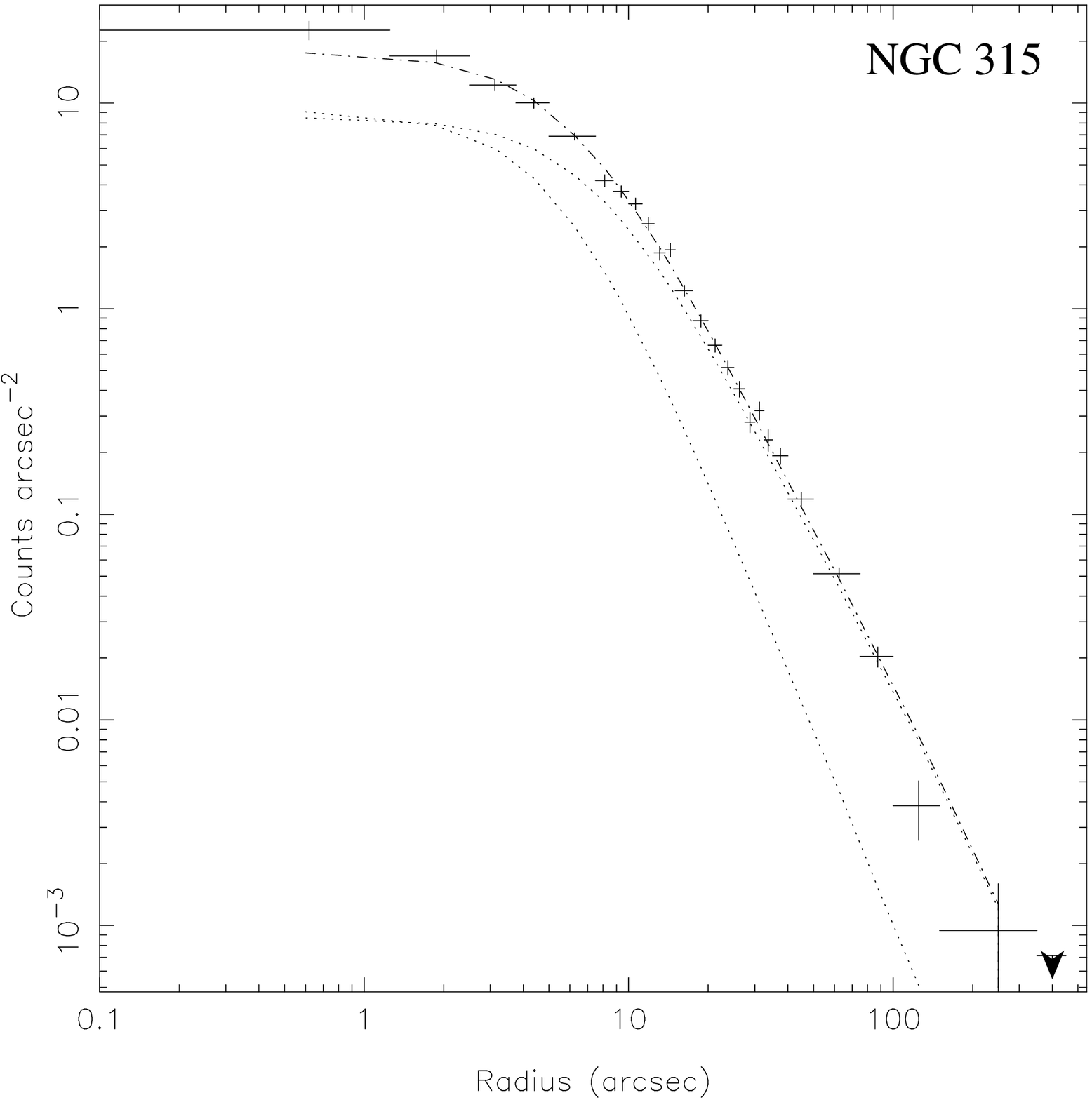,height=5.0cm}
\epsfig{figure=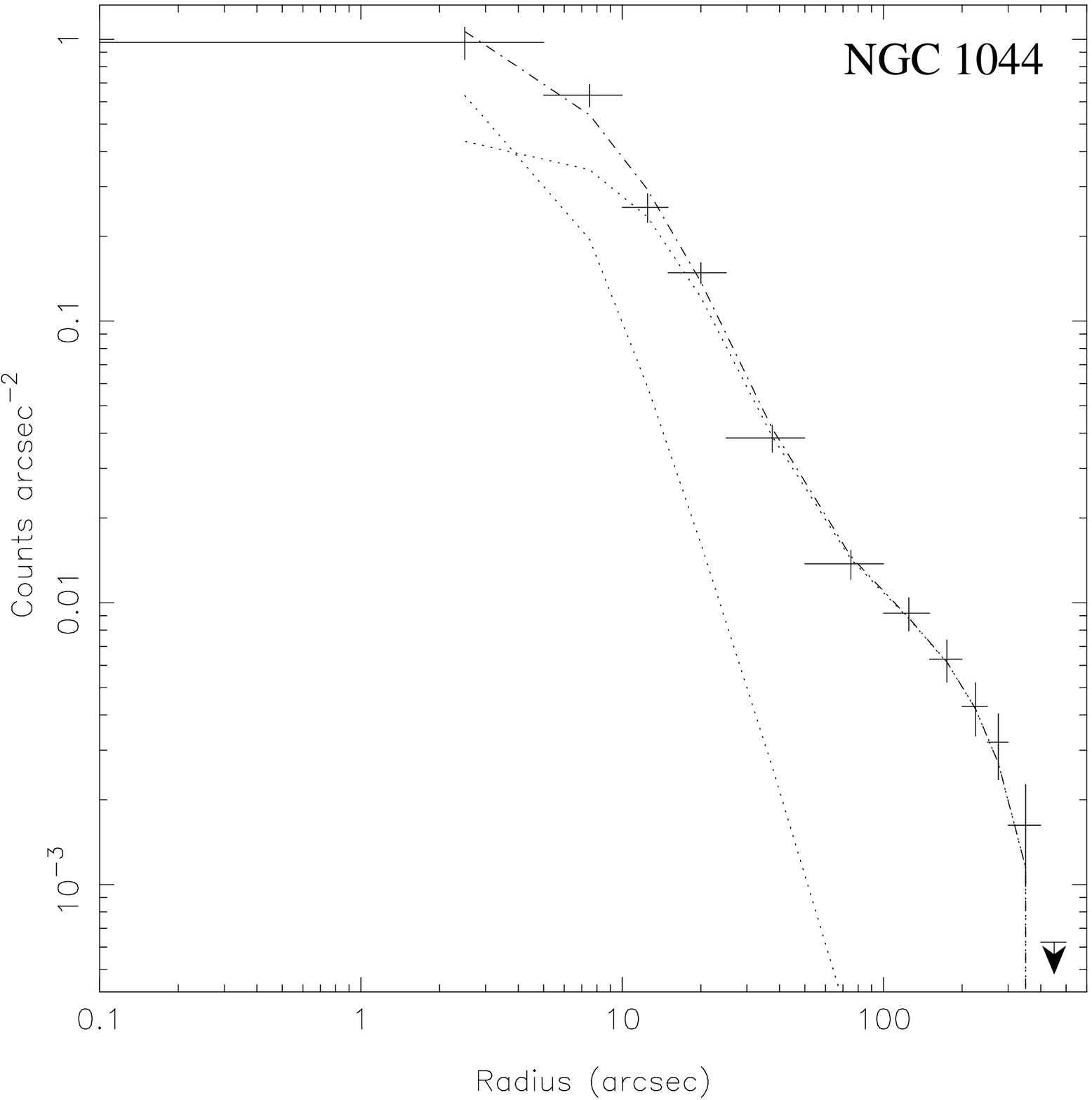,height=5.0cm}}}}
\caption{ EPIC pn surface brightness profile with best-fitting point
  source plus $\beta$ model or $projb$ model, as discussed in the
  text, for the newly observed galaxies. Top row (l-r): the bridged
  sources 3C\,296, 3C\,76.1, NGC\,4261; bottom row: the plumed sources
  NGC\,315 and NGC\,1044. (The somewhat poor fit in the outer regions
  for NGC\,4261 is because the model is from the best joint fit, rather
  than the best fit to the pn profile alone).}
\label{sxprofs}
\end{figure*}

3C\,296 has previously been observed with {\it Chandra} \citep{h05}.
These observations revealed the presence of a galaxy-scale hot-gas
component with $kT = 0.7$ keV extending to distances of $\sim 1$
arcmin. This component dominated over the compact emission associated
with the AGN beyond $\sim 1$ arcsec. Our {\it XMM-Newton} surface
brightness profile for 3C\,296 (top lefthand panel of
Fig.~\ref{sxprofs}) is dominated by the galaxy-scale component to a
distance of $\sim 60$ arcsec, beyond which the profile flattens
slightly. 

As for NGC\,315, we carried out a joint fit to the {\it Chandra} and
{\it XMM-Newton} profiles for 3C\,296, so as to provide the best
possible constraints on both the inner and outer gas properties. A
single $\beta$ model is an unacceptable fit to the profiles, due to
the excess of emission at large radii. We therefore used the
point-source plus $projb$ model to model both the inner and outer
components, finding good fits, as listed in Table~\ref{sxfits}. The
inner parameters we measure are in good agreement with the
best-fitting parameters of a single $\beta$ model fitted to the {\it
Chandra} profile alone, $\beta = 0.56\pm0.01$ and $r_{c} =
0.7^{+0.14}_{-0.10}$ arcsec \citep{h05}. Table~\ref{spectra} lists the
results of a spectral fit to the group emission in the region between
50 and 400 arcsec. As the fit is comparatively poor, we extracted
spectra in concentric annuli to investigate whether a temperature
gradient is present. The temperature was found to decrease from $\sim
1.4$ keV towards the centre to $\sim 0.9$ keV in the outer regions;
however, the temperatures are consistent within the errors, so that no
significant gradient can be claimed. We used the fitted surface
brightness profile model to determine profiles of gas density and
pressure, assuming an isothermal atmosphere. Density and pressure
profiles, and their uncertainties were determined as for the other
sources. While the {\it Chandra} observation of 3C\,296 showed a
significant temperature gradient in the inner regions of the
galaxy-scale emission, the inner temperature beyond 4 arcsec differs
by at most 10 per cent from the {\it XMM-Newton} global temperature, and so
this difference is accounted for by the comparatively large
uncertainty of the {\it XMM-Newton} global temperature, which we
include in the pressure profile uncertainties.

We also extracted a spectrum for the core region and for the nearby
group galaxy NGC\,5531 ($z=0.026$), shown to the north-west in
Fig.~\ref{296im}. As found by \citet{h05} a power-law with only
Galactic absorption was not a good fit to the core spectra. In good
agreement with the {\it Chandra} results, we found a good fit to an
absorbed power-law plus {\it mekal} model, with $kT =
0.74^{+0.3}_{-0.5}$ keV, $N_{H}= (2\pm1) \times 10^{21}$ cm$^{-2}$,
and $\Gamma=1.6\pm0.4$. The gas temperature is somewhat higher than
that measured by {\it Chandra}, which is not surprising as our region
encompasses more of the galaxy atmosphere, which is hotter on scales
of tens of kpc. For NGC\,5531, we found a good fit to a thermal model,
with $kT=0.60^{+0.12}_{-0.14}$ keV and $Z = 0.2^{+0.7}_{-0.1}
Z_{\sun}$ ($\chi^{2} = 17.2$ for 14 d.o.f.), in contrast to the
results with {\it Chandra}, which required the inclusion of the
power-law component. When a power-law component was included, its
normalization tended to negligible values, and the fit was not
improved. We measure an unabsorbed 2-10 keV flux density a factor of
$\sim 17$ lower than that reported by \citet{h05}, and so variability
of an AGN component may be the explanation for the differing spectral
results.


We can use the observed cavity associated with the eastern lobe to
 obtain an estimate for the jet power for 3C\,296. If we use the sound
 crossing time to the edge of the lobe ($2 \times 10^{8}$ yr) as an
 upper limit on the source age, and calculate the energy required to
 inflate the lobe as $4P_{mid}V$, where $P_{mid}$ is the external
 pressure at the midpoint of the lobe (see Table~\ref{properties}),
 and $V$ is the volume estimated from the extent of radio-lobe
 emission (this is preferable to estimating the cavity size directly
 from the X-ray data, due to {\it XMM-Newton}'s spatial resolution and
 the need for smoothing of the X-ray data to reveal the cavity
 structure), we find that the energy required to inflate the lobe is
 $\sim 3.9 \times 10^{51}$ J ($3.9 \times 10^{58}$ ergs),
 corresponding to a jet power of $1.2 \times 10^{36}$ W ($1.2
 \times 10^{43}$ erg s$^{-1}$) for the two jets, assuming the western
 jet has the same power. This inferred jet power is roughly an order
 of magnitude higher than the X-ray luminosity of the 3C\,296 group,
 so that significant heating is occuring in addition to the energy
 input required to balance radiative losses. The cavity properties are
 within the range shown in Fig.~8 of \citet{mcn07}, but lie to the
 high cavity power side of the cavity power/X-ray luminosity
 distribution. This jet power estimate is an order of magnitude lower
 than the estimate for 3C\,31 from the kinematic model of
 \citet{lb02a}, which is somewhat surprising given that 3C\,296 has a
 higher luminosity at 178 MHz (see Table~\ref{properties}). The sound
 crossing time is likely to be an overestimate of the lobe expansion
 time, as the lobes appear to be slightly supersonic in their outer
 regions at the present time, and were probably more highly
 overpressured at earlier times. Hence our calculation of the energy
 input from 3C\,296 discussed above may be an underestimate of the
 true value.

\section{Results for 3C\,66B, 3C\,449 and NGC\,6251}

For the three radio galaxies where we have previously published
detailed analyses of the environmental properties as probed by {\it
XMM-Newton} we chose to make use of the results published in
\citet{c03b} and \citet{ev05}, where very similar analysis procedures
were used. In Table~\ref{spectra}, we list the global gas properties
obtained from single thermal model fits to the group gas as reported
in \citet{c03b} and \citet{ev05}. To enable a fair comparison with the
pressure profiles for the newly observed objects, we refitted the
surface brightness profiles for the three previously observed sources
with the point-source plus $projb$ model where an inner component is
detected (3C\,66B and NGC\,6251) or a point-source plus $\beta$ model
(3C\,449) and used the Markov-Chain Monte Carlo method, which
improves the sampling of parameter space. For NGC\,6251, we included
the {\it Chandra} profile of \citet{ev05} to better constrain the
small core-radius inner component. The results of these fits are
listed in Table~\ref{sxfits}. We note that the quality of the fits for
3C\,66B and 3C\,449 is poor, as was the case for the single $\beta$
model fits to the outer regions reported in \citet{c03b}. We attribute
this to the very high signal-to-noise data for these two sources, so
that real small-scale deviations in the gas distributions may be
affecting the quality of the fit.

\begin{figure*}
\centerline{\vbox{\hbox{
\epsfig{figure=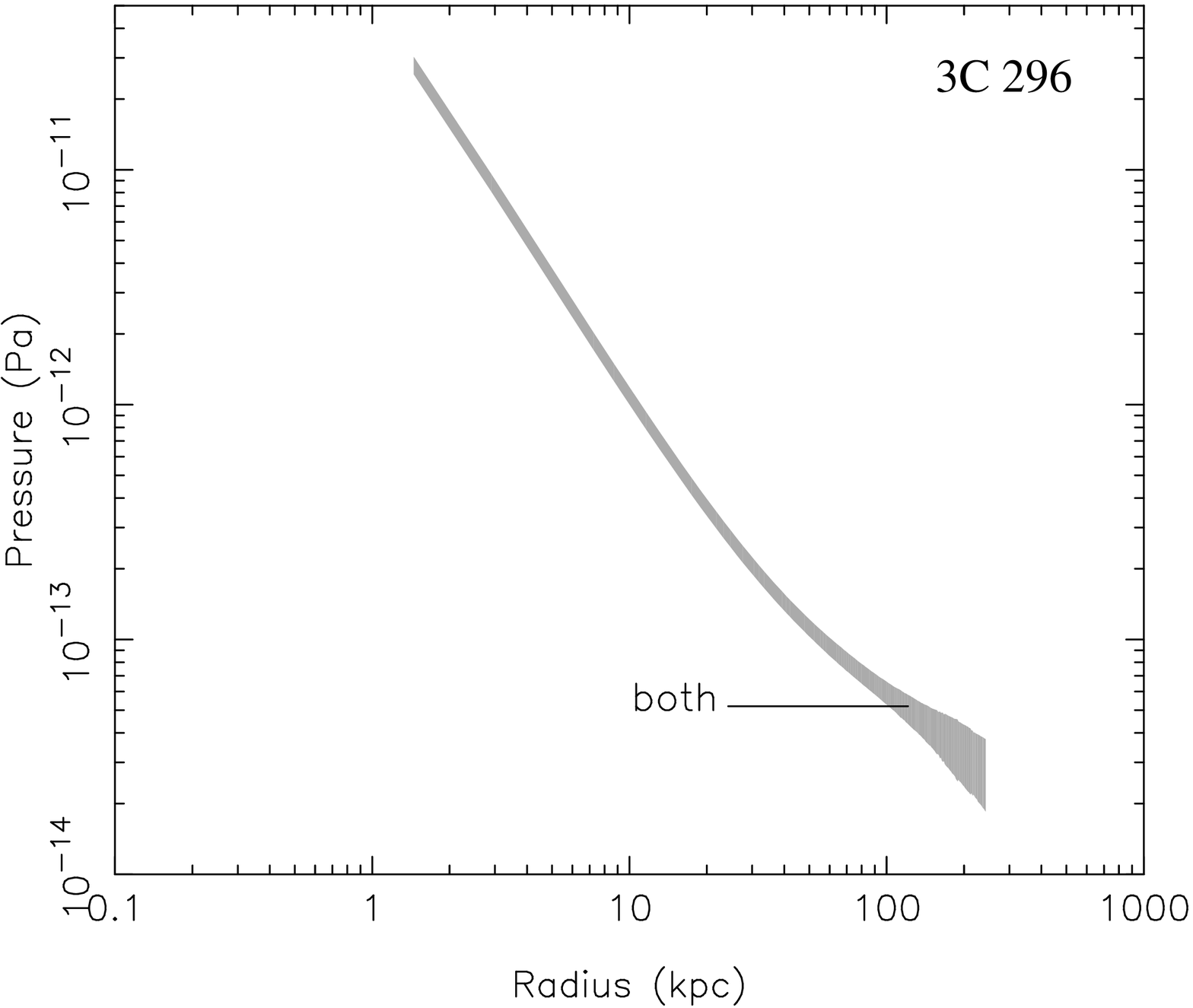,height=5.0cm}
\epsfig{figure=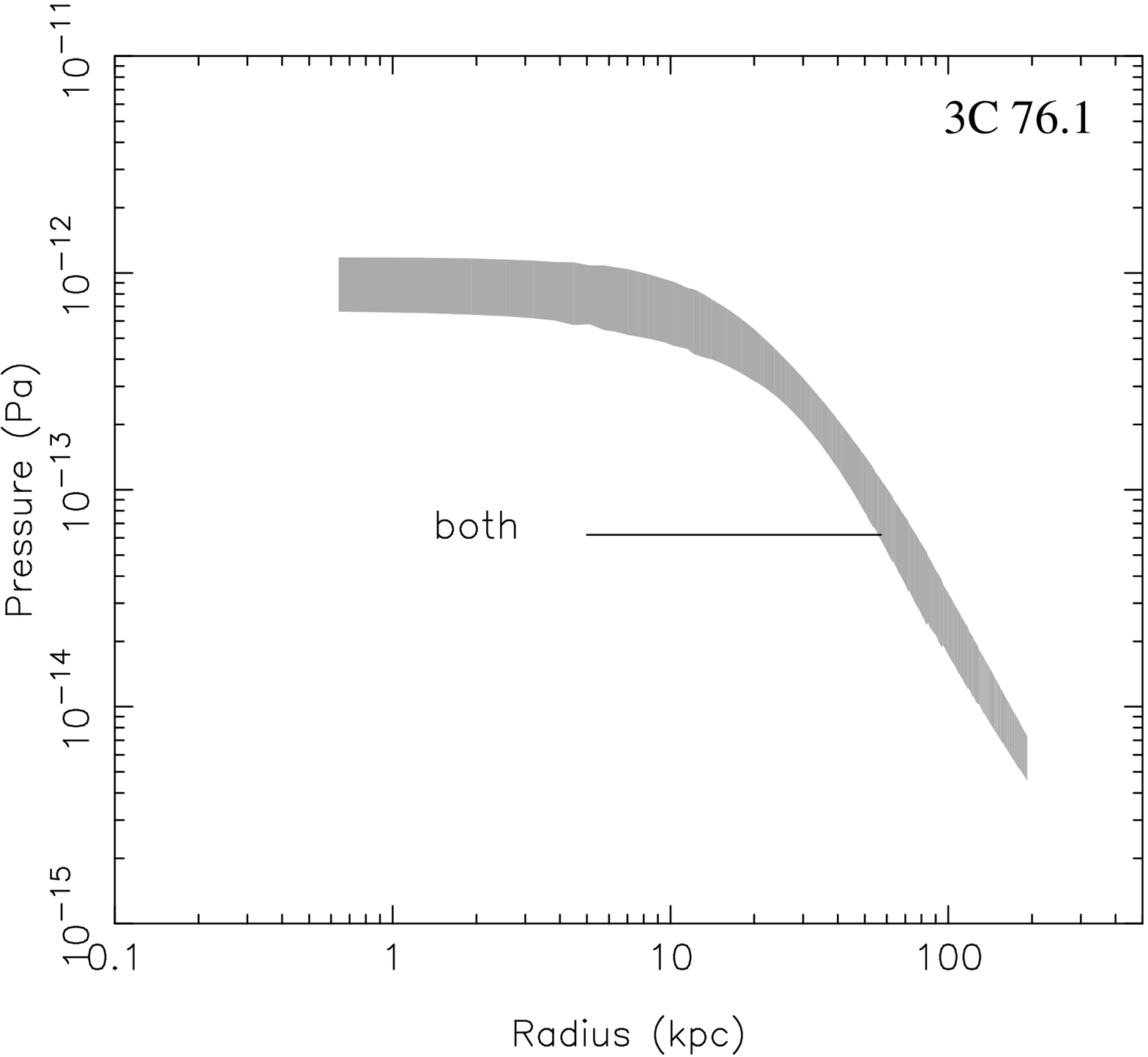,height=5.2cm}
\epsfig{figure=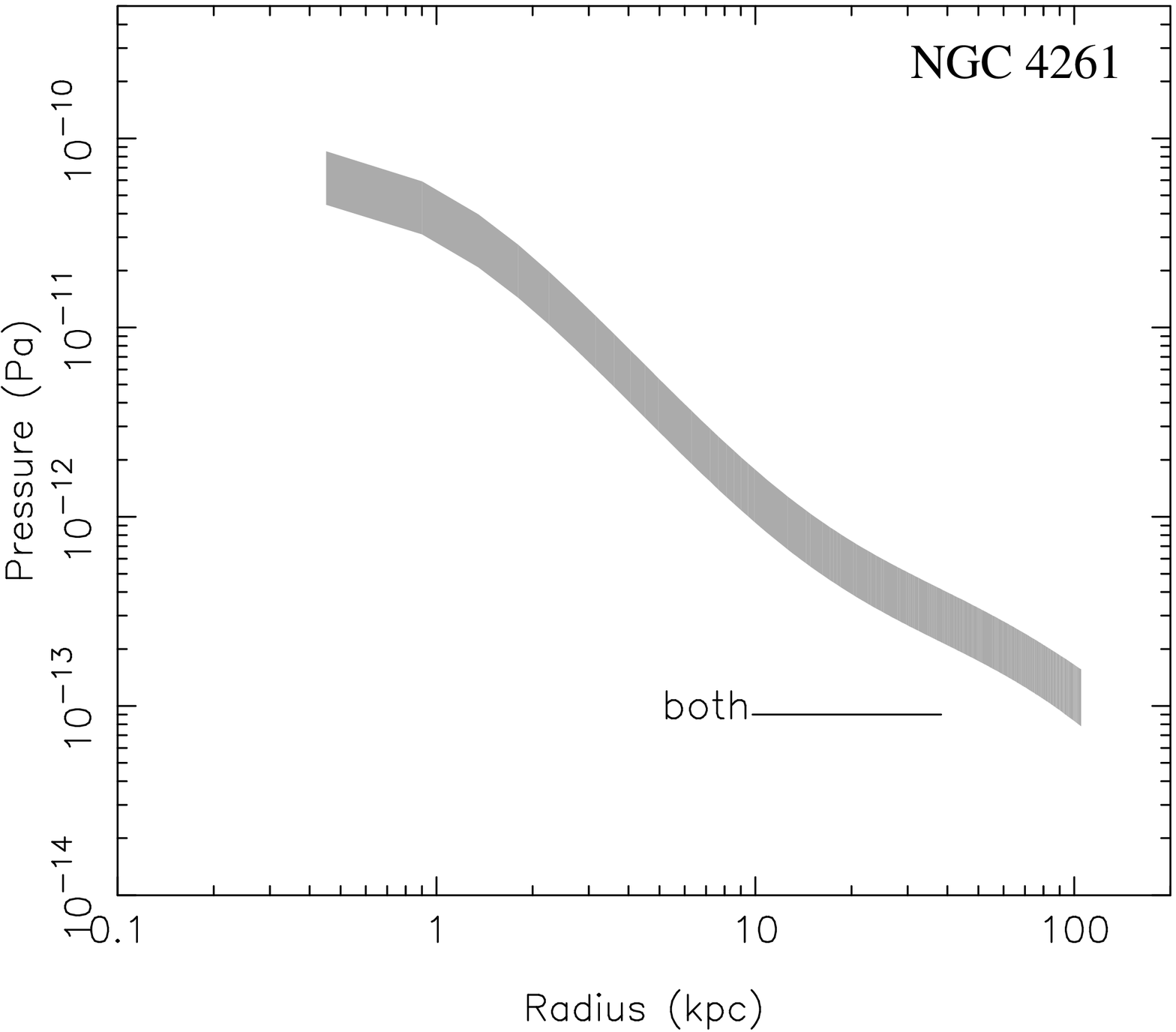,height=5.0cm}}
\hbox{
\epsfig{figure=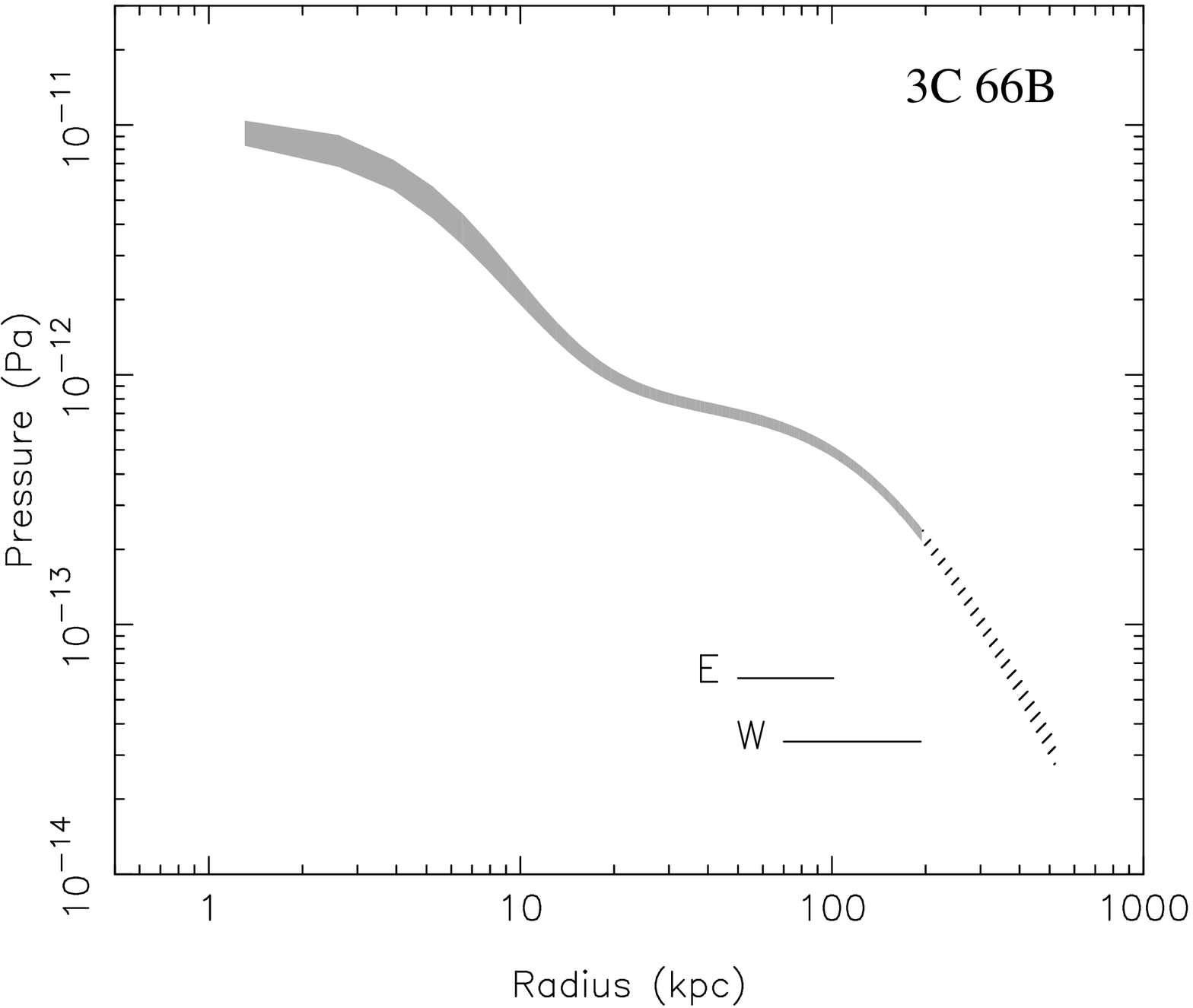,height=5.0cm}
\epsfig{figure=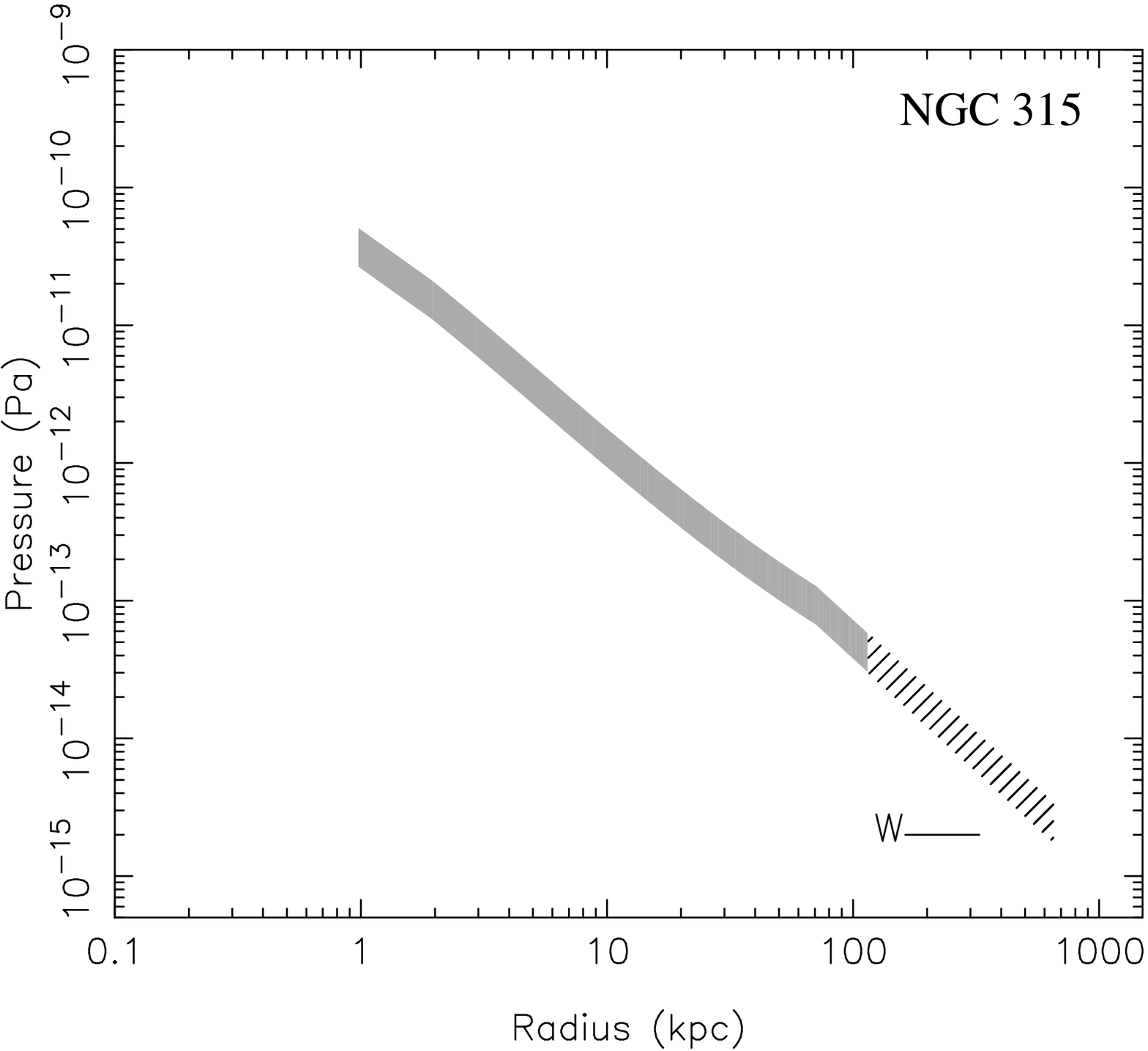,height=5.0cm}
\epsfig{figure=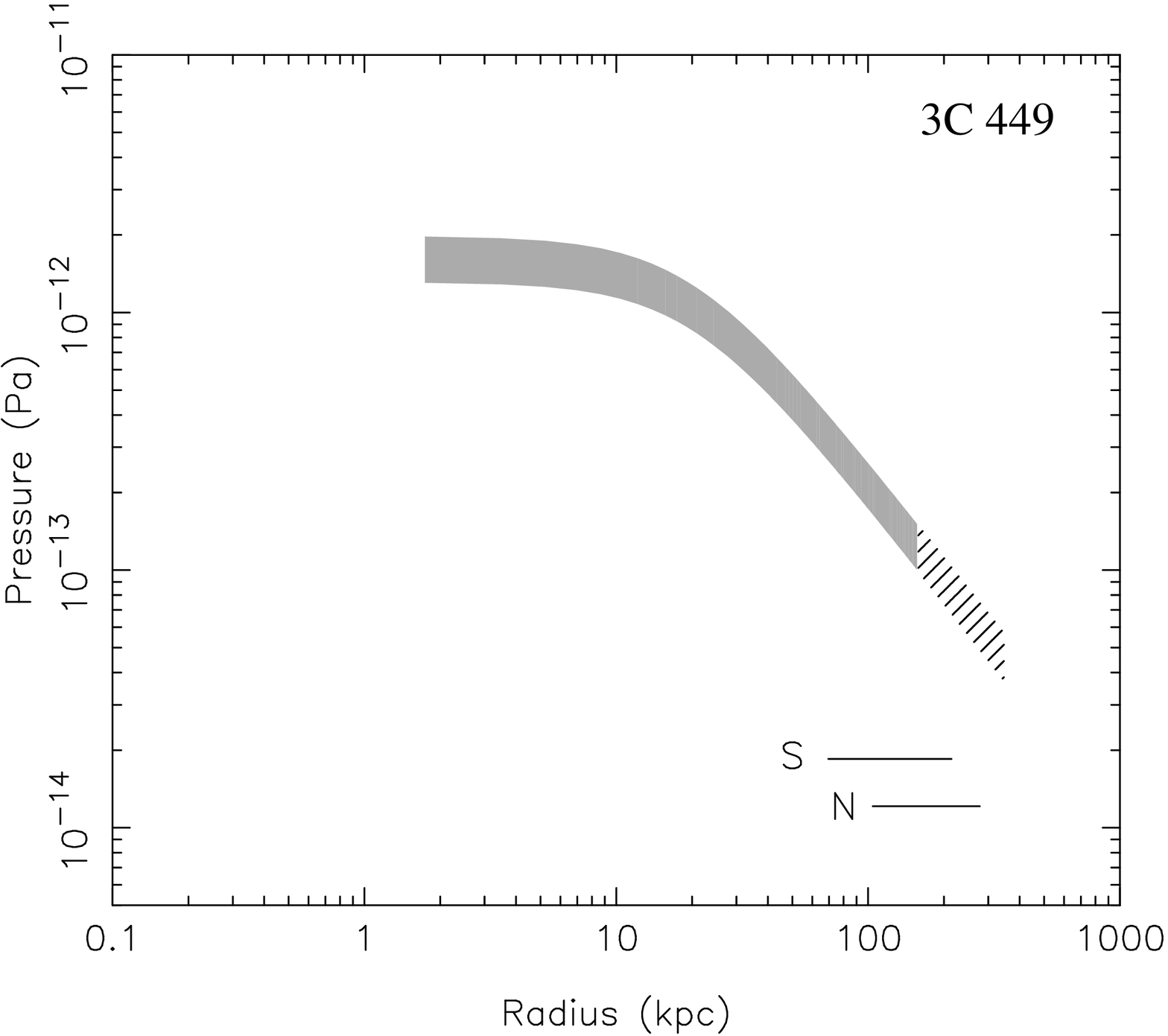,height=5.0cm}}
\hbox{
\epsfig{figure=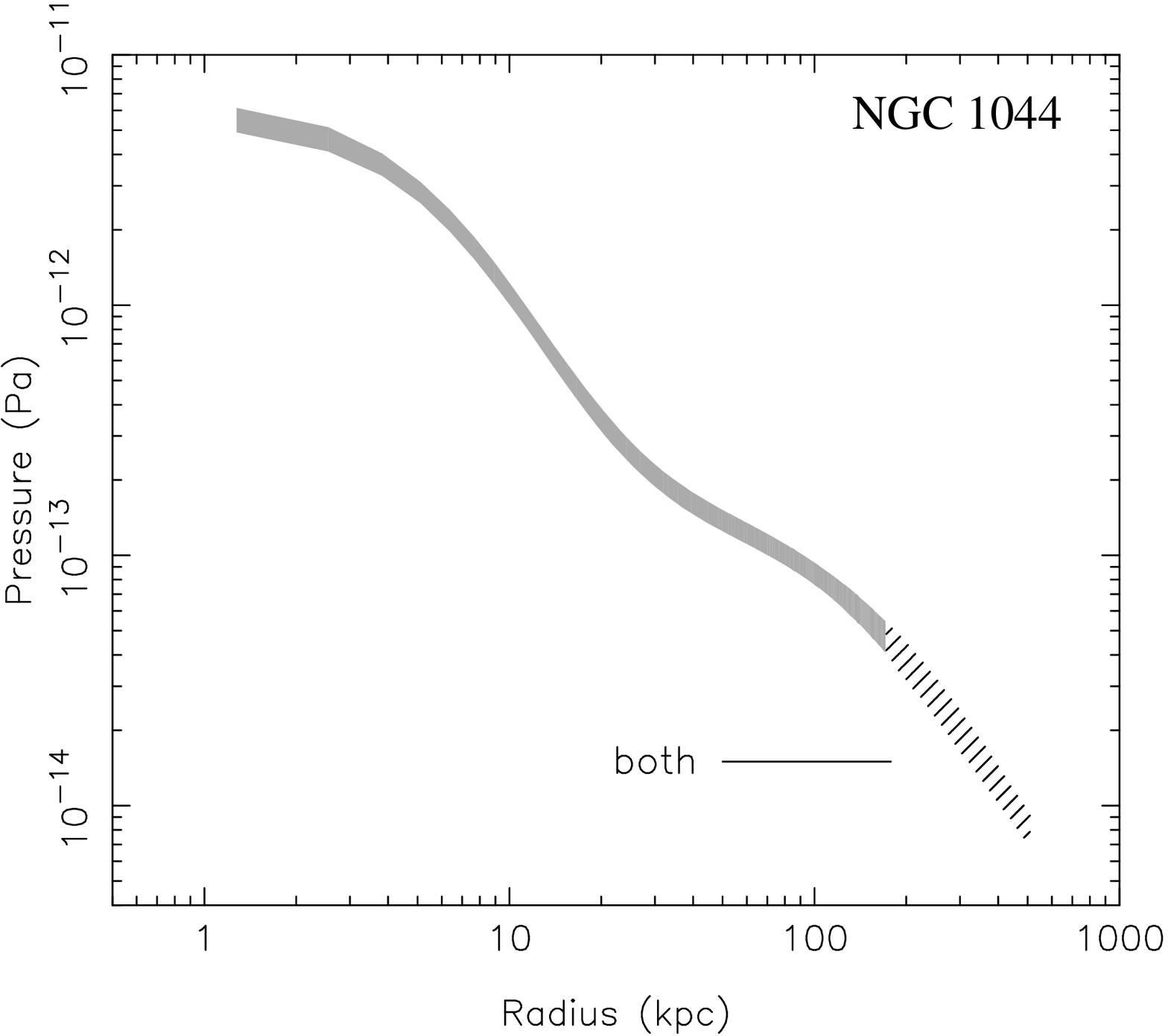,height=5.2cm}
\epsfig{figure=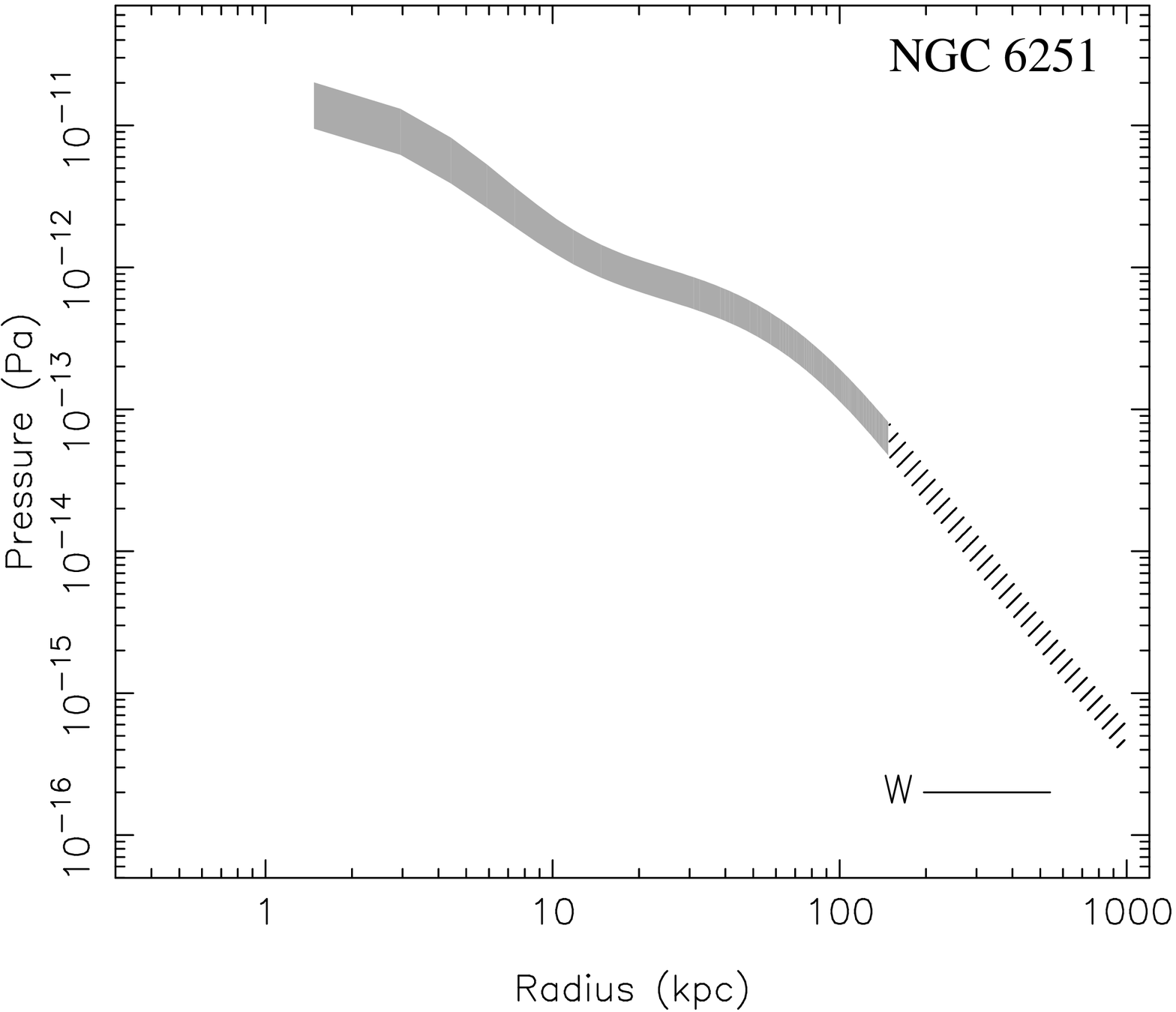,height=5.0cm}}}}
\caption{Pressure profiles for the {\it XMM}-observed radio-galaxy
  environments. The grey shaded area indicates the $1\sigma$
  uncertainty, which takes into account uncertainties in all the model
  parameters and in the temperature measurement. Hatched regions
  indicate an extrapolation of the pressure profile model beyond the
  region covered by the data. Horizontal lines indicate internal
  pressures of radio lobes. Top row (l-r): the bridged sources
  3C\,296, 3C\,76.1, NGC\,4261; middle row: the plumed sources
  3C\,66B, NGC\,315, and 3C\,449; bottom row: NGC\,1044, and
  NGC\,6251.}
\label{pressures}
\end{figure*}

\begin{table*}
\caption{Spectral fits to the large-scale group emission for each
  environment. Results are for joint fits to the three {\it
  XMM-Newton} cameras in the energy range 0.3 -- 7.0 keV. Luminosities
  are the Bayesian estimates for the bolometric luminosities to a
  radius of $r_{500}$ (see Table~\ref{properties}) with $1\sigma$
  errors, as determined from our surface brightness profile model
  fits.}
\label{spectra}
\vskip 10pt
\begin{tabular}{lrrrrrrr}
\hline
Object&$z$&$N_{H}$&$kT$&$Z$&$\chi^{2}$ (d.o.f.)&$\log(L_{X})$&Notes\\
&&($10^{20}$ cm$^{-2}$)&(keV)&($Z_{\sun}$)&&(erg s$^{-1}$)&\\
\hline
NGC\,315&0.016&5.88&$1.3^{+0.5}_{-0.3}$&0.3 (fixed)&117 (84)&$41.89\pm0.01$&\\
3C\,31&0.017&5.39&$1.5\pm0.1$&0.35 (fixed)&25 (22)&$43.10\pm0.01$&Results from OP04\\
3C\,66B&0.0215&8.36&$1.73^{+0.03}_{-0.04}$&$0.28\pm0.03$&819 (718)&$43.03^{+0.03}_{-0.11}$&See
C03 for details\\
NGC\,1044&0.021&8.76&$1.13^{+0.14}_{-0.06}$&$0.64^{+1.3}_{-0.22}$& 375
(330)&$42.13\pm0.08$&\\
3C\,76.1&0.032&10.8&$0.91^{+0.25}_{-0.14}$&0.3 (fixed) &26.6 (25)&$41.39^{+0.02}_{-0.12}$&\\
NGC\,4261&0.0073&1.55&$1.45^{+0.23}_{-0.01}$&$0.26^{+0.15}_{-0.04}$&1186 (819)&$42.07^{+0.21}_{-0.40}$&\\
3C\,296&0.024&1.86&$0.90^{+0.10}_{-0.04}$&$0.14^{+0.06}_{-0.08}$&1071 (728)&$42.40^{+0.13}_{-0.29}$&\\
NGC\,6251&0.0244&5.82&$1.6^{+0.5}_{-0.3}$&$0.4^{+0.7}_{-0.3}$&119
(109)&$41.93^{+0.02}_{-0.04}$& See E05 for details\\
3C\,449&0.0171&11.8&$0.98\pm0.02$&$0.13^{+0.01}_{-0.02}$& 881
(441)&$43.08\pm0.04$&See C03 for details\\
\hline
\end{tabular}
\begin{minipage}{17cm}
References: OP04 refers to \citet{op04}; C03 refers to \citet{c03b};
E05 refers to \citet{ev05}.
\end{minipage}
\end{table*}

\begin{table*}
\caption{Results of surface brightness profile fits. For each
  parameter we list the parameter value from the fit with the minimum
  $\chi^{2}$ and the Bayesian parameter estimate together with
  credible intervals taking into account all fit parameters. Column 2
  gives the model normalizations in electron density for the best fit
  in units of $10^{3}$ m$^{-3}$ -- note that for the $projb$ model this does
  not correspond exactly to the density at zero radius. Radii are
  given in units of arcsec.}
\label{sxfits}
\vskip 10pt
\begin{tabular}{|l|rrrrrrr|rrrrr|}
\hline
Object&\multicolumn{7}{|c|}{Best fit}&\multicolumn{5}{|c|}{Bayesian
  estimate}\\
&$n_{e,0}$&$\beta$&$r_{c}$&$\beta_{in}$&$r_{c,in}$&$N$&$\chi^{2}$ (dof)&$\beta$&$r_{c}$&$\beta_{in}$&$r_{c,in}$&$N$\\
\hline
NGC\,315&400&0.54&2.12&-&-&-&186 (126)&$0.54^{+0.02}_{-0.01}$&$2.13^{+0.45}_{-0.39}$&-&-&-\\
3C\,66B&12&1.2&428.2&1.2&17.6&0.083&470 (230)&$0.91^{+0.29}_{-0.47}$&$367^{+103}_{-155}$&$1.1^{+0.1}_{-0.4}$&$16.4^{+3.6}_{-10.1}$&$0.094^{+0.3}_{-0.07}$\\
NGC\,1044&14&1.2&416&0.84&13.3&0.0225&38.2 (28)&$0.81^{+0.39}_{-0.51}$&$356^{+640}_{-255}$&$0.88^{+0.32}_{-0.35}$&$14.2^{+5.8}_{-9.3}$&$0.0243^{+0.0254}_{-0.0185}$\\
3C\,76.1&2.6&0.94&42.1&-&-&-&11.3 (16)&$0.9^{*}$&$43^{*}$&-&-&-\\
NGC\,4261&77&0.30&243.6&0.64&7.0&0.00443&156
(100)&$0.62^{*}$&$482^{*}$&$0.64^{+0.35}_{-0.14}$&$7.0^{+3.0}_{-4.3}$&$0.0046^{*}$\\
3C\,296&720&0.31&77.5&0.60&0.95&0.0000446&80.5 (92)&$0.67^{*}$&$545^{*}$&$0.58^{+0.07}_{-0.03}$&$0.86^{+0.44}_{-0.31}$&$0.00021^{+0.0008}_{-0.0001}$\\
NGC\,6251&38&1.1&150&1.2&9.9&0.044&160 (139)&$1.0^{+0.2}_{-0.5}$&$138^{+76}_{-82}$&$0.96^{*}$&$8.1^{*}$&$0.053\pm0.050$\\
3C\,449&3.7&0.42&57.1&-&-&-&457 (273)&$0.42^{+0.06}_{-0.05}$&$58^{+19}_{-16}$&-&-&-\\
\hline
\end{tabular}
\begin{minipage}{17cm}
$^{*}$ Unconstrained within the parameter range explored. See
  Table~\ref{priors}.
\end{minipage}
\end{table*}

\begin{table*}
\caption{Bayesian priors for the surface brightness profile fits.
  Priors in core radius and relative normalisation were determined
  individually for each source by finding extreme fits that gave poor
  $\chi^{2}$ values.}
\label{priors}
\vskip 10pt
\begin{tabular}{lrrrrrrrrrr}
\hline
Object&$\beta_{min}$&$\beta_{max}$&$r_{c,min}$&$r_{c,max}$&$\beta_{in,min}$&$\beta_{in,max}$&$r_{cin,min}$&$r_{cin,max}$&$N_{min}$&$N_{max}$\\
\hline
NGC\,315&0.3&1.2&1&20&-&-&-&-&-&-\\
3C\,66B&0.3&1.2&75&1000&0.3&1.2&1&30&0.001&0.5\\
NGC\,1044&0.3&1.2&50&1000&0.3&1.2&0.5&20.0&0.0005&0.02\\
3C\,76.1&0.3&1.2&2&100&-&-&-&-&-&-\\
NGC\,4261&0.3&1.2&50.0&2000&0.3&1.2&0.1&10.0&0.00001&0.001\\
3C\,296&0.3&1.2&50&1000&0.3&1.2&0.1&10.0&0.00001&0.001\\
NGC\,6251&0.3&1.2&15&500&0.3&1.2&15&500&0.0001&0.1\\
3C\,449&0.3&1.2&10&200&-&-&-&-&-&-\\
\hline
\end{tabular}
\end{table*}
\begin{table*}
\caption{Properties of the radio galaxies and their group environments}
\label{properties}
\vskip 10pt
\begin{tabular}{lrrrrrrrrrrrrrrr}
\hline
Object&$L_{178}$&$D$&$r_{ax}$&$V$&$r_{500}$&$P_{E}$&$P_{M}$&$P_{int}$&$R_{E}$&$R_{M}$&$E_{req}$&$E_{rg}$\\
\hline
NGC\,315&23.38&1000&0.14&65.35&512&-14.13&-13.64&-14.70&3.7&11.4&$52.86^{+0.22}_{-0.51}$&52.31\\
3C\,31&23.96&307&0.16&63.85&558&-13.16&-12.60&-13.53&2.3&8.5&$52.99^{+0.2}_{-0.5}$&51.85\\
3C\,66B&24.34&296&0.35&64.27&603&-12.46&-12.22&-13.33&7.4&12.9&$52.83^{+0.15}_{-0.23}$&52.65\\
NGC\,1044&23.26&306&0.30&63.82&473&-13.25&-12.98&-13.82&3.7&6.9&$52.11^{+0.11}_{-0.16}$&51.44\\
3C\,76.1&24.40&126&0.71&63.98&418&-13.12&-12.58&-13.21&1.2&4.3&$51.99^{+0.14}_{-0.21}$&52.00\\
NGC\,4261&23.67&81&0.50&63.00&545&-12.51&-12.24&-13.04&3.4&6.3&$51.92^{+0.08}_{-0.09}$&51.36\\
3C\,296&24.16&212&0.46&64.25&415&-13.25&-12.98&-13.28&1.1&2.0&$<52.11$&51.87\\
NGC\,6251&24.08&1900&0.24&66.08&577&-14.63&-13.83&-15.70&12.0&74&$52.80^{+0.15}_{-0.22}$&52.85\\
3C\,449&23.81&533&0.26&64.19&436&-13.17&-12.80&-13.81&4.4&10.2&--&51.99\\
\hline
\end{tabular}
\begin{minipage}{17cm}
Columns: 1. Radio-source or galaxy name; 2. Logarithm of the 178-MHz
radio luminosity (W Hz$^{-1}$ sr$^{-1}$); 3. Radio source projected
physical size (kpc); 4. Axial ratio at widest part of radio source; 5.
Logarithm of the sum of the volumes of the two lobes (m$^{3}$),
determined by assuming a spherical or cylindrical volume based on the
extent of the radio emission; 6. $r_{500}$, the radius enclosing a
mean density of 500 times $\rho_{crit}$, determined using the relation
of \citet{app05} (kpc); 7.
Logarithm of the thermal gas pressure at the outer edge of the radio
lobes , averaged for the two lobes except in the cases of NGC\,315 and
NGC\,6251 (Pa); 8. Logarithm of the thermal gas pressure at the
midpoint of the radio lobes, averaged for the two lobes except in the
cases of NGC\,315 and NGC\,6251 (Pa); 9. Logarithm of the internal
equipartition pressure, averaged for the two lobes except in the cases
of NGC\,315 and NGC\,6251 (Pa); 10. Average pressure ratio at the lobe
endpoints; 11. Average pressure ratio at the lobe midpoint; 12.
Logarithm of the energy required to produce the observed temperature
offset from the RQ $L_{X}/T_{X}$ relation (J) (errors take into
account uncertainties in the observed and predicted temperatures); 13.
Logarithm of the energy available from the current episode of
radio-galaxy activity (4$P_{M} V$, J).
\end{minipage}
\end{table*}
\section{Discussion}
\label{disc}

Table~\ref{properties} summarizes the group properties as measured
from the {\it XMM-Newton} data. The three previously studied
radio-galaxy environments are included, with properties taken from
\citet{c03b} and \citet{ev05}. Properties of the radio galaxies
obtained from the radio maps and from the literature are also included
for comparison. It is clear from these results and from
Table~\ref{spectra} (and can also be seen from the X-ray images) that
the low-power radio galaxies in this ``representative'' sample inhabit
a wide range of environments, with bolometric X-ray luminosities
spanning nearly two orders of magnitude from $\sim 10^{41}$ erg
s$^{-1}$ to $10^{43}$ erg s$^{-1}$, consistent with earlier results
\citep{wb00}. The result is important for theoretical models of
radio-galaxy evolution and environmental impact, as it has sometimes
been assumed by modellers that environmental properties do not vary
significantly.

In the following sections, we discuss the relationship between
environmental properties and radio structure, source dynamics and
particle content.

\begin{table}
\caption{Correlations between radio-source and group gas properties,
  determined using the Spearman rank test. Column 3 is the value of
  Student's $t$ for the given comparison, and column 4 is the null
  hypothesis probability, i.e. the probability of obtaining the given
  value of Student's $t$ from an uncorrelated dataset.}
\label{spearman}
\vskip 10pt
\begin{tabular}{llrr}
\hline
Parameter 1&Parameter 2&$t$&NHP\\
\hline
$L_{R}$&$L_{X}$&0.088241&0.466\\
&$T_{X}$&0.044102&0.483\\
&$\beta$&1.396017&0.103\\
&$r_{c}$&0.683130&0.258\\
&$P_{ext}$&0.883277&0.203\\
&$P_{ext}/P_{int}$&0.447&0.334\\
&d$P$/d$r$&0.293695&0.389\\
$D$&$L_{X}$&0.221249&0.416\\
&$T_{X}$&0.883277&0.203\\
&$\beta$&0.221249&0.416\\
&$r_{c}$&1.272136&0.12\\
&$P_{ext}$&2.853810&0.012\\
&$P_{ext}/P_{int}$&2.263&0.029\\
&d$P$/d$r$&1.705606&0.070\\
$r_{ax}$&$L_{X}$&0.634866&0.273\\
&$T_{X}$&1.098087&0.154\\
&$\beta$&1.098087&0.154\\
&$r_{c}$&0.988538&0.178\\
&$P_{ext}$&1.819606&0.056\\
&$P_{ext}/P_{int}$&1.396&0.103\\
&d$P$/d$r$&1.506237&0.0914\\
$P_{int}$&$L_{X}$&0.088241&0.466\\
&$T_{X}$&0.781661&0.23\\
&$\beta$&0.176777&0.432\\
&$r_{c}$&0.935414&0.190\\
&$P_{ext}$&3.743997&0.0036\\
&d$P$/d$r$&1.326473&0.113\\
\hline
\end{tabular}
\end{table}

\begin{table}
\caption{Comparison of environmental properties for bridged and plumed
  sources. Column 2 gives the KS $D$ value for a comparison between
  the bridged and plumed subsamples, and column 3 gives the null
  hypothesis probability for the subsamples being drawn from the same
  parent population. For the pressure ratio comparison each radio lobe
  was treated separately, whereas the other comparisons are for the
  source as a whole. $E_{ratio}$ is the ratio between $E_{req}$ and
  $E_{rg}$ (see Table~\ref{properties}).}
\label{ks}
\vskip 10pt
\begin{tabular}{lrr}
\hline
Parameter&$D$&NHP\\
\hline
$L_{X}$&0.50&0.37\\
$T_{X}$&0.40&0.65\\
$\beta$&0.50&0.37\\
$r_{c}$&0.60&0.18\\
$P_{ext}$&0.47&0.45\\
$P_{ext}/P_{int}$&0.78&0.005\\
$P_{int}$&0.67&0.10\\
d$P$/d$r$&0.50&0.37\\
$E_{req}$&0.60&0.26\\
$E_{rg}$&0.43&0.55\\
$E_{ratio}$&0.60&0.26\\
\hline
\end{tabular}
\end{table}

\subsection{Relationship between properties of the radio galaxy and of the environment}
\label{xraydio}

The importance of jet/environment interactions for group and cluster
gas has been highlighted by the large number of X-ray cavity systems
now known, ranging from elliptical galaxies to the richest clusters
\citep[e.g.][]{boh93,dun04, bir04,mcn07}. Of the nine systems studied here
with {\it XMM-Newton}, 6 have significant detections of cavities
associated with the radio lobes [3C\,296, NGC\,1044, NGC\,4261,
3C\,66B \citep{c03b}, 3C\,449 \citep{c03b} and NGC\,6251
\citep{ev05}]. In several cases where a cavity is not identifiable in
the X-ray data, the emission is nevertheless elongated perpendicular
to the direction of the radio axis (3C\,76.1, NGC\,315), which
suggests a link between the structure of the X-ray emitting gas and
the radio source.

\subsubsection{Correlations between X-ray and radio properties}

While the images presented in the previous sections clearly indicate
an important relationship between the radio and X-ray properties of
these systems, the extent to which the environmental properties
control the large-scale radio morphology remains difficult to
establish observationally. We looked for correlations between the
radio luminosity, radio-source projected physical size, axial ratio,
internal pressure, and the environmental properties as determined from
our X-ray observations (X-ray luminosity, gas temperature, $\beta$,
$r_{c}$, external pressure, ratio of external to internal pressure,
and pressure gradient at the radius of the lobes).
Table~\ref{spearman} lists the results of Spearman rank tests for each
pair of parameters. The strongest correlations are plotted in
Fig.~\ref{correlations}.

We found that radio luminosity is uncorrelated with any of the group
gas properties; however, some correlations were found between external
pressure (measured at the midpoint of the source) and the projected
radio-source size, axial ratio and external pressure. In particular,
correlations that are significant at the $\sim 90$ per cent confidence
level or higher were found between radio source size (in kpc) and both
external pressure and pressure gradient (null hypothesis probabilities
of 1.2 per cent and 7 per cent, respectively), and between axial ratio
and both external pressure and pressure gradient (null hypothesis
probabilities of 5.6 per cent and 9.1 per cent, respectively).
Although these results are not highly significant (and are somewhat
dominated by the two largest sources, NGC\,315 and NGC\,6251: see
Fig.~\ref{correlations}), they suggest a link between radio-source
evolution and group gas properties. However, the correlation of
radio-source size with external pressure and pressure gradient may
simply result from a selection effect: we might expect the largest
sources to probe further out into the group atmospheres, where the
pressure is lower and the gradient likely to be steeper. To test this
explanation, we determined the external pressures at a fixed physical
radius of 300 kpc (comparable to the projected distances probed by the
lobes of NGC\,315 and NGC\,6251), which we consider to be an effective
pressure normalization for the outer regions of each group. If the
correlation between current source size and axial ratio with external
pressure acting on the lobe is due solely to the fact that the larger
sources can reach lower pressure regions of the groups, then we would
expect no systematic difference in the pressure normalization at a
fixed physical radius as a function of source size or axial ratio.
Indeed, this is what we find, and so we conclude that the correlations
with external pressure and its gradient are most likely to be a
selection effect. The lack of correlations between radio source
structure and $\beta$ and $r_{c}$ support this explanation. As radio
source size and axial ratio are themselves strongly correlated (larger
sources tend to have small axial ratios), it is plausible that the
correlations with axial ratio have the same explanation; however, the
size-axial ratio relationship itself is likely to be telling us
something about FRI evolution, and so the different external pressures
and pressure gradients experienced by the smaller, and typically
fatter sources (high pressures and flatter pressure gradients)
compared to the larger, and typically thinner sources (low pressures
and steeper pressure gradients), may be a determining factor in
evolving a large FRI source.

We also find significant correlations between both radio-source size
and axial ratio and the pressure ratio, $P_{ext}/P_{int}$, with null
hypothesis probabilities of 2.9 per cent and 10 per cent,
respectively. We will discuss the relationship between pressure ratio
(which we consider to be a measure of the size of the energetic
contribution required from non-radiating particles -- see below) and
radio-source structure in more detail in Section~\ref{particles}.

The lack of correlations between X-ray luminosity/temperature and any
of the radio properties of these systems indicates that radio-source
structure is independent of group environment properties. Our sample
covers the full range of typical FRI radio morphologies, slightly more
than an order of magnitude in radio luminosity, and as discussed in
the previous section, nearly two orders of magnitude in X-ray
luminosity. We can therefore conclude that, while all of the FRI radio
sources in our sample have been found to inhabit group-scale
environments, the evolution of their radio structure appears to be
unrelated to the richness of the group environment, as parametrized by
the X-ray luminosity and temperature. The only environmental
properties that may be relevant to the source evolution are the group
pressure and pressure distribution.

\begin{figure*}
\centering{\hbox{
   \epsfig{figure=296_radio.ps,height=7cm}
   \epsfig{figure=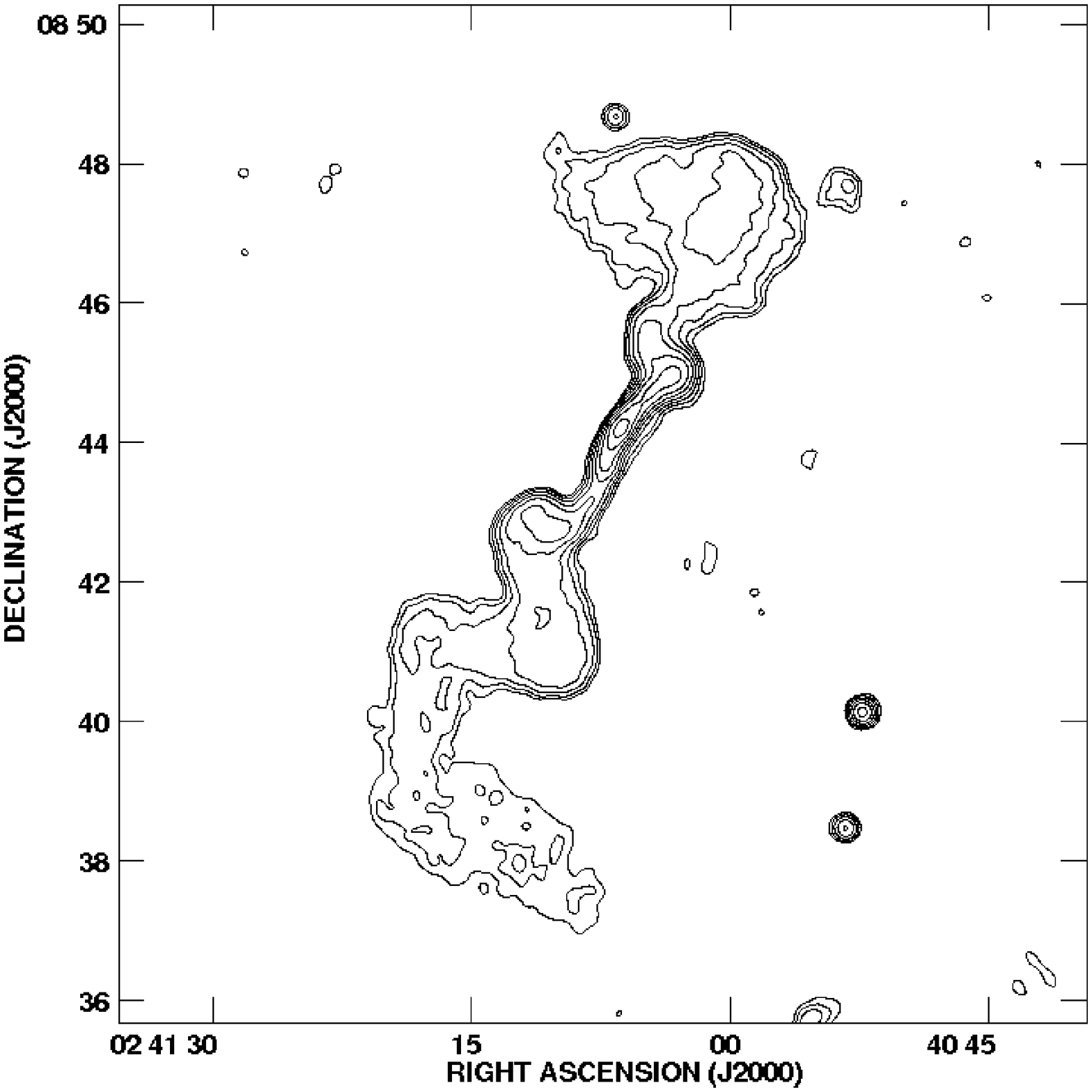,height=7cm}}}
\caption{Examples of the two morphological classes of FRI radio
  galaxies, as described in the text. L: a typical ``bridged'' source,
  3C\,296, in which the lobe material is primarily between the nucleus
  and the end of the collimated jet; R: a typical ``plumed'' source,
  NGC\,1044, in which the lobe material is primarily beyond the end of
  the collimated jet.}
\label{morph}
\end{figure*}

\subsubsection{Radio morphological class and environment}

To investigate the relationship between radio structure and
environment further, we divide the radio sources based on their
large-scale morphology into the two classes of ``plumed'' and
``bridged'' sources, as defined by Leahy et al.
(1998)\footnote{http://www.jb.man.ac.uk/atlas/anatomy.html}. By this
definition, ``plumed'' FRIs are those in which the majority of lobe
material is beyond the end of the collimated jet, whereas in
``bridged'' FRIs, the majority of lobe material is between the end of
the collimated jet and the nucleus. Fig.~\ref{morph} shows radio maps
of a typical member of each class. Based on this definition, we
classified NCG\,1044, 3C\,31, NGC\,315, NGC\,6251, and 3C\,449
as ``plumed'' sources, and 3C\,296, 3C\,76.1, 3C\,66B and NGC\,4261 as
``bridged'' sources. 

We first tested whether the jet power is the controlling parameter in
determining whether a source shows a bridged or plumed morphology by
comparing the 178-MHz radio luminosities of the two subsamples. We
found no difference in the radio luminosity distributions of the two
classes, and so we conclude that environment must play at least some
role in determining the large-scale radio structure. We note that for
this sample the size distributions of the two classes are
significantly different (Fig.~\ref{hists}), with plumed sources being
typically larger than bridged sources. Results of Kolmogorov-Smirnov
(KS) tests comparing the X-ray-measured group properties are shown in
Table~\ref{ks}. We find no significant differences between the
intrinsic group properties of the two subsamples, but do find a
significant difference between the pressure ratios ($P_{ext}/P_{int}$)
for the two subsamples, as discussed in Section~\ref{particles}.

We conclude that the current X-ray data do not conclusively
demonstrate any close relationship between environmental conditions
and radio-source structure. It appears that both plumed and bridged
sources can be produced in a range of environments, although the
difference in typical sizes (and axial ratios) of the two classes,
together with the results discussed in the previous section, suggests
that the two classes may represent an evolutionary sequence in which
the environment plays some role in turning bridge structures into
plumes. An evolutionary model of this sort would require a transition
from one morphology to the other. A combination of buoyancy effects and
collapse due to the end of over-pressuring could produce such a
transition \citep[e.g.][]{h99b}; however, while sources exist in which
one side is bridged and the other plumed (e.g. 3C\,66B), objects in a
state of transition between bridged and plumed structures do not
appear to be common in the FR-I population, which suggests that the
timescale for the transition would need to be short,

\begin{figure*}
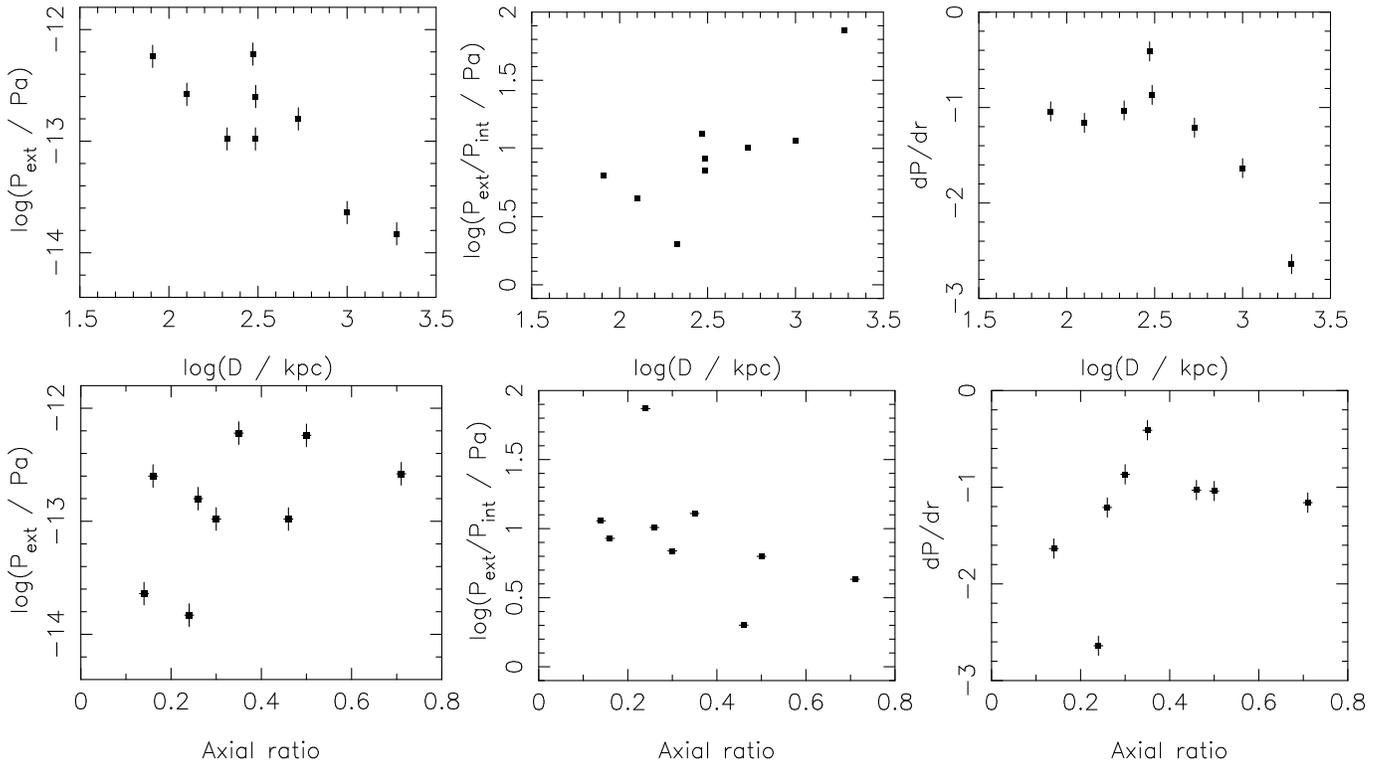

\centerline{\vbox{
\hbox{
\epsfig{figure=dpext.ps,height=5.0cm}
\epsfig{figure=dprat.ps,height=5.0cm}
\epsfig{figure=dpgrad.ps,height=5.0cm}
}
\hbox{
\epsfig{figure=axpext.ps,height=5.0cm}
\epsfig{figure=axprat.ps,height=5.0cm}
\epsfig{figure=axpgrad.ps,height=5.0cm}
}}}
\caption{Correlations between radio structure and group gas
  properties. Top: Size vs. external pressure (left), pressure ratio
  ($P_{ext}/P_{int}$) (middle), and pressure gradient (right); bottom:
  Axial ratio vs. external pressure (left), pressure ratio
  ($P_{ext}/P_{int}$) (middle), and pressure gradient (right).}
\label{correlations}
\end{figure*}

\begin{figure}
\epsfig{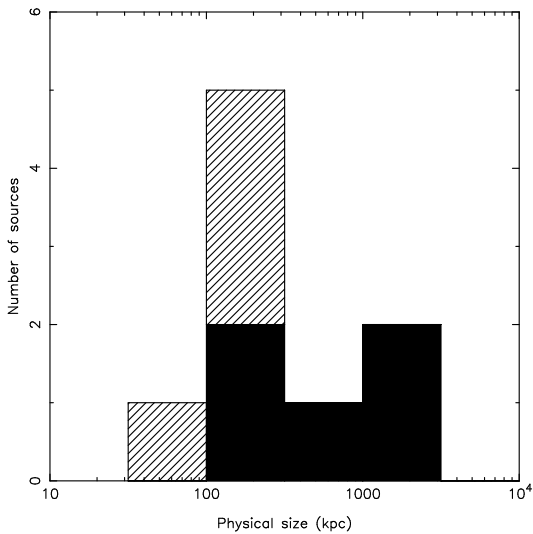}
\caption{Histograms of radio-source size for the bridged (hatched)
  and plumed (solid black) sources, showing a statistically significant
  difference in their distributions.}
\label{hists}
\end{figure}

\subsection{Source dynamics and particle content}
\label{particles}

Fig.~\ref{pressures} shows external pressure profiles for the eight
sources observed with {\it XMM-Newton}, determined from the most
probable pressure at each radius based on the MCMC probability
distribution as discussed in the previous section, with internal,
equipartition pressures overplotted. The plotted uncertainty range
takes into account uncertainties in both the temperature and the
surface brightness model. Internal equipartition pressures were
determined for each radio source, and for different components of the
source where appropriate, using the code of \citet{hbw98}. The
electron energy spectrum was normalized using flux densities measured
from the 1.4-GHz maps of Table~\ref{radiomaps} with 178-MHz flux
densities from the 3CRR catalogue \citep{lai83} included where
available. We assumed $\gamma_{min} = 10$, $\gamma_{max} = 10^{6}$, an
energy injection index of $\delta = 2$ (corresponding to radio
spectral index $\alpha = 0.5$) with a break in the spectrum in the
region of $\gamma = 5 \times 10^{3} - 5 \times 10^{4}$ as appropriate
to fit the radio data points. The equipartition pressures assume no
contribution from protons.

\subsubsection{Particle content and radio morphology}

\begin{figure*}
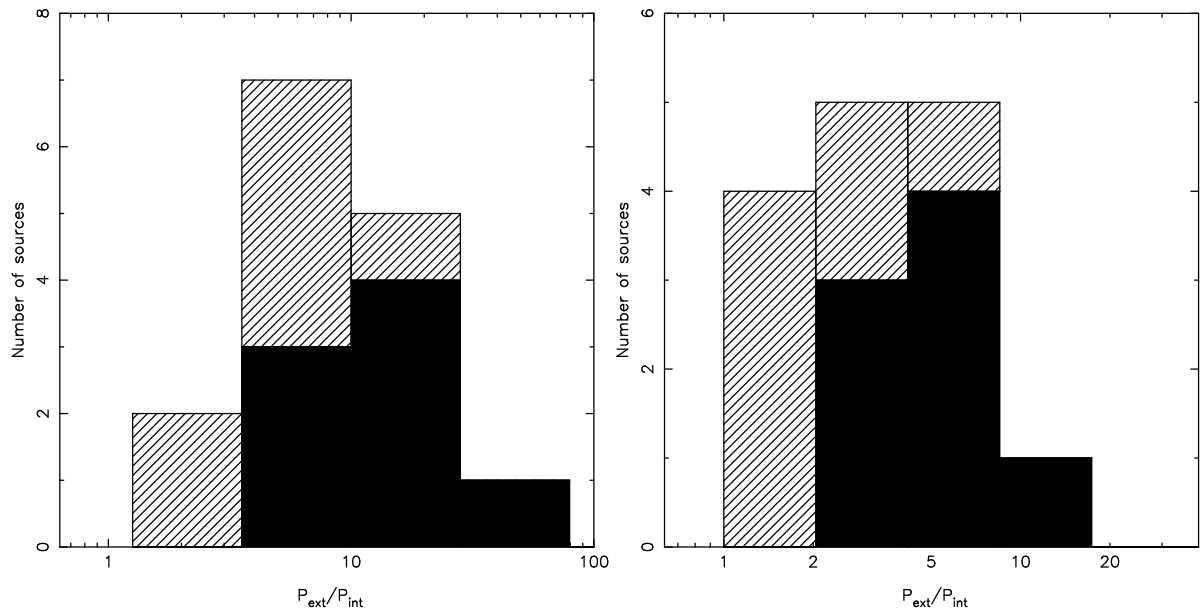

\centerline{\hbox{
\epsfig{figure=pratios_mid.ps,height=8cm}
\epsfig{figure=pratios_edge.ps,height=8cm}}}
\caption{Histograms of the $P_{ext}/P_{int}$, the ratio of external to
internal (equipartition) pressure for the full sample. The left panel
is for pressures at the midpoint of the radio lobe and the right panel
for pressures at the outermost of edge of the radio lobe. The sample
is subdivided into bridged (hatched) and plumed (solid black) lobees.
For all except NGC\,6251 and NGC\,315, both lobes are plotted.}
\label{phists}
\end{figure*}

In agreement with previous studies using earlier X-ray observatories
\citep[e.g.][]{mor88,wb00} and using {\it XMM-Newton}
\citep[e.g.][]{c03b}, we found that the measured external pressures
from the hot gas environments were in many cases significantly higher
than the minimum internal pressures for the radio lobes (if it is
assumed that the sources are in the plane of the sky; relaxation of
this assumption does not affect our conclusions, as discussed below).
Fig.~\ref{phists} (top) shows histograms of the pressure ratios
($P_{ext}/P_{int}$) for the 16 lobes in the sample (only one lobe is
within the {\it XMM-Newton} field-of-view for NGC\,315 and NGC\,6251)
showing that the lobes range from being apparently underpressured by a
factor of $\sim 70$ to approximate pressure balance.

This is the first time that such a study has been carried out for a
sample that includes a wide range of observed FR-I radio morphologies,
and so a primary aim of our work was to investigate whether the amount
of ``missing'' pressure is related to the properties of the
radio-source and/or its environmental interaction. If we consider two
extreme cases, 3C\,296, which is apparently slightly overpressured,
and 3C\,449, which is apparently underpressured by a factor of 10, it
is clear that 3C\,296 is a typical bridged source, whereas 3C\,449 is
plumed. This morphological difference, together with the correlations
between radio-source structure and pressure ratio noted in the
previous section, suggests a model in which jet/environment
interactions affect the particle and/or energy content of the radio
lobes; for example, the dominant factor in determining the ratio of
radiating to non-radiating particles may be the efficiency of the
entrainment process. We would expect that the entrainment rate for
plumed sources is higher, as a larger fraction of the jet surface is
in direct contact with the external medium, whereas a jet that is
embedded in a lobe is not in direct contact with the group gas (except
perhaps very close to the nucleus), and so its entrainment rate might
be expected to be lower. If the amount of entrained material is
sufficient to affect significantly the internal energy density of the
radio lobes (which is likely to require heating of the entrained
matter), then we might expect to see a larger apparent imbalance in
the plumed sources.

The lower panel of Fig.~\ref{phists} shows histograms of the pressure
ratios for the two categories, which demonstrate that there is indeed
a signficant difference in the two populations in the expected sense:
the bridged sources are typically closer to pressure balance at
equipartition than the plumed sources. Using a KS test, we can exclude
the hypothesis that the two subsamples are drawn from the same parent
population at a confidence level of $>97$ per cent. As shown in
Table~\ref{ks} this does not appear to be solely due to differences in
the internal pressures for the two subsamples. We note that the most
extreme pressure ratio in the ``bridged'' category is the eastern lobe
of 3C\,66B, which has a ratio of 11.5. The classification of this
source as bridged is not unambiguous, because the eastern jet does not
enter the lobe for some distance from the nucleus (see, e.g.
\citealt{h96}). All of the other bridged lobes have ratios $<5$, and
so the distinction between the two populations is considerably more
significant if the eastern lobe of 3C\,66B is excluded (a single
parent population can be excluded at the 99.8 per cent confidence
level).

This comparison therefore supports the hypothesis that entrainment of
the jet's surroundings on scales of tens to hundreds of kpc is
responsible for the apparent pressure imbalance of FR-I radio
galaxies. Unfortunately there is no obvious observational proxy for
jet deceleration or entrainment that can be used to test this
hypothesis further with the current data. However, comparisons with
the models of \citet[][and in prep.]{lb02b} show that plumed sources
such as 3C\,31 have an entrainment rate that increases with distance
from the nucleus, whereas bridged sources such as 3C\,296 may not.
Clearly the entrainment characteristics of jets that are directly in
contact with the external medium will be different from those of jets
that propagate inside lobe material. The strong correlations between
pressure ratio and axial ratio discussed in the previous section also
provide support for this model.

\subsubsection{Projection effects}

The pressure comparisons discussed here use projected distances, and
hence the true pressure ratios are likely to be somewhat different.
For a radio source at a small angle to the line of sight, the source
volume will have been underestimated, leading to an overestimate of
$P_{int}$. At the same time, the external pressure acting at the
midpoint and outer edge of the lobe ($P_{ext}$) will be overestimated,
because the radio source will extend further out into the atmosphere
which is decreasing in pressure with radius. Hence the extent to which
the pressure ratio is affected depends on the particular system. In
\citet{c03b}, we argued that for 3C\,449 an angle of $< 11^{\circ}$ to
the line of sight would be required to obtain pressure balance,
whereas the two-sided nature of its jets imply $\theta > 75^{\circ}$
\citep{fer99}. For $\theta = 75^{\circ}$, the pressure ratio decreases
by only 3 per cent from that obtained for the assumption that the
source is in the plane of the sky (from 12.0 to 11.6 for the northern
plume). It is clear that projection alone cannot be reponsible for the
large apparent pressure imbalance in this plumed source. The source
likely to be most affected by projection effects is NGC\,6251, which
has a one-sided jet on large scales. \citet{jon02} argue that $\theta
< 40^{\circ}$, based on parsec-scale limits. We find that an angle of
$< 12^{\circ}$ is required for the pressure ratio at the midpoint of
the western plume of NGC\,6251 to be consistent with the median value
for the bridged sources ($P_{ext}/P_{int} = 3.3$). Such a small angle
cannot be ruled out, but given the large projected size of this
source, it is implausible (an angle of $<12^{\circ}$ over the entire
length of the source leads to a physical size $> 10$ Mpc -- see e.g.
\citealt{bw93}). An angle of $\sim 40^{\circ}$, consistent with the
pc-scale constraints, would give a more modest reduction in pressure
ratio, resulting in $P_{ext}/P_{int} \sim 29$. In addition, estimates
for the line-of-sight angles for other plumed sources (3C\,31 and
NGC\,315) from kinematic modelling of the radio data are not extreme
\citep{lb02a,lai06b}. We conclude that the difference between the
apparent pressure imbalances of bridged and plumed sources is not
solely the result of projection effects.

\subsubsection{Implications}

If the model we propose here to explain the apparent pressure deficit
is correct for this sample of FR-Is, then an obvious question is
whether it is consistent with the pressure offsets observed in cluster
centre radio sources \citep[e.g.][]{mor88,dun04,bir04}, which
typically do not possess well-collimated jets on large scales. Given
the amorphous structures of many cluster centre sources, suggestive of
strong environmental influence, it seems plausible that there is
considerable mixing of cluster gas with radio-lobe plasma (and
heating, as required in order for the cavities to be detected) before
the bubbles rise in the cluster atmosphere. In addition, \citet{dun06}
argue, based on a comparison of the leptonic content of the
small-scale jet and the energetics of the cluster bubbles, that even
in bubbles with large apparent pressure offsets such as those of the
Perseus cluster, the jet is initially leptonic, so that the pressure
offset on large scales is more likely to be due to entrainment rather
than a relativistic proton population, consistent with the picture we
present here. Hence, while the categorization of group-scale sources
into bridged and plumed morphologies with different energetics may not
be directly applicable to cluster centre sources, these results may
nevertheless provide some insight into the solution of the pressure
balance problem for these systems as well.

An alternative model could be proposed in which the strength of the
interaction dictates the level of entrainment. For example, sources
that are overpressured may be less easily disrupted, so that
entrainment levels are low, while sources close to pressure balance
are more susceptible to turbulence and higher entrainment rates. It is
unclear how such a picture would work in detail; however, our data do
not allow us to rule out such a model.

\subsection{Environmental impact}
\label{envimp}

As discussed in Section~\ref{xraydio}, it is clear that the group gas
surrounding FR-I radio galaxies is strongly affected by their
presence. Earlier {\it XMM-Newton} studies of the group environments
of FR-Is found direct evidence for localized heating of the group gas,
and indirect evidence for more generalized heating \citep{c03b}. In
\citet{c05a} we reported an apparent systematic difference in the
radio properties of galaxy groups with and without central radio
sources, in the sense that radio-loud groups appear to be
systematically hotter for a given luminosity. A recent study comparing
REFLEX/NORAS clusters with NVSS has found similar results
\citep{mag07}, while a {\it Chandra} study of a subset of the
\citet{c05a} sample by \citet{jet07} has shown that the gas properties
in the innermost regions of the group do not appear to be
significantly affected by the presence of a radio source, so that any
heating processes must be most significant on larger scales,
comparable to the size of the radio lobes.

\begin{figure*}
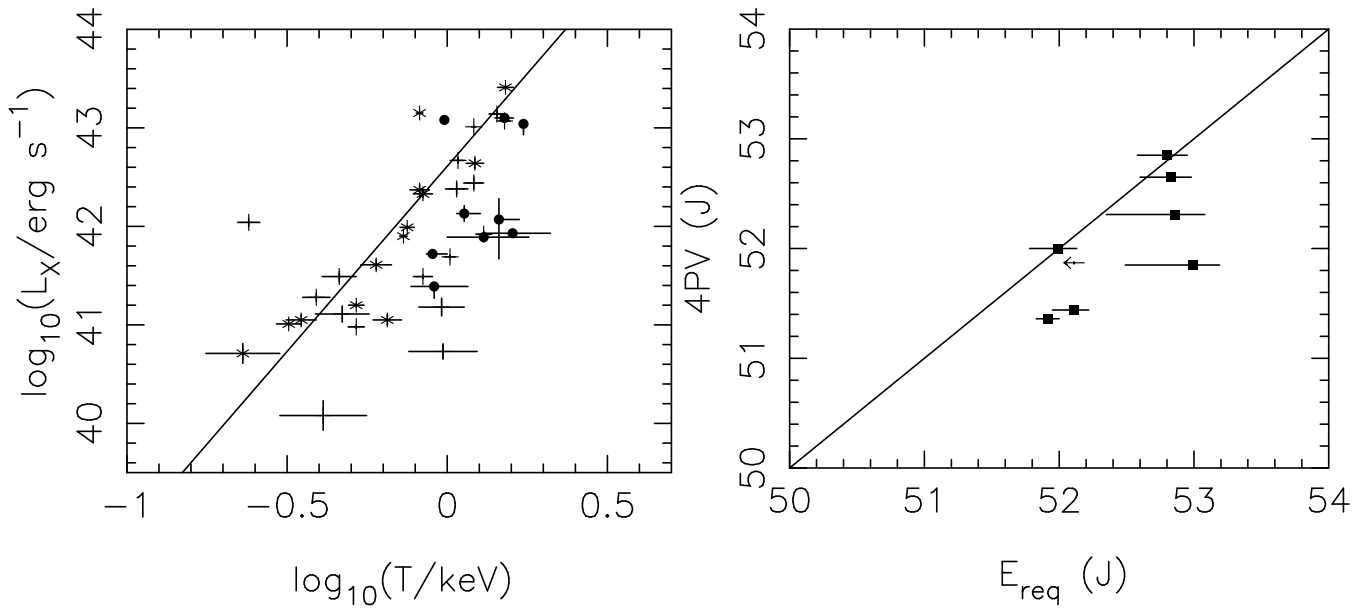

\centering{
 \hbox{
\epsfig{figure=lt_r500.ps,height=8cm}
\epsfig{figure=4pv_ereq.ps,height=8cm}}}
\caption{Left: The $L_{X}/T_{X}$ relation for the {\it XMM-Newton}
    radio-galaxy environments sample (filled circles) compared with
    the radio-quiet (x symbols) and radio-loud (+ symbols) samples of
    \citet{c05a}. Solid line indicates the best-fitting relation for
    radio-quiet groups. Right: The energy required to produce the
    observed temperature offsets ($C \Delta T$) vs. energy available
    from the current radio source ($4PV$). Solid line is the line of
    equality.}
\label{lt}
\end{figure*}

In Fig.~\ref{lt} we show the X-ray luminosity-temperature relation for
the current sample of 9 radio galaxy group environments (filled
circles). The $\times$ and $+$ symbols indicate the radio-quiet and
radio-loud samples of \citet{c05a}, respectively, and the solid line
indicates the best-fitting relation for the radio-quiet groups. The
points for the {\it XMM-Newton} radio-galaxy sample all lie to the
high-temperature side of the radio-quiet relation, thus confirming the
earlier result: groups containing extended radio sources scatter to
higher temperatures for a given X-ray luminosity. In \citet{c05a}, we
argued, based on a lack of difference in the X-ray-to-optical
luminosity ratios for the two subsamples, that this effect must be due
to a temperature increase, rather than the luminosity decrease
expected. The origin of this effect therefore remains unclear. It
would be expected that any energy injected into the environment would
alter both its luminosity and temperature, with proportionate
increases in both if the gas remains in hydrostatic equilibrium. This
is consistent with the position of the FRI groups on the
$L_{X}$--$T_{X}$ diagram. It has been argued that the slope of the
$L_{X}$--$T_{X}$ relation is set mainly by the effects of radiative
cooling \citep{voi04}; the offset associated with radio-loud AGN
activity that we observe here may therefore have the primary effect of
increasing the scatter in the $L_{X}$--$T_{X}$ relation, rather than
altering its slope.

We checked whether it was possible that the presence of extended
non-thermal emission in the radio-galaxy environments could be biasing
our temperature estimates by fitting a combined {\it mekal} plus
power-law model to the groups with the largest temperature excesses,
NGC\,1044 and NGC\,4261, with the temperature fixed at that predicted
by the radio-quiet $L_{X}/T_{X}$ relation. In both cases acceptable
fits with power-law indices between 1.6 and 2 were obtained. In the
case of NGC\,4261 the fit statistic was considerably worse for this
model ($\chi^{2} = 1124$ for 626 d.o.f.), whereas for NGC\,1044 the
fit statistic was very similar to that of the {\it mekal}-only model
($\chi^{2} = 160$ for 129 d.o.f.). The X-ray luminosities of the
non-thermal components were $\sim 6 \times 10^{41}$ erg s$^{-1}$ in
both cases, comparable to the total luminosity of the group. We
cannot, therefore, rule out the presence of extended non-thermal X-ray
emission in all of the groups in the sample from spectral data alone.
To explain the entire ``temperature excess'' in this way would,
however, require a large fraction of the measured X-ray luminosity to
originate in extended non-thermal emission.

Column 14 of Table~\ref{properties} lists the energy required to raise
the temperature of all the group gas within the spherical region
enclosing the radio galaxy by $\Delta T$, the difference between the
observed gas temperature in that region and the temperature predicted
for a group of the same luminosity using the radio-quiet $L_{X}/T_{X}$
relation of \citet{c05a}. To try to understand whether the radio
source is capable of producing these effects, we estimated the total
energy available from the current episode of radio activity as 4$PV$
(this includes the energy stored in the lobes, which cannot yet have
had an effect on the environment), using the total lobe volume and the
pressure at the lobe midpoint. Fig.~\ref{lt} (right) compares the
required and available energies for each source. As shown in
Table~\ref{ks}, there is no significant difference in the ratios of
required to available energy for the bridged and plumed sources. In
five of the eight sources with temperature excesses the radio source
appears to be sufficiently powerful to heat the gas by the required
amount, but in three cases (3C\,31, NGC\,4261 and NGC\,315) the
available energy is significantly less than that required. It has been
argued previously (e.g. Nipoti \& Binney 2006) that if AGN radio
oubursts are an intermittent process, then it is most probable that we
will observe outbursts whose energy injection rate is lower than the
average for the system. However, if the higher than expected
temperatures in these systems are the result of radio-source energy
injection, then they would be expected to disappear on timescales
comparable to the sound crossing time, so it is not clear that
previous outbursts from these AGN can be responsible for the effect we
observe. Another possibility, which is difficult to rule out, is that
radio-loud AGN prefer to inhabit hotter than average group
environments.

\section{Conclusions}

We have carried out the first {\it XMM-Newton} study of a sample of low-power (FR-I) radio galaxies and their
environments. We find the following results:
\begin{itemize}
\item The radio galaxies inhabit group-scale hot-gas environments with
luminosities ranging from $2 \times 10^{41}$ erg s$^{-1}$ to $10^{43}$
erg s$^{-1}$, consistent with earlier studies of FR-I environments.
\item The radio properties of the FR-Is, including radio luminosity,
source size and axial ratio, are uncorrelated with the richness of the
group environments, as parametrized by either X-ray luminosity or
temperature.
\item Marginally significant correlations were found between source
size and external pressure and pressure gradient, and between axial
ratio and both external pressure and pressure gradient, with long,
narrow sources associated with lower pressures and steeper gradients;
however, this could simply be a selection effect, as longer sources
(which tend to be narrower) are able to probe the outer regions of
groups, where the pressure gradient is steeper.
\item In agreement with earlier work, we find that FR-I radio galaxies
are acted on by external pressures that are significantly higher than
their internal, equipartition pressures, with apparent pressure
imbalances ranging from a factor of $70$ times underpressured to rough
pressure balance.
\item We report for the first time that the apparent pressure
imbalance is linked to radio-source morphology: ``plumed'' FR-I
sources typically have large pressure deficits, whereas ``bridged''
FR-Is are closer to pressure balance. We interpret this result as
evidence that ``plumed'' sources have a higher ratio of non-radiating
to radiating particles than ``bridged'' sources, which is likely to be caused by the higher entrainment rate expected for ``plumed'' sources.
\item We find that the temperatures of the FR-I group environments are
typically significantly higher than predicted for their luminosity,
which supports our earlier claim for a temperature excess in
radio-loud groups. In 5/8 sources, the current radio source may be
sufficiently powerful to have produced this excess via heating;
however, in three cases the current source is too weak. We conclude
that radio-source heating remains the most plausible explanation for
this result; however, we cannot rule out an alternative explanation in
which hotter groups are more favourable for the generation of an FR-I
radio source.
\end{itemize}

\section*{Acknowledgments}
We thank Etienne Pointecouteau, Monique Arnaud and Gabriel Pratt for
developing the background subtraction method used in this paper, for
providing filter-wheel closed background datasets, and for useful
discussions about {\it XMM-Newton} background subtraction. We also
thank the referee for helpful comments. MJH thanks the Royal Society
for a Research Fellowship. This work is partly based on observations
obtained with {\it XMM-Newton}, an ESA science mission with
instruments and contributions directly funded by ESA Member States and
the USA (NASA). The National Radio Astronomy Observatory is a facility
of the National Science Foundation operated under cooperative
agreement by Associated Universities.


\label{lastpage}
\end{document}